\documentclass[twoside,french]{book}
\usepackage{babel}
\usepackage{graphicx}
\usepackage{amsmath}
\usepackage[latin1]{inputenc}

\input{tcilatex}

\begin{document}

\title{{\Huge La cin\'{e}matique relativiste }\\
{\Huge sous-jacente }\\
{\Huge \`{a} l'ellipse de Poincar\'{e}}}
\author{{\Large Yves Pierseaux}}
\date{{\large S\'{e}minaires donn\'{e}s \`{a} l'ULB }\\
{\large le vendredi 13 mai 2005}\\
{\large (''Poincar\'{e}'s relativistic kinematics'', iihe) }\\
{\large et le mardi 20 d\'{e}cembre 2005 }\\
{\large (''La cinématique sous-jacente à l'ellipse de Poincaré'')}\\
\textit{Publié dans le bulletin de l'iihe (Inter-University institut for
high energies, iihe ,ULB-VUB) et soumis à l'Académie des Sciences de Paris.}}
\maketitle
\tableofcontents


\section{R\'{e}sum\'{e}}

Ceci n'est pas un essai historique sur l'origine de la relativit\'{e}
restreinte car la cin\'{e}matique d'Einstein date de 1905 et la
cin\'{e}matique sous-jacente \`{a} l'ellipse de Poincar\'{e} date de 2005.
Mieux encore : le front d'onde ellipso\"{i}dal \'{e}mis par une source en
mouvement, d\'{e}crit par Poincar\'{e} en 1908, est impensable dans le cadre
de la cin\'{e}matique einsteinienne car il est observ\'{e} dans le
syst\`{e}me de la source ! La seule mani\`{e}re de d\'{e}fendre l'opinion
selon laquelle Poincar\'{e} aurait trouv\'{e} la cin\'{e}matique d'Einstein
avant Einstein ( ?) est de ne pas parler de l'ellipse allong\'{e}e dont les
seuls ingr\'{e}dients (vitesse de la lumi\`{e}re, temps, espace) sont
pourtant pr\'{e}cis\'{e}ment ceux d'une cin\'{e}matique.

Poincar\'{e} s'est-il tromp\'{e} ? Nous avons fait l'hypoth\`{e}se que
Poincar\'{e} ne s'est pas tromp\'{e} mais qu'il y a une cin\'{e}matique
sous-jacente \`{a} l'ellipse de Poincar\'{e} qui n'est pas celle d'Einstein.
Aucun sp\'{e}cialiste de Poincar\'{e}, jusqu'\`{a} pr\'{e}sent, ne s'est
avis\'{e} d'examiner en profondeur cette \'{e}ventualit\'{e}. Cela est
d'autant plus consternant que Poincar\'{e} n'a jamais accus\'{e} l'Ecole
allemande de relativit\'{e} (Einstein, Planck, Laue, Sommerfeld, Minkowski,
Born) de l'avoir plagi\'{e} mais qu'il a, au contraire soulign\'{e} que
cette derni\`{e}re avait adopt\'{e} une autre convention relativiste que la
sienne (conf\'{e}rence de Londres).

Rendre justice \`{a} Poincar\'{e} consiste donc \`{a} aller dans la
direction qu'il a lui-m\^{e}me indiqu\'{e}e en montrant que la convention
(la d\'{e}finition des unit\'{e}s d'espace et de temps) qu'il a adopt\'{e}e
est parfaitement compatible avec la transformation de Lorentz et donc avec
l'invariance de la vitesse de la lumi\`{e}re. Cet essai est donc un essai
scientifique dont l'objectif est de d\'{e}montrer que la physique a
\'{e}t\'{e} amput\'{e}e d'une th\'{e}orie importante, voire m\^{e}me
fondamentale, en grande partie \`{a} cause de la pol\'{e}mique st\'{e}rile
des priorit\'{e}s.

Notre point de d\'{e}part est une question physico-math\'{e}matique : quelle
est l'image (dans K') par la transformation de Lorentz (TL) d'un front
d'onde sph\'{e}rique (dans K, en translation uniforme par rapport \`{a} K'),
autrement dit d'un front d'\'{e}v\'{e}nements simultan\'{e}s ? La
simultan\'{e}it\'{e} \'{e}tant relative, le front image est un front
ellipso\"{i}dal spatio-temporel. Les sph\`{e}res einsteiniennes sont des
\guillemotleft sph\`{e}res de synchronisation \guillemotright\ qui
d\'{e}finissent pr\'{e}alablement la simultan\'{e}it\'{e} dans les deux
syst\`{e}mes K et K'(annexe 1).

Nous \'{e}tablissons alors (\`{a} deux dimensions spatiales), sur la seule
base de la TL, toutes les caract\'{e}ristiques g\'{e}om\'{e}triques de cette
\guillemotleft\ ellipse d'observation \guillemotright\ (observateur au
foyer, 1\`{e}re partie, voir le plan). Nous retrouvons ainsi notamment les
formules einsteiniennes de transformation relativiste des angles.

Nous utilisons ensuite la tangente \`{a} l'ellipse pour montrer que le
raisonnement sur les ondes sph\'{e}riques (circulaires) peut \^{e}tre
\'{e}tendu aux ondes planes. Il suffit de mettre en contraste la tangente
\`{a} l'ellipse pour le front spatio-temporel image poincar\'{e}en
(relativit\'{e} de la simultan\'{e}it\'{e}) et la tangente au cercle
(perpendiculaire \`{a} la direction de propagation dans les deux
syst\`{e}mes) pour le front image einsteinien (double transversalit\'{e} et
double simultan\'{e}it\'{e}).

On voit ainsi que l'on peut mettre en question les fronts einsteiniens
(sph\'{e}riques ou plans) purement spatiaux (\'{e}v\'{e}nements
simultan\'{e}s dans K et K') exactement comme on peut mettre en question la
rigidit\'{e} pr\'{e}relativiste (\'{e}v\'{e}nements simultan\'{e}s dans K et
K').

Une objection peut alors \^{e}tre formul\'{e}e \`{a} l'\'{e}gard des fronts
spatio-temporels de Poincar\'{e} ?

Quelle est leur signification physique ? Nous montrons alors que l'on peut
d\'{e}duire rigoureusement, sur la base de ces fronts plans (pour les objets
lointains), une formule Doppler (relativiste) qui n'est pas la m\^{e}me que
celle d'Einstein. Les r\'{e}alisations exp\'{e}rimentales r\'{e}alis\'{e}es
\`{a} ce jour sur l'effet Doppler du second ordre ne permettent pas de
trancher entre les deux formules.

Nous montrons \'{e}galement, en pla\c{c}ant un vecteur sur la tangente \`{a}
l'ellipse, que la repr\'{e}sentation einsteinienne de l'onde plane revient
\`{a} annuler la composante longitudinale de ce vecteur, laquelle correspond
pr\'{e}cis\'{e}ment \`{a} \guillemotleft\ l'\'{e}cart \`{a} la
simultan\'{e}it\'{e} \guillemotright\ $\Delta t^{\prime }$sur le front
d'onde image poincar\'{e}en (c'est ce que nous avons appel\'{e}
\guillemotleft\ le th\'{e}or\`{e}me de Poincar\'{e} \guillemotright ).

Les deux conventions diff\'{e}rentes (front spatial selon Einstein et front
spatio-temporel selon Poincar\'{e}) correspondent \`{a} deux conditions de
jauge diff\'{e}rentes au niveau de la th\'{e}orie \'{e}lectromagn\'{e}tique
(\guillemotleft\ structure fine \guillemotright\ de
l'\'{e}lectromagn\'{e}tisme, G. Rousseaux) : Einstein a (implicitement)
adopt\'{e} la jauge transverse (d\^{i}te de Coulomb compl\'{e}t\'{e}e)
tandis que Poincar\'{e} a adopt\'{e} (explicitement) la jauge de Lorenz qui
est coupl\'{e}e avec le quadrivecteur potentiel.

Apr\`{e}s avoir utilis\'{e} la tangente \`{a} l'ellipse d'observation, nous
utilisons alors le second foyer. Nous montrons alors que l'on observe,
depuis le syst\`{e}me en mouvement, (en faisant une moyenne arithm\'{e}tique 
$t_{M}^{\prime }$ \guillemotleft\ aller-retour \guillemotright\ tr\`{e}s
concr\`{e}te sur le plan physique) non seulement une dilatation de la
dur\'{e}e propre (comme Einstein) mais aussi une dilatation de la distance
propre (en contradiction avec Einstein). Nous d\'{e}couvrons alors la source
profonde de cette diff\'{e}rence irr\'{e}ductible dans l'utilisation
respectivement asym\'{e}trique par Einstein et sym\'{e}trique par
Poincar\'{e} de la TL. La contraction einteinienne suppose l'utilisation
d'une seule TL (la TL spatiale) et l'annulation du m\^{e}me terme $\Delta
t^{\prime }$ qui avait \'{e}t\'{e} annul\'{e} avec la composante
longitudinale du front d'onde plan.

La d\'{e}finition poincar\'{e}enne in\'{e}dite de la distance impropre
(\'{e}valu\'{e}e par le temps dilat\'{e} de parcours de la lumi\`{e}re) est
conforme avec la d\'{e}finition de la distance en cosmologie. Nous
remarquons \'{e}galement qu'\`{a} trois dimensions la distance allong\'{e}e
se traduit par une expansion isotrope. La convention de Poincar\'{e} n'est
donc pas autre chose que ces unit\'{e}s spatio-temporelles (propres et
impropres) induite par l'invariance de la vitesse de la lumi\`{e}re
\guillemotleft\ $c=1$ \guillemotright\ dans chaque syst\`{e}me K et K'.

Nous proposons ensuite une synth\`{e}se entre la formule Doppler de
Poincar\'{e} pour les objets lointains d'une part (1-2) et l'expansion
isotrope de l'espace d'autre part (1.3). Un rapprochement heuristique est
sugg\'{e}r\'{e} avec la cosmologie.

Dans la 2\`{e}me partie de l'essai consacr\'{e} \`{a} l'ellipse dans le
syst\`{e}me de la source (d\'{e}duite math\'{e}matiquement par la TL
inverse), nous montrons que dans la convention poincar\'{e}enne de
synchronisation \guillemotleft\ en onde ellipso\"{i}dale \guillemotright ,
la diff\'{e}rence de temps $\Delta t$ entre l'aller et le retour n'est pas a
priori nulle comme elle l'est chez Einstein (dans les deux syst\`{e}mes K et
K', annexe 1). Nous montrons que la somme arithm\'{e}tique $t_{M}$ permet
d'expliquer imm\'{e}diatement le r\'{e}sultat nul de l'exp\'{e}rience de
Michelson tandis que la diff\'{e}rence arithm\'{e}tique $\Delta t$ permet
d'expliquer imm\'{e}diatement le r\'{e}sultat non nul de l'exp\'{e}rience de
Sagnac (annexe 3).

Nous d\'{e}montrons enfin que la contraction de Lorentz poincar\'{e}enne
(les ellipso\"{i}des sont allong\'{e}s parce que les m\`{e}tres sont
contract\'{e}s) est la seule qui est enti\`{e}rement compatible avec la TL
(les deux TL, spatiale et temporelle).

Nous concluons que la \guillemotleft\ structure fine \guillemotright\ de la
relativit\'{e} restreinte est d\'{e}sormais d\'{e}montr\'{e}e. La
cin\'{e}matique d'Einstein avec ses horloges atomiques identiques est
ins\'{e}parable de la th\'{e}orie quantique de la lumi\`{e}re autrement dit
du photon d\'{e}fini en jauge de Coulomb.

Elle r\`{e}gne et r\`{e}gnera encore longtemps sur l'infiniment petit. Par
contre nous d\'{e}veloppons une r\'{e}flexion sur la sym\'{e}trie
spatio-temporelle parfaite de Poincar\'{e} qui pourrrait bien avoir une
application au niveau de l'infiniment grand.

Et nous sugg\'{e}rons ainsi un rapprochement heuristique in\'{e}dit avec le
principe cosmologique de Hoyle.

\section{Introduction historique: front d'onde sph\'{e}rique d'Einstein et
ellipso\"{i}dal de Poincar\'{e}}

Einstein \'{e}crit en 1905 dans le \S 3 de son c\'{e}l\`{e}bre article\cite
{4}:

\begin{quotation}
Soit envoy\'{e}e \`{a} l'instant $t=\tau =0$ \textbf{une onde sph\'{e}rique}
de l'origine commune, \`{a} cet instant, des coordonn\'{e}es des deux
syst\`{e}mes, qui se propage dans le syst\`{e}me K avec la vitesse c. Si (x,
y, z) est un point entra\^{i}n\'{e} par cette onde, on aura

\begin{equation}
x^{2}+y^{2}+z^{2}=c^{2}t^{2}  \tag{a}
\end{equation}

En transformant cette \'{e}quation \`{a} l'aide de nos \'{e}quations de
transformation, nous aurons, apr\`{e}s un calcul simple,

\begin{equation}
\xi ^{2}+\eta ^{2}+\zeta ^{2}=c^{2}\tau ^{2}  \tag{b}
\end{equation}

L'onde en question reste par cons\'{e}quent aussi dans le syst\`{e}me en
mouvement \textbf{une onde sph\'{e}rique} dont la vitesse de propagation est
c.
\end{quotation}

\bigskip Poincar\'{e} \'{e}crit en 1908 dans son deuxi\`{e}me article sur
''la dynamique de l'\'{e}lectron'' sous la rubrique ''le principe de
relativit\'{e}'' \cite{21}:

\begin{quotation}
Supposons un observateur et une source entra\^{i}n\'{e}s ensemble dans la
translation: les surfaces d'ondes \'{e}man\'{e}es de la source seront des
sph\`{e}res ayant pour centres les positions successives de la source; la
distance de ce centre \`{a} la position actuelle de la source sera
proportionnelle au temps \'{e}coul\'{e} depuis l'\'{e}mission,
c'est-\`{a}-dire au rayon de la sph\`{e}re. Toutes ces sph\`{e}res seront
donc homoth\'{e}tiques l'une de l'autre, par rapport \`{a} la position
actuelle F de la source. Mais, pour notre observateur, \`{a} cause de la
contraction,\textit{\ \textbf{toutes ces sph\`{e}res para\^{i}tront des
ellipso\"{i}des allong\'{e}s}, }et tous ces ellipso\"{i}des seront encore
homoth\'{e}tiques par rapport au point F.
\end{quotation}

On trouve alors dans la suite du texte de Poincar\'{e} l'\'{e}quation \`{a}
deux dimensions d'une \textit{ellipse allong\'{e}e} dont l'observateur ''au
repos'' (appelons le O') occupe le centre et dont la source S (avec ''notre
observateur'', appelons-le O) en mouvement occupe le foyer (appelons le F).
Poincar\'{e} conclut que \textit{''cette fois, la compensation est
rigoureuse, et c'est ce qui explique l'exp\'{e}rience de Michelson.'' \cite
{21}.}

\bigskip

\textit{\ }Il consid\`{e}re ainsi l'exp\'{e}rience fameuse o\`{u}
l'observateur Michelson O se trouve avec la source S sur la Terre (K)
suppos\'{e}e en mouvement par rapport \`{a} l'\'{e}ther (K'). Dans le
syst\`{e}me K' o\`{u} l'\'{e}ther est (\textit{par d\'{e}finition }selon
Poincar\'{e}) au repos les ondes \'{e}mises seront sph\'{e}riques. Pour
l'observateur O \guillemotleft\ Michelson \guillemotright\ elles
para\^{i}tront ellipso\"{i}dales allong\'{e}es (\'{e}tant donn\'{e} la
contraction de la tige en mouvement). Poincar\'{e} utilise alors une
propri\'{e}t\'{e} g\'{e}om\'{e}trique de l'ellipse pour montrer que les
temps \guillemotleft\ aller-retour \guillemotright\ (voir annexe 1) dans les
deux directions perpendiculaires seront les m\^{e}mes.

\bigskip Le contraste entre les analyses respectives des deux grands
relativistes, Einstein et Poincar\'{e}, d'une exp\'{e}rience qui semble
\^{e}tre la m\^{e}me est saisissant: selon Einstein l'image d'un front
d'onde sph\'{e}rique est un front d'onde sph\'{e}rique tandis que selon
Poincar\'{e} l'image d'un front d'onde sph\'{e}rique est un front d'onde
ellipso\"{i}dal. Les ondes ellipso\"{i}dales allong\'{e}es de Poincar\'{e}
ont \'{e}t\'{e} presque compl\`{e}tement ignor\'{e}es pendant un si\`{e}cle%
\footnote{%
Il faut signaler cependant un bref \'{e}change \'{e}pistolaire entre Keswani
et Born \cite{2} au sujet, non pas directement de l'ellipse de Poincar\'{e},
mais des deux points longitudinaux de cette derni\`{e}re. Keswani faisait
remarquer que les temps de parcours n'\'{e}tait pas les m\^{e}mes et que
c'\'{e}tait difficile \`{a} expliquer avec l'image sph\'{e}rique
einsteinienne. Nous donnons la partie scientifique (et aimable) de la
r\'{e}ponse de Born en conclusion.}. Constatons d'embl\'{e}e que d'un point
de vue historique \cite{22, 23, 24}:

1) Poincar\'{e} n'utilise pas l\textbf{a transformation de Lorentz (TL)} et
d\'{e}duit le caract\`{e}re ellipso\"{i}dal du front de la contraction des
unit\'{e}s de longueur, \textit{\'{e}tant donn\'{e} que la vitesse de la
lumi\`{e}re est invariante }(il est important de ne pas confondre l'ellipse
de Poincar\'{e}-1904 et l'ellipse de Poincar\'{e}-1906)\footnote{%
Les fronts d'ondes ellipso\"{i}daux de Poincar\'{e} se trouvent d\'{e}j\`{a}
dans son cours de 1905-1906 ''Les limites de la loi de Newton'' \cite{20}.
On les trouve aussi dans la ''M\'{e}canique Nouvelle'' \cite{23} de 1909 et
enfin un cours de Poincar\'{e} de 1912 \cite{24}, tr\`{e}s peu de temps
avant sa mort. Notons que Poincar\'{e} d\'{e}crit d\'{e}j\`{a} un front
ellipso\"{i}dal allong\'{e} dans la conf\'{e}rence St Louis de 1904. Dans ce
dernier cas toutefois l'ellipso\"{i}de allong\'{e} est pr\'{e}sent\'{e}
comme une \textit{alternative} \`{a} la contraction des unit\'{e}s de
longueur et non pas comme une \textit{cons\'{e}quence }de la d\'{e}finition
relativiste des unit\'{e}s. Cet ellipso\"{i}de non-relativiste a fait
l'objet des travaux de Guillaume \cite{23}, Dive \cite{3} et Leroux \cite{9}.%
}.

2) Poincar\'{e} montre la compensation de l'aberration stellaire au 2\`{e}me
ordre sans \'{e}tablir une formule relativiste de changement d'angle.

3) Poincar\'{e} consid\`{e}re au d\'{e}part une s\'{e}rie d'ondes \'{e}mises
par la source sans \'{e}tablir une formule Doppler relativiste.

Nous proposons tout d'abord de pr\'{e}ciser et de d\'{e}velopper tous les
points qui pr\'{e}c\`{e}dent. Signalons que, d'un point de vue historique,
les seules formules relativistes que l'on trouve dans l'article d'Einstein
de 1905 (application de la cin\'{e}matique \cite{4}) et qu'on ne trouve
nulle part dans l'oeuvre de Poincar\'{e} (avant et apr\`{e}s 1905) sont
pr\'{e}cis\'{e}ment les formules d'optique relativiste (effet Doppler et
aberration).

Le cas de figure ci-dessus trait\'{e} par Poincar\'{e} est pour le moins
''non-standard'' (image \textit{elliptique} dans le syst\`{e}me de la\textit{%
\ source, 2\`{e}me partie}) et c'est s\^{u}rement une des raisons pour
laquelle personne ne s'est risqu\'{e} \`{a} \textit{d\'{e}velopper} ce qui
\'{e}tait sous-jacent \`{a} cette ellipse historique de Poincar\'{e}. En
\'{e}tudiant \`{a} pr\'{e}sent le cas standard (image d'un front d'onde 
\textit{circulaire }dans le r\'{e}f\'{e}rentiel de la \textit{source}) nous
quittons le sol de l'histoire pour celui de la physique math\'{e}matique. 
\mainmatter

\part{Ellipse dans le syst\`{e}me de l'observateur (observateur au foyer)}

\chapter{Relativit\'{e} de la simultan\'{e}it\'{e} et ellipse allong\'{e}e}

\section{Image par la transformation de Lorentz d'\'{e}v\'{e}nements sur un
front d'onde circulaire}

Supposons le cas de figure standard o\`{u} la source est au repos en O dans
K et o\`{u} le syst\`{e}me K' est en translation uniforme par rapport \`{a}
K. Quelle est l'image par la TL, dans le syst\`{e}me K', d'un front d'onde
circulaire dans K, \'{e}mis en $t^{\prime }=t=0$\ par cette source S?

Poincar\'{e} \'{e}crit la TL de mani\`{e}re parfaitement sym\'{e}trique (en
x, t) \cite{18}:

\begin{equation}
x^{\prime }=k(x+\varepsilon t)\qquad \qquad \qquad y^{\prime }=y\qquad
\qquad t^{\prime }=k(t+\varepsilon x)
\end{equation}

Les notations historiques de Poincar\'{e} $\varepsilon ,k$\ ($\varepsilon $
repr\'{e}sente la vitesse de translation) correspondent aux notations $\beta
,\gamma $\ d'Einstein-Planck \'{e}tant donn\'{e} que\ Poincar\'{e} choisit ''%
\textit{les unit\'{e}s d'espace et de temps de telle fa\c{c}on que }$c=1$'' 
\cite{18}. Pour la facilit\'{e} du lecteur actuel nous adoptons les
notations standard $\beta ,\gamma $ mais nous maintenons les ''unit\'{e}s
spatiotemporelles $c=1$'' de Poincar\'{e}\footnote{%
Sur la base de l'invariance de la vitesse de la lumi\`{e}re ($c=1$), il est
possible de d\'{e}finir les unit\'{e}s d'espace et de temps de deux
mani\`{e}res diff\'{e}rentes: celle bas\'{e}e sur la th\'{e}orie de la
contraction de Poincar\'{e} et celle bas\'{e}e sur la th\'{e}orie de la
contraction d'Einstein.
\par
{}}:

\begin{equation}
x^{\prime }=\gamma (x+\beta t)\qquad \qquad \qquad y^{\prime }=y\qquad
\qquad t^{\prime }=\gamma (t+\beta x)
\end{equation}

La TL \'{e}tant une transformation ponctuelle (une transformation
d'\'{e}v\'{e}nements), cherchons les images $(x^{\prime },y^{\prime
},t^{\prime })$ dans K' de diff\'{e}rents points ''\'{e}v\'{e}nements'' $%
(x,y,t)$ appartenant au front d'onde circulaire fix\'{e} par $t=t_{0}$ (voir 
\textbf{figure 1}, avec $t_{0}=r_{0}=1)$:

\begin{equation}
x^{2}+y^{2}=r_{0}^{2}=t_{0}^{2}
\end{equation}

Donnons quatre exemples (repris sur la figure1).

\textbf{Points longitudinaux} (voir note 1 et conclusion):
l'\'{e}v\'{e}nement E $(1,0,1)$ a pour image E', $\gamma (1+\varepsilon
),0,\gamma (1+\beta ),$ et se trouve sur l\textit{e plus grand cercle}
pointill\'{e} $x^{\prime 2}+y^{\prime 2}=t^{\prime 2}$ de rayon $r^{\prime
}=t^{\prime }=\gamma (1+\beta );$ l'\'{e}v\'{e}nement A $(-1,0,1)$ a pour
image A', $\gamma (\beta -1),0,\gamma (1-\beta ),$ et se trouve sur \textit{%
le plus petit cercle} pointill\'{e} $x^{\prime 2}+y^{\prime 2}=t^{\prime 2}$
de rayon $r^{\prime }=t^{\prime }=\gamma (1-\beta ).$

\textbf{Point transversal}: l'\'{e}v\'{e}nement C$(0,1,1)$ a pour image C', (%
$\gamma \beta ,1,\gamma ),$ et se trouve sur le cercle pointill\'{e} $%
x^{\prime 2}+y^{\prime 2}=t^{\prime 2}$ de rayon $r^{\prime }=t^{\prime }=$ $%
\gamma .$

\textbf{Point quelconque: }l'\'{e}v\'{e}nement D ($\frac{\sqrt{2}}{2}$, $%
\frac{\sqrt{2}}{2},1)$ a pour image D', $\gamma (\frac{\sqrt{2}}{2}+\beta ),%
\frac{\sqrt{2}}{2},\gamma (1+\frac{\sqrt{2}}{2}\beta ),$ et se trouve sur le
cercle en pointill\'{e} $x^{\prime 2}+y^{\prime 2}=t^{\prime 2}$ de rayon $%
r^{\prime }=t^{\prime }=$ $\gamma (1+\frac{\sqrt{2}}{2}\beta ).$

On constate ainsi que les points-images dans K', contrairement \`{a} une
id\'{e}e tr\`{e}s r\'{e}pandue, ne sont pas situ\'{e}s sur un m\^{e}me front
d'onde circulaire mais, en raison \ m\^{e}me de l'invariance de la forme
quadratique par la TL, dans une ''couronne circulaire'' (cercles \
pointill\'{e}s $x^{\prime 2}+y^{\prime 2}=r^{\prime 2}$ avec $\gamma
(1-\beta )\leq r^{\prime }\leq \gamma (1+\beta ),$ \textbf{figure1}). Nous
avons mis ces cercles en pointill\'{e}s car ils ne repr\'{e}sentent rien de
physique (voir cependant l'onde plane einsteinienne, \S\ 2.2.3). Ce qui est
physique c'est la \textit{simultan\'{e}it\'{e}} des \textit{%
\'{e}v\'{e}nements}. Les quatre \'{e}v\'{e}nements choisis ci-dessus sont
simultan\'{e}s dans $K$ (t est fix\'{e}) mais ils ne sont pas simultan\'{e}s
dans $K^{\prime }$ (t' n'est pas fix\'{e}).

On g\'{e}n\'{e}ralise la repr\'{e}sentation par la TL de tous les points du
front d'onde en introduisant, dans le r\'{e}f\'{e}rentiel K, l'angle $\theta 
$ form\'{e} par le rayon vecteur \textbf{r} et l'axe Ox, avec $x=r\cos
\theta $ et $y=r\sin \theta $ (\textbf{figure 1}). L'image du front est
alors donn\'{e}e soit par une \textit{\'{e}quation temporelle:}

\begin{equation}
t^{\prime }=\gamma t_{0}(1+\beta cos\theta )
\end{equation}

ou par une \textit{\'{e}quation radiale}:

\begin{equation}
r^{\prime }=\gamma r_{0}(1+\beta cos\theta )
\end{equation}

Ces deux derni\`{e}res \'{e}quations (1.4 \& 1.5) constituent une TL pour
les fronts d'onde qui laisse invariante la forme quadratique (3)

\begin{equation}
x^{\prime 2}+y^{^{\prime }2}=r^{\prime 2}=t^{\prime 2}
\end{equation}
Sur les figures nous normalisons ($t_{0}=r_{0}=1)$ toujours: $t^{\prime
}=r^{\prime }=\gamma (1+\beta cos\theta ).$

\section{Relativit\'{e} de la simultan\'{e}it\'{e} et front image elliptique}

A quelle forme g\'{e}om\'{e}trique correspondent ces \'{e}quations (1.4 \&
1.5)? Cherchons l'\'{e}quation de l'ellipse allong\'{e}e de Poincar\'{e} en
coordonn\'{e}es cart\'{e}siennes. Nous cherchons la forme de l'onde $t=t_{0}$
dans K, \'{e}tant donn\'{e} que la forme quadratique prim\'{e}e (1.6) est
invariante. Si t' \'{e}tait fix\'{e}, on aurait \'{e}videmment une onde
circulaire; mais t' d\'{e}pend de x (de l'abscisse).\textbf{\ }Si on
\'{e}crit t' en fonction de x, on n'aura pas la forme ou l'image par la TL
dans K'. Il faut donc \'{e}crire \textbf{t' en fonction de x'. }Un calcul
\'{e}l\'{e}mentaire (1.2) donne: 
\begin{equation}
t^{\prime }=\gamma ^{-1}t_{0}+\beta x^{\prime }
\end{equation}

On obtient donc imm\'{e}diatement l'\'{e}quation prim\'{e}e:

\begin{equation}
x^{\prime 2}+y^{^{\prime }2}=(\gamma ^{-1}t_{0}+\beta x^{\prime })^{2}
\end{equation}

qui est celle d'une ellipse allong\'{e}e (\textbf{figure 2}, ellipse
allong\'{e}e normalis\'{e}e):

L'image dans K' (x', y',\ \textbf{et t'}) du front circulaire de K' est
elliptique parce que t' d\'{e}pend de x' (de l'abscisse) par la TL (\textbf{%
figure 2}). Cette ellipse allong\'{e}e (1.8) (\textbf{figure2}), dont
l'observateur O' est au Foyer F et source S au centre C, peut \'{e}galement
s'\'{e}crire

\begin{equation}
(\gamma ^{-1}x^{\prime }-\beta t_{0})^{2}+y^{\prime 2}=t_{0}^{2}
\end{equation}

Ce qui revient \`{a} remplacer x dans (1.3) par:

\begin{equation*}
x=\gamma ^{-1}x^{\prime }-\beta t_{0}
\end{equation*}

Le front image spatiotemporel est d\'{e}fini par l'ensemble des
\'{e}v\'{e}nements images (la TL est une transformation ponctuelle). L'image
elliptique par la TL traduit directement \textbf{la relativit\'{e} de la
simultan\'{e}it\'{e}} pour tous les \'{e}v\'{e}nements du front (sauf ceux
qui ont la m\^{e}me abscisse, \textit{ce qui prend une importance
consid\'{e}rable dans l'analyse de l'onde plane}, voir chapitre 2).

On peut ''mat\'{e}rialiser'' le front d'onde en imaginant qu'il excite des
atomes dispos\'{e}s en cercle (ABC etc..). autour de la source S dans K: les
atomes seront tous excit\'{e}s en m\^{e}me temps dans K mais pas dans K' (on
peut aussi placer en chaque point un miroir tangent parfaitement
r\'{e}fl\'{e}chissant, voir chapitre 3).

L'\'{e}quation (a) du jeune Einstein (voir introduction historique) est une
sph\`{e}re mais (b) n'est une sph\`{e}re que si t' est fix\'{e} (le
''m\^{e}me temps''). Mais le temps image t' n'est pas fix\'{e} (le
''m\^{e}me temps'') par la TL (\'{e}quation temporelle, 4), il est fix\'{e}
par la convention einsteinienne de synchronisation (voir annexe 1).

Si on affirme que l'image d'un front sph\'{e}rique (autour de O) est un
front sph\'{e}rique (autour de O'), cela implique que deux
\'{e}v\'{e}nements simultan\'{e}s (le ''m\^{e}me temps'') dans K sont 
\textit{encore }simultan\'{e}s\textit{\ par rapport \`{a} K'}
(ind\'{e}pendamment de leur abscisse). Il y a violation flagrante de la
relativit\'{e} de la simultan\'{e}it\'{e}: l'image d'une onde sph\'{e}rique
par la TL \textbf{n'est donc pas} une onde sph\'{e}rique mais une onde
ellipso\"{i}dale\cite{16}.

Il semble d\`{e}s lors y avoir, \`{a} premi\`{e}re vue, une contradiction
entre les ''sph\`{e}res einsteiniennes'' (simultan\'{e}it\'{e} absolue) et
la ''simultan\'{e}it\'{e} relative''. Pourtant ce n'est pas ici que le
probl\`{e}me se pose (voir onde plane, chapitre 2) car les \textbf{%
sph\`{e}res einsteiniennes} ne sont rien d'autre que la \textbf{%
synchronisation einsteinienne} dans les deux syst\`{e}mes K et K'. Le jeune
Einstein, ne pr\'{e}cise pas dans quel syst\`{e}me la source est au repos
(voir citation, introduction historique). Il consid\`{e}re implicitement
deux sources identiques et, comme l'\'{e}ther est supprim\'{e}, il normalise
dans les deux syst\`{e}mes en fixant t \textbf{et en fixant t'}: $%
t_{0}=t_{0}^{\prime }$ $=1$ (voir annexe 1, \cite{14}. Il y a donc une
parfaite coh\'{e}rence entre les fronts d'onde rigides et les tiges rigides 
\textit{identiques} du jeune Einstein: $r_{0}=r_{0}^{\prime }$ $=1$
(d\'{e}finies par deux \'{e}v\'{e}nements simultan\'{e}s dans les deux
syst\`{e}mes). Il n'y a pas de contradiction interne car la convention
einsteinienne de simultan\'{e}it\'{e} (de ''rigidit\'{e}'') est \textit{%
pr\'{e}alable} (son \S 1 et 2) \`{a} la relativit\'{e} de la
simultan\'{e}it\'{e} (de la ''rigidit\'{e}'') d'un syst\`{e}me K (K') \`{a}
l'autre K' (K) (son \S 4). En d'autres termes, l'interpr\'{e}tation
einsteinienne de la TL pr\'{e}suppose une double (dans dans chaque
syst\`{e}me K-K' du \textit{couple} relativiste ) normalisation
(sph\'{e}rique) des unit\'{e}s spatiotemporelles ($r_{0}=r_{0}^{\prime }$ $%
=1 $, $t_{0}=t_{0}^{\prime }$ $=1)$ (voir annexe 1 et chapitre 3).

\section{{Transformation relativiste des angles et }ellipso\"{i}des
allong\'{e}s}

Jusqu'\`{a} pr\'{e}sent on s'est servi de l'\'{e}quation temporelle (1.4)
pour montrer que l'ellipse de Poincar\'{e} n'est rien d'autre qu'une
repr\'{e}sentation \textit{rigoureuse et g\'{e}n\'{e}rale} de la
relativit\'{e} de la simultan\'{e}it\'{e}. Nous pouvons maintenant nous
servir de l'\'{e}quation radiale (1.5) afin d'\'{e}tablir les formules
relativistes de transformation des angles. Etablissons tout d'abord
l'\'{e}quation polaire de l'ellipse \ Cette derni\`{e}re se d\'{e}duit
ais\'{e}ment de l'\'{e}quation cart\'{e}sienne avec le p\^{o}le $O^{\prime }$%
, le foyer $F$\ et l'angle polaire $\theta ^{\prime }$\ d\'{e}fini dans K'.
On a en effet: 
\begin{equation*}
r^{\prime }=\frac{p}{1{\LARGE -}e\cos \theta ^{\prime }}
\end{equation*}

avec les deux param\`{e}tres standard de l'ellipse $e,p:p=a(1-\beta
^{2})=r_{0}\gamma \gamma ^{-2}=r_{0}\gamma ^{-1}.$ L'\'{e}quation
\'{e}quation polaire de l'ellipse allong\'{e}e s'\'{e}crit: 
\begin{equation}
r^{\prime }=\frac{r_{0}\sqrt{1-\beta ^{2}}}{1-\beta \cos \theta ^{\prime }}=%
\frac{r_{0}}{\gamma (1-\beta \cos \theta ^{\prime })}
\end{equation}

avec l'excentricit\'{e} $e=\frac{f}{a}=\frac{\ \gamma \beta }{\ \gamma }%
=\beta $\ et les deux param\`{e}tres standard de la relativit\'{e} $\beta ,$%
\ $\gamma $ on a $a^{2}-f^{2}=b^{2}$ puisque $\gamma ^{2}-\beta ^{2}\gamma
^{2}=1.$

(\textbf{figure 2 }l'ellipse est normalis\'{e}e: $r^{\prime }=\frac{1}{%
\gamma (1-\beta \cos \theta ^{\prime })})$

Il suffit maintenant de comparer avec l'\'{e}quation radiale (1.5) $%
r^{\prime }=\gamma r_{0}(1+\beta cos\theta )$ afin de d\'{e}duire la
transformation relativiste des angles (objectif n$°2$ voir introduction
historique, \cite{16}): 
\begin{equation}
\cos \theta ^{\prime }=\frac{\cos \theta +\beta }{1+\beta \cos \theta }%
\qquad \sin \theta ^{\prime }=\frac{\sqrt{1-\beta ^{2}}}{1+\beta \cos \theta 
}\sin \theta \qquad tg\text{ }\theta ^{\prime }=\frac{\sqrt{1-\beta ^{2}}}{%
\cos \theta +\beta }\sin \theta
\end{equation}

\textit{L'\'{e}quation radiale (5) repr\'{e}sente donc bien l'\'{e}quation
d'une ellipse allong\'{e}e (cqfd)}$.$ Le lien indissociable entre la TL (1.4
\& 1.5) et l'ellipse allong\'{e}e est ainsi compl\`{e}tement \'{e}tabli. On
peut d\'{e}finir l'ellipse allong\'{e}e aussi bien par les espaces parcourus
que par les temps de parcours:

\begin{equation}
r^{\prime }=\frac{r_{0}}{\gamma (1-\beta \cos \theta ^{\prime })}\text{%
\qquad\ \ \ \ }t^{\prime }=\frac{t_{0}}{\gamma (1-\beta \cos \theta ^{\prime
})}  \tag{1.10bis}
\end{equation}

L'\'{e}quation temporelle de l'ellipse (1.4) traduit physiquement la
relativit\'{e} de la simultan\'{e}it\'{e} tandis que l'\'{e}quation radiale
(1.5) traduit physiquement la transformation relativiste de l'angle.

La g\'{e}n\'{e}ralisation \`{a} trois dimensions des ellipses en
ellipso\"{i}des est \'{e}vidente puisqu'il suffit de passer des
coordonn\'{e}es polaires aux coordonn\'{e}es sph\'{e}riques et que l'angle
azimutal n'est pas modifi\'{e} par la TL $(\phi =\phi ^{\prime })$. Les
ellipso\"{i}des \textit{d'observation }de Poincar\'{e} traduisent
directement la transformation relativiste de\textit{\ l'angle solide }$%
d\Omega $ d'ouverture du c\^{o}ne de lumi\`{e}re:

\begin{equation*}
d\Omega ^{\prime }=\frac{1-\beta ^{2}}{(1+\beta \cos \theta )^{2}}d\Omega
\end{equation*}

L'ellipso\"{i}de \textit{d'observation }donne ainsi une repr\'{e}sentation\
physique, \textit{globale et concr\`{e}te\footnote{%
On se contente g\'{e}n\'{e}ralement de donner quelques transformations
d'angles. Ainsi on voit imm\'{e}diatement que dans le cas o\`{u} $\theta
^{\prime }=\frac{\pi }{2}$ (ouverture d'une demi-sph\`{e}re), on trouve que
l'angle $\theta $ du c\^{o}ne de lumi\`{e}re est donn\'{e} $\cos \theta =%
\frac{v}{c})$},} du retr\'{e}cissement \ \textit{observ\'{e} }de l'angle
d'ouverture d'un c\^{o}ne de lumi\`{e}re emis par une source S en mouvement
autrement dit ''l'effet de t\^{e}te de la lumi\`{e}re'' ou \textbf{''l'effet
headlight''}(ondes sortantes isotropes de la source deviennent non-isotropes
pour l'observateur). Ces effets physiques de l'ellipso\"{i}de,
sp\'{e}cifiquement relativistes (\textit{second ordre}), sont observ\'{e}s
par exemple dans le rayonnement de freinage ou le rayonnement synchrotron.

\section{Ellipses isotropes ''c=1'' et 4-vecteurs du genre lumi\`{e}re}

Pr\'{e}cisons d'embl\'{e}e ce que nous entendons par ellipse ''isotrope'' en
rappelant que l'ellipse de Poincar\'{e} (avec source au foyer) ''1906'' ne
doit pas \^{e}tre confondue avec celle de 1904 (voir note 2, qui traduit une
anistropie au niveau de la vitesse de la lumi\`{e}re). Insistons sur le fait
que l'ellipse allong\'{e}e est enti\`{e}rement construite sur l'invariance $%
c=1$ de la vitesse de la lumi\`{e}re. Il s'agit d'une ellipse ISOTROPE au
sens o\`{u} la vitesse de la lumi\`{e}re est identique dans toutes les
directions dans les deux syst\`{e}mes car on a par CONSTRUCTION:

\begin{equation}
\frac{r_{0}}{t_{0}}=\frac{r^{\prime }}{t^{\prime }}=c=1
\end{equation}

L'ellipse est isotrope aussi bien au sens ''one-way speed of light'' que
''two-ways speed of light'' (chapitre 3) dans la terminologie anglo-saxone.
Les effets ''anisotropes'' se manifestent\ physiquement au niveau du temps
(relativit\'{e} de la simultan\'{e}it\'{e}) et de l'espace (r\'{e}duction du
c\^{o}ne d'ouverture). Voyons comment l'on peut traduire
math\'{e}matiquement l' isotropie poincar\'{e}enne. Nous avons montr\'{e}
dans le paragraphe pr\'{e}c\'{e}dent que les points d'un front d'onde
circulaire$\ (r=r_{0}=t=t_{0\text{ }},$ $c=1)$ deviennent \textit{\ par la TL%
} les points d'un front d'onde elliptique, d'\'{e}quation (1.5) $r^{\prime
}=\gamma r(1+\beta \cos \theta )$. Cela revient \`{a} \'{e}crire que le
quadrivecteur isotrope du genre lumi\`{e}re (intervalle nul ou isotrope:$\
r^{2}-t^{2}=0$) (1.3), 
\begin{equation}
(r\cos \theta ,\text{ }r\sin \theta ,\text{ }0,\text{ }t)
\end{equation}
se transforme par la TL (directe, source au centre \textbf{figure 3}) 
\begin{equation}
r^{\prime }\cos \theta ^{\prime }=\gamma (r\cos \theta +\beta t)\qquad
r^{\prime }\sin \theta ^{\prime }=r\sin \theta \qquad t^{\prime }=\gamma
(t+\beta r\cos \theta )  \tag{1.13bis}
\end{equation}
en le quadrivecteur isotrope ($r^{\prime 2}-t^{\prime 2}=0$) (6) du genre
lumi\`{e}re: 
\begin{equation}
(r^{\prime }\cos \theta ^{\prime },\text{ }r^{\prime }\sin \theta ^{\prime },%
\text{ }0,\text{ }t^{\prime })
\end{equation}

On v\'{e}rifie sur base de la transformation des composantes (1.13bis) du
4-vecteur que l'on obtient avec l'\'{e}quation du front elliptique (1.5), la
formule de transformation relativiste de l'angle (1.11). En prenant la TL 
\textit{inverse} (on intervertit prime et non-prime et on change la signe de
\ la vitesse $\beta $) on \textit{inverse} les r\^{o}les respectifs de la
source S et de l'observateur O'. On obtient alors l'ellipse allong\'{e}e, 
\begin{equation}
r=\gamma r^{\prime }(1-\beta \cos \theta ^{\prime })\qquad \qquad t=\gamma
t^{\prime }(1-\beta \cos \theta ^{\prime })
\end{equation}
dans le syst\`{e}me de la source (l'ellipse historique de Poincar\'{e}),
laquelle se trouve au deuxi\`{e}me foyer de l'ellipse. L'interpr\'{e}tation
quadrivectorielle de l'ellipse allong\'{e}e est fondamentale: la \textit{%
transformation} de Lorentz (directe $\beta $ ou inverse $-\beta $) d'un
cercle est une \textit{transformation} ''elliptique''. Cette \textbf{%
sym\'{e}trie} math\'{e}matique entre les deux ellipses (figure 3 et 4, le
cas standard et le cas non-standard) sera\ examin\'{e}e physiquement, dans
tous les d\'{e}tails, dans la deuxi\`{e}me partie de la pr\'{e}sente
\'{e}tude.

\textbf{RESUME} (chapitre 1) y a deux interpr\'{e}tations physiques
diff\'{e}rentes de l'\'{e}quation (6) $x^{\prime 2}+y^{^{\prime
}2}=r^{\prime 2}=t^{\prime 2}:$

1) On ne fixe pas t' et (1.6) est l'\'{e}quation d'un front d'onde
elliptique (ellipsoidal) spatiotemporel (un ensemble d'\'{e}v\'{e}nements
simultan\'{e}s ne reste pas un ensemble d'\'{e}v\'{e}nements
simultan\'{e}s). C'est la m\'{e}trique de Poincar\'{e}.

2) On fixe t' et (1. 6) est l'\'{e}quation d'un front d'onde circulaire
(sph\'{e}rique) spatial ou rigide (un ensemble d'\'{e}v\'{e}nement
simultan\'{e}s reste un ensemble d'\'{e}v\'{e}nements simultan\'{e}s). C'est
la m\'{e}trique d'Einstein (1905). Nous montrerons que ce contraste pour les
fronts sph\'{e}riques s'applique aussi au fronts plans et qu'il se traduit
au niveau de l'\'{e}criture de la phase d'une onde plane (voir le
r\'{e}sum\'{e} onde plane). Cette distinction est importante car il ne faut
pas confondre ''m\'{e}trique'' au sens de l'invariance (1.3 \& 1.6) et
''m\'{e}trique'' au sens (''R\'{e}volution fran\c{c}aise'') de la
d\'{e}finition du m\`{e}tre (chapitre 3). Remarquons que l'ellipse a
\'{e}t\'{e} enti\`{e}rement induite par la TL avec les unit\'{e}s $c=1$ et
donc il faut s'attendre \`{a} ce que la sym\'{e}trie spatiotemporelle de
cette derni\`{e}re se manifeste \`{a} travers les propri\'{e}t\'{e}s de
l'ellipse.

\chapter{Onde plane et formule Doppler de Poincar\'{e}}

Tout ce qui suit est enti\`{e}rement d\'{e}duit des propri\'{e}t\'{e}s
g\'{e}om\'{e}triques de l'ellipse allong\'{e}e standard d\'{e}finie au
chapitre1. Nous allons utiliser dans le chapitre 2 essentiellement les
propri\'{e}t\'{e}s de la tangente \`{a} une ellipse et dans le chapitre 3
les propri\'{e}t\'{e}s des foyers (et des directrices) d'une ellipse.

\section{''Quadrivecteur d'Onde'' et formule ''Doppler'' de Poincar\'{e}}

Il est bien connu que la transformation relativiste des angles (l'aberration
stellaire) est coupl\'{e}e chez Einstein \`{a} la formule Doppler
relativiste (1905, \S 7). Il est en outre \'{e}vident que la situation 
\textit{physique} envisag\'{e}e (l'\'{e}mission de fronts d'onde
sph\'{e}riques par une source vus par un observateur en mouvement) doit
donner un ''Doppler''.

Supposons une succession de $n$ fronts d'onde \'{e}mis par la source de K.
Il est alors naturel de caract\'{e}riser cette succession de $n$ fronts
d'onde par une longueur d'onde $\frac{r}{n}=\lambda $ et par une p\'{e}riode$%
\frac{t}{n}=T=\frac{1}{\nu }.$ On peut associer \`{a} ces ondes un
quadrivecteur $(\mathbf{\Lambda },T)$ que l'on appelera ''le 4-vecteur
d'onde de Poincar\'{e}'' $(c\neq 1)$:

\begin{equation}
(\lambda \cos \theta ,\text{ }\lambda \sin \theta ,\text{ }0\text{, }%
cT)_{poincar\acute{e}}
\end{equation}

de norme nulle

\begin{equation*}
\left\| (\lambda \cos \theta ,\text{ }\lambda \sin \theta ,0\text{, }%
cT)\right\| =\lambda ^{2}-c^{2}T^{2}=0
\end{equation*}

Le 4-vecteur spatiotemporel du genre lumi\`{e}re de Poincar\'{e} (1.13) est
donc directement traduit en ''\textit{quadrivecteur d'onde'' (2.1) }(\textbf{%
figure 5}). Nous avons mis ''quadrivecteur d'onde'' entre guillemets pour
des raisons qui sont explicit\'{e}es \`{a} la fin du pr\'{e}sent paragraphe
(voir ''restriction''). C'est du reste \'{e}vident que l'on doit associer 
\textit{d'un point de vue classique} une ''droite de lumi\`{e}re''
(direction de propagation $\mathbf{r=\Lambda }$, dimension inverse du
vecteur d'onde $\mathbf{k}$) \`{a} une ONDE de lumi\`{e}re.

Rappelons que le ph\'{e}nom\`{e}ne Doppler est fond\'{e} sur le fait que le
temps de parcours entre deux ondes successives n'est pas le m\^{e}me selon
que l'observateur est en mouvement (T' dans K') par rapport \`{a} la source
ou au repos par rapport \`{a} la source (T dans K). \textit{La fr\'{e}quence
Doppler\ pr\'{e}relativiste est alors donn\'{e}e par l'inverse de ce temps}
de parcours pour l'observateur au repos et l'observateur en mouvement
(l'angle $\theta $ ou $\theta ^{\prime }$ est d\'{e}fini entre la
''direction du mouvement'', $Ox$ ou $Ox^{\prime }$, et la ''direction de
propagation'' $\mathbf{r=\Lambda }$ ou $\mathbf{r}^{\prime }\mathbf{=\Lambda 
}^{\prime }$).

Le ''quadrivecteur d'onde'' de Poincar\'{e} dans K se transforme en le
''quadrivecteur d'onde'' dans K', $c=1$:

\begin{equation}
(\mathbf{\Lambda },T)\rightarrow (\mathbf{\Lambda }^{\prime },T^{\prime }) 
\tag{2.1bis}
\end{equation}

Autrement dit les fronts circulaires successifs, se transforment selon
l'observateur O', en fronts elliptiques successifs. On a, avec $\frac{%
r^{\prime }}{n}=\lambda ^{\prime }$ et $\frac{t^{\prime }}{n^{\prime }}%
=T^{\prime }=\frac{1}{\nu ^{\prime }},$ (voir figure, $r^{\prime }=2\lambda
^{\prime },$...$n^{\prime }\lambda ^{\prime })$: 
\begin{equation}
n=n^{\prime }
\end{equation}
Les composantes du 4-vecteur (14) se transforment alors:

\begin{equation}
\lambda ^{\prime }\cos \theta ^{\prime }=\gamma (\lambda \cos \theta +\beta
T)\qquad \lambda ^{\prime }\sin \theta ^{\prime }=\lambda \sin \theta \qquad
T^{\prime }=\gamma (T+\beta \lambda \cos \theta )  \tag{2.1ter}
\end{equation}

qui donnent imm\'{e}diatement $(\lambda \nu =\frac{\lambda }{T}=c=1)$ les
formules relativistes de transformation de la longueur d'onde et de la
p\'{e}riode d'une onde (voir 1.4 et 1.5).

\begin{equation}
\lambda ^{\prime }=\gamma \lambda (1+\beta \cos \theta )\qquad T^{\prime
}=\gamma T(1+\beta \cos \theta )  \notag
\end{equation}

Autrement dit au niveau des fr\'{e}quences:

\begin{equation}
\nu ^{\prime }=\nu \frac{1}{\gamma (1+\beta \cos \theta )}
\end{equation}

\textit{C'est la formule ''Doppler'' de Poincar\'{e}}. La transformation de
la fr\'{e}quence (longueur d'onde) d'une onde lumineuse ($\lambda ,$ $\nu )$
est donc imm\'{e}diatement impos\'{e}e par la TL temporelle (spatiale) (1.2$%
) $ $(r,$ $t)$.

Evidemment puisque c'est l'invariance de la vitesse de l'\textbf{ONDE}
lumineuse ($\lambda $, $T),$ (\textit{''la longueur-p\'{e}riode''} d'onde)
qui impose une nouvelle structure relativiste de $(r,$ $t)$ ''\textit{%
l'espace-temps}''.

En adoptant l'angle de l'observateur $\theta ^{\prime }$ qui \textit{re\c{c}%
oit }(ce qui est naturel pour un Doppler) les fronts d'onde
ellipso\"{i}daux, on a alors, compte tenu de la transformation relativiste
de l'angle (1.11) (source au repos et observateur en mouvement):

\begin{equation}
\nu ^{\prime }=\gamma \nu (1-\beta \cos \theta ^{\prime })
\end{equation}

La sym\'{e}trie source-observateur est parfaite puisque en supposant que
c'est la source qui est en mouvement on a par (1.15) $\nu =\gamma \nu
^{\prime }(1+\beta \cos \theta ),$ qui est identique \`{a} (2.4). (\textbf{%
figure 4}). Pour \'{e}tablir le caract\`{e}re relativiste de la formule
''Doppler'' de Poincar\'{e}, il faut prendre en compte l'ellipse dans le
syst\`{e}me de la source (1.15 et deuxi\`{e}me partie).

Notre objectif numero 3 est ainsi accompli (voir introduction historique) et
on s'attend \`{a} trouver la m\^{e}me formule Doppler que celle d'Einstein.

Or la formule (2.4) \textbf{n'est pas la m\^{e}me formule que celle
d'Einstein} (2.5) o\`{u} l'on a, avec une formule structurellement
identique, non pas l'angle $\theta ^{\prime }$ du syst\`{e}me de
l'observateur, mais l'angle $\theta $ du syst\`{e}me de la source 
\cite[paragraphe 7]{4}: 
\begin{equation}
\nu ^{\prime }=\gamma \nu (1-\beta \cos \theta )
\end{equation}

\textit{Pourquoi cette diff\'{e}rence?} Einstein d\'{e}finit,
ind\'{e}pendamment du 4-vecteur du genre lumi\`{e}re (1.13), un autre
4-vecteur d'onde (\textit{renvers\'{e}} par rapport \`{a} celui de
Poincar\'{e}): 
\begin{equation}
(\lambda ^{-1}\cos \theta ,\lambda ^{-1}\sin \theta ,\text{ }0,\text{ }%
c^{-1}T^{-1})_{einstein}\qquad ou\qquad (\frac{1}{\lambda }\cos \theta ,%
\frac{1}{\lambda }\sin \theta ,\text{ }0,\text{ }\frac{\nu }{c})_{einstein}
\end{equation}
qu'il peut associer \`{a} son \textit{hypoth\`{e}se quantique} sur la
lumi\`{e}re (la fr\'{e}quence se transforme \textit{comme le temps} et non
pas comme l'inverse du temps (voir les travaux de Louis de Broglie),
autrement dit le \textbf{PHOTON} $"E=h\nu "$ (proportionalit\'{e} du
4-vecteur impulsion avec le 4-vecteur d'onde einsteinien (2.6)).

Einstein fait donc un choix quantique \cite{13} qui s'\'{e}loigne d'une
conception classique purement ondulatoire de la lumi\`{e}re; le choix
einsteinien suppose en outre une certaine mani\`{e}re d'introduire la
fr\'{e}quence de la source correspondant \`{a} une certaine mani\`{e}re
(compatible avec 21) de d\'{e}finir math\'{e}matiquement la phase invariante
d'une onde plane (produit scalaire voir chapitre 2). Pour le doppler
longitudinal, les formules d'Einstein et de Poincar\'{e} sont les m\^{e}mes:

\begin{equation}
\nu ^{\prime }=\nu \sqrt{\frac{1-\beta }{1+\beta }=}\gamma (1-\beta )
\end{equation}

Les deux formules (2.4 \& 2.5) sont compatibles avec les exp\'{e}riences
\`{a} ce jour r\'{e}alis\'{e}es \cite{10}, \cite{6}. Ces derni\`{e}res
mesurent l'effet quadratique $\gamma $ \textit{au voisinage du doppler
longitudinal} $(1-\beta )$ pour lequel les deux formules sont identiques. En
fait les exp\'{e}riences r\'{e}alis\'{e}es v\'{e}rifient (2.7)\footnote{%
Mandelberg \cite{10} \'{e}crit en 1962 ''The wavelength of the light emitted
from an oncomming atome of velocity \ v viewed at a small angle $\theta ,$
near 0 to the beam direction by an observer stationary in the laboratory
reference ($\theta \sim $0,075 rad). Les formules d'Einstein et de
Poincar\'{e} donnent quasiment la m\^{e}me correction quadratique. ''There
are many reasons why a perpendicular observation of the beam is not feasible
among these being the difficulty ... to reduce the first order Doppler
shift''. Plus r\'{e}cemment (1985) voir Kaivola \cite{7}, le constat
dress\'{e} par Mandelberg reste valable. Parmi les raisons \'{e}voqu\'{e}es
par Mandelberg, il y a aussi une raison th\'{e}orique (voir Pierseaux,
''Einstein's relativistic Doppler formula'' \cite{17}).}. Elle ne permettent
pas de trancher entre (2.4) et (2.5).

En r\'{e}sum\'{e}, de la m\^{e}me mani\`{e}re que l'on d\'{e}duit du
4-vecteur d'onde einsteinien (2.6) des transformations relativistes de
l'angle et de la fr\'{e}quence (2.5), on d\'{e}duit du 4-vecteur d'onde
poincar\'{e}en (2.1) des transformations relativistes de l'angle et de la
fr\'{e}quence (2.4). Celles-l\`{a} sont les m\^{e}mes tandis que celles-ci
ne sont pas les m\^{e}mes.

\bigskip \textbf{Restriction.} \ Nous avons mis ''Quadrivecteur d'onde'' et
''Doppler'' entre guillemets pour exprimer la restriction suivante (voir
annexe 2).

Le quadrivecteur d'onde einsteinien d\'{e}crit la transformation d'une onde
plane en une onde plane, autrement dit la direction de propagation de l'onde
\'{e}tant constante, on obtient une formule Doppler au sens physique le plus
\'{e}vident.

Le ''quadrivecteur d'onde de Poincar\'{e}'' d\'{e}crit la transformation
d'une s\'{e}rie d'ondes sph\'{e}riques en ondes ellipso\"{i}dales, et, comme
la direction de propagation de l'onde n'est pas constante (l'angle $\theta $
ou $\theta ^{\prime })$, il faut faire une hypoth\`{e}se suppl\'{e}mentaire
pour avoir une formule Doppler au sens strict du terme . Nous montrerons que 
\textit{la tangente \`{a} l'ellipse} permet de d\'{e}finir l'image
spatiotemporelle d'un front plan et que par cons\'{e}quent le 4-vecteur de
Poincar\'{e} (2.1 sans guillemets) correspond bien \`{a} une serie de fronts
plans et doit donc donner une formule Doppler sans aucune restriction (2.4
sans guillemets).

\section{Image par la TL d'un front d'onde plane: plan tangent \`{a} un
ellipso\"{i}de (Poincar\'{e}) ou \`{a} une sph\`{e}re (Einstein)}

Jusqu'\`{a} pr\'{e}sent nous avons seulement trait\'{e} la question de
l'image d'une onde sph\'{e}rique. Qu'en est-il maintenant de l'image de
l'onde plane? Supposons une deuxi\`{e}me source S$_{2}$ \`{a} l'infini (on a
n$_{2}\rightarrow \infty ,$ voir 17$)$ au repos dans K dans la direction $%
\theta $ (\textbf{figure 6}) qui \'{e}met au m\^{e}me rythme que S$_{1}$

Supposons que le front plan consid\'{e}r\'{e} passe en O \`{a} l'instant $%
t=t^{\prime }=0$ (\`{a} l'instant o\`{u} S$_{1}$ \'{e}met l'onde
sph\'{e}rique). Il sera en $t=1$ tangent au front circulaire, \'{e}mis par
la source S$_{1},$ $t=r=1$ (\textbf{figure 6}).

Cherchons tout d'abord l'\'{e}quation du front d'onde plan (ici \`{a} 2
dimensions spatiales: \textit{une droite d'onde}) autrement dit le front
objet dans K (\textbf{figure 6})\bigskip . L'\'{e}quation du rayon passant
par E du cercle $t=1$ (angle $\theta )$ est $x\sin \theta -y\cos \theta =0.$
La tangente au cercle au point E est (figure 6 normalis\'{e}e on a $t=r=1)$:

\begin{equation}
x\cos \theta +y\sin \theta =t
\end{equation}

avec le coefficient angulaire $a=-\cot g$ $\theta .$ On a donc \`{a} ce
stade d\'{e}fini \ notre ''objet'' autrement dit le front d'onde plan dans
le syst\`{e}me K (qui est le m\^{e}me chez Einstein et chez Poincar\'{e}).

Cherchons \`{a} pr\'{e}sent l'image par la TL du front dans le syst\`{e}me
K'. Comme la TL est une transformation ponctuelle (d'\'{e}v\'{e}nements),
cherchons d'abord les images des points-\'{e}v\'{e}nements (exactement comme
pour le front d'onde circulaire, \S 1.1). L'image par la TL de
l'\'{e}v\'{e}nement E \ ($x_{E}=t\cos \theta ,$ $y_{E}=t\sin \theta ,$ t)
est l'\'{e}v\'{e}nement E' (sur l'ellipse):

\begin{equation*}
x_{E}^{\prime }=\gamma t(\cos \theta +\beta )\qquad y_{E}^{\prime }=t\sin
\theta \qquad t_{E}^{\prime }=\gamma t(1+\beta \cos \theta )
\end{equation*}
Cherchons l'image des \'{e}v\'{e}nements G ($x_{G}=\frac{t}{\cos \theta }%
,y_{G}=0,$ $t)$ et de D($x_{D}=0,$ $y_{D}=\frac{t}{\sin \theta },$ $t)$ par
la TL:

\begin{equation*}
x_{G}^{\prime }=\gamma t(\frac{1}{\cos \theta }+\beta )\qquad y_{G}^{\prime
}=0\qquad x_{D}^{\prime }=\gamma \beta t\qquad y_{D}^{\prime }=\frac{t}{\sin
\theta }
\end{equation*}

On constate ainsi que les \'{e}v\'{e}nements DEG, \'{e}v\'{e}nements
simultan\'{e}s dans K ($t_{D}=t_{E}=t_{G}=t=1$), \textit{ne sont pas des
\'{e}v\'{e}nements simultan\'{e}s} (D'E'G') dans K' (\textbf{figure 6})$.$

\begin{equation}
t_{E}^{\prime }=\gamma t(1+\beta \cos \theta )\text{ \ \ \ \ \ \ }%
t_{G}^{\prime }=\gamma t(1+\frac{\beta }{\cos \theta })\qquad t_{D}^{\prime
}=\gamma t
\end{equation}

sauf s'ils ont m\^{e}me abscisse, autrement dit si le front plan est
longitudinal $\theta =\theta ^{\prime }=0$ (\textbf{figure 6bis})

La non-simultan\'{e}it\'{e} (2.9) ne doit gu\`{e}re constituer une surprise
si on adh\`{e}re \`{a} un point de vue relativiste (image \textit{%
spatio-temporelle }d'un front d'onde). Ces \'{e}v\'{e}nements D'E'G' sont
sur la tangente \`{a} l'ellipse qui s'\'{e}crit \`{a} partir de

\begin{equation}
y^{\prime }\sin \theta ^{\prime }+x^{\prime }\cos \theta ^{\prime
}=t^{\prime }
\end{equation}

\textbf{\'{e}tant donn\'{e} que t' n'est pas fix\'{e}} par la TL et donc
avec $t^{\prime }=\gamma (t+\beta x)=\gamma ^{-1}t+\beta x^{\prime }$, on
obtient l'\'{e}quation de la tangente \`{a} l'ellipse:

\begin{equation}
y^{\prime }\sin \theta ^{\prime }+x^{\prime }(\cos \theta ^{\prime }-\beta
)=\gamma ^{-1}t
\end{equation}

Le \textit{front d'onde poincar\'{e}en (\textbf{figure 6)}} a comme
coefficient angulaire en fonction de $\theta ^{\prime }$ (angle dans K'):

\QTP{Body Math}
\begin{equation}
a_{poincar\acute{e}}^{\prime }=tg\alpha ^{\prime }=\frac{\beta -\cos \theta
^{\prime }}{\sin \theta ^{\prime }}
\end{equation}

Einstein d\'{e}finit en 1905 l'angle image:

\begin{quotation}
En appelant $\theta ^{\prime }$ l'angle form\'{e} par \textbf{la normale de
l'onde} (direction radiale) dans le syst\`{e}me en mouvement et la
''direction du mouvement''\footnote{%
En fait dans le texte original, Einstein s'est tromp\'{e} et \`{a} \'{e}crit
\`{a} la place de ''la direction du mouvement'', ''la ligne
observateur-source''. Il a corrig\'{e} sur son exemplaire personnel (voir
Balibar, Einstein, Oeuvres choisies, Relativit\'{e}s I, page 49 note 56).} 
\cite[paragraphe 7]{4}
\end{quotation}

\textit{Le front image einteinien est transversal (''la normale de l'onde'',
\ \textbf{figure 6}) par rapport \`{a} la direction radiale ou la direction
de propagation.}

\textit{Le front einsteinien est donc tangent au cercle ''pointill\'{e}'' }%
(voir chapitre 1, \textbf{figure 1, 2}) de centre O' et de rayon, $t^{\prime
}=\gamma (1+\beta \cos \theta ).$ Si\textbf{\ t' est fix\'{e}}, alors (2.10)
est l'\'{e}quation d'une droite dont le coefficient angulaire est:

\begin{equation}
a_{einstein}^{\prime }=-\cot g\theta ^{\prime }
\end{equation}

La \textbf{double transversalit\'{e}} (ou \textbf{double simultan\'{e}it\'{e}%
}) einsteinienne se traduit au niveau du front image par:\textit{\ } 
\begin{equation}
a^{\prime }=-\cot g\theta ^{\prime }\qquad \Leftrightarrow \qquad (\Delta
t^{\prime })_{front}=0
\end{equation}

Comme pour la convention de synchronisation d'Einstein (annexe 1), la
vitesse relative n'intervient $\beta $ pas dans la d\'{e}termination\ (2.12)
de l'angle image $\theta ^{\prime }.$\textit{\ }

\textbf{RESUME} En r\'{e}sum\'{e} il y a deux interpr\'{e}tations physiques
diff\'{e}rentes de l'\'{e}quation (2.10) $y^{\prime }\sin \theta ^{\prime
}+x^{\prime }\cos \theta ^{\prime }=t^{\prime }$ (nous invitons le lecteur
\`{a} comparer ce r\'{e}sum\'{e} avec le r\'{e}sum\'{e} ''ondes
sph\'{e}riques'')

1) On ne fixe pas t' et (2.10) est l'\'{e}quation\ de la tangente \`{a}
l'ellipse ou d'un front d'onde plan spatiotemporel (un ensemble
d'\'{e}v\'{e}nements simultan\'{e}s ne reste pas un ensemble
d'\'{e}v\'{e}nements simultan\'{e}s). C'est la cin\'{e}matique sous-jacente
\`{a} l'ellipse de Poincar\'{e} (2005).

2) On fixe t' et (2.10) est l'\'{e}quation\ de la tangente au cercle ou d'un
front d'onde plan spatial ou rigide (un ensemble d'\'{e}v\'{e}nements
simultan\'{e}s reste un ensemble d'\'{e}v\'{e}nements simultan\'{e}s). C'est
la cin\'{e}matique d'Einstein (1905).

Nous devons maintenant \'{e}tablir que le ''quadrivecteur d'onde'' de
Poincar\'{e} ne d\'{e}finit pas seulement univoquement un point E' du front
ellipso\"{i}dal image mais d\'{e}finit aussi univoquement la tangente T'
\`{a} l'ellipse en ce point autrement dit l'onde plane. Il est d\`{e}s lors
g\'{e}om\'{e}triquement \'{e}vident que l'on peut enlever les guillemets et
associer \textit{de plein droit }le quadrivecteur d'onde de Poincar\'{e}
\`{a} la tangente \`{a} l'ellipse comme on associe le quadrivecteur d'onde
d'Einstein \`{a} la tangente au cercle.

Toutefois d'un point de vue physique, on ne s'int\'{e}resse pas tant aux
droits qu'aux faits. Que se passe-t-il au niveau de la phase d'une onde
plane? Jusqu'\`{a} pr\'{e}sent nous avons fait une analyse
quasi-exclusivement \textit{temporelle} du front d'onde plane en
d\'{e}veloppant respectivement l'image poincar\'{e}enne (tangente \`{a}
l'ellipse, le temps t' n'est pas fix\'{e}, relativit\'{e} de la
simultan\'{e}it\'{e}) et l'image einsteinienne (tangente au cercle, le temps
t' est fix\'{e}, simultan\'{e}it\'{e} absolue).Tout d\'{e}pendra de la fa\c{c%
}on dont chacun des auteurs introduit \textit{le rayon vecteur image}\textbf{%
\ }$\mathbf{r}^{\prime }$ dans K', \'{e}tant donn\'{e} que le front objet
est donn\'{e} de la m\^{e}me mani\`{e}re dans K.

\section{Phase d'une onde plane et formule Doppler d'Einstein}

Selon Einstein, si le temps $t^{\prime }$ est fix\'{e} sur le front par
l'abscisse $x_{T}^{\prime }$ du point de tangence T'(ou par x$_{T}$ du point
de tangence T, \textbf{figure 7}), il n'en est pas de m\^{e}me pour le rayon
vecteur $\mathbf{r}^{\prime }$ qui va \textit{de l'origine} O' vers un point
quelconque du front $P^{\prime }$\textbf{,} ($\mathbf{O}^{\prime }\mathbf{P}%
^{\prime })$:

\bigskip Il y a donc une \textbf{asym\'{e}trie} einsteinienne entre la
transformation de t en t' et de r en r' qui semble justifi\'{e}e par le fait
que cette asym\'{e}trie r\'{e}side aussi dans la d\'{e}finition du front
objet (t est fix\'{e} et r varie pour chaque point du front objet). Voyons
ceci dans les d\'{e}tails.

L'\'{e}criture einsteinienne de la phase, avec la notation standard du
vecteur d'onde $\mathbf{k}=\frac{2\pi }{\lambda }\mathbf{1}_{k\text{ }}=%
\frac{2\pi \nu }{c}\mathbf{1}_{k\text{ }}$ainsi que de la fr\'{e}quence $\nu
=\frac{\omega }{2\pi }\ $est la suivante \cite[§7 ]{4}: 
\begin{equation}
\omega t-\mathbf{k.r=}\text{ }\phi =\omega ^{\prime }t^{\prime }-\mathbf{k}%
^{\prime }\mathbf{.r}^{\prime }=\phi ^{\prime }
\end{equation}

Ce qui revient \`{a} \'{e}crire \`{a} deux dimensions \cite{4}: 
\begin{equation}
2\pi \nu (t-\frac{x}{c}\cos \theta -\frac{y}{c}\sin \theta )=\phi =2\pi \nu
^{\prime }(t^{\prime }-\frac{x^{\prime }}{c}\cos \theta ^{\prime }-\frac{%
y^{\prime }}{c}\sin \theta ^{\prime })=\phi ^{\prime }
\end{equation}
Si on fixe le temps t pour le front objet et le temps t' pour le front
image, l'\'{e}quation du front einsteinien, objet et image, s'\'{e}crit
habituellement \`{a} l'aide du produit scalaire: 
\begin{equation}
\mathbf{k.r}=k.\mathbf{1}_{n}.\mathbf{r}=Cte\qquad \qquad \mathbf{k}^{\prime
}\mathbf{.r}^{\prime }=k^{\prime }.\mathbf{1}_{n}^{\prime }.\mathbf{r}%
^{\prime }\text{ }\mathbf{=}\text{ }Cte^{\prime }
\end{equation}
$\mathbf{1}_{n}$ \'{e}tant le vecteur unit\'{e} sur la normale de l'onde
objet et, selon les propres termes d'Einstein, $\mathbf{1}_{n}^{\prime }$ le
vecteur unit\'{e} sur la normale de l'onde image. On retrouve alors avec les
produits scalaires $\mathbf{1}_{n}.\mathbf{r}$ et$\ \mathbf{1}_{n}^{\prime }.%
\mathbf{r}^{\prime }$ (23 et 25): 
\begin{equation*}
y\sin \theta +x\cos \theta =t\qquad \qquad y^{\prime }\sin \theta ^{\prime
}+x^{\prime }\cos \theta ^{\prime }=t_{fix\acute{e}}^{\prime }
\end{equation*}

\bigskip\ Ecrivons maintenant le produit scalaire einsteinien en fonction de
l'angle $\delta $ (\textbf{figure 7}) entre les vecteurs $\mathbf{r}$\textbf{%
\ }et\textbf{\ }$\mathbf{k,}$ autrement dit entre $\mathbf{r}_{T}$ $\mathbf{%
=OT}$ et\ $\ \mathbf{r}_{P}$ $\mathbf{=OP}$, et l'angle $\delta ^{\prime }$
entre $\mathbf{r}^{\prime }$\textbf{\ }et\textbf{\ }$\mathbf{k}^{\prime }$,
autrement dit entre $\mathbf{r}_{T}^{\prime }$ $\mathbf{=O}^{\prime }\mathbf{%
T}^{\prime }$ et\ $\mathbf{r}_{P}^{\prime }$ $\mathbf{=O}^{\prime }\mathbf{P}%
^{\prime }$){\footnotesize \ }\textbf{.} L'invariant einsteinien (2.15)
s'\'{e}crit alors:

\begin{equation*}
\omega t-k.r\cos \delta \text{ }\mathbf{=}\text{ }\phi =\omega ^{\prime
}t^{\prime }-k^{\prime }.r^{\prime }\cos \delta ^{\prime }=\phi ^{\prime }
\end{equation*}

\begin{equation}
\nu (t-r\cos \delta )\text{ }\mathbf{=}\text{ }\phi =\nu ^{\prime
}(t^{\prime }-r^{\prime }\cos \delta ^{\prime })=\phi ^{\prime }
\end{equation}

La phase constante $\phi ^{\prime }$ en chaque point \textbf{r' }du front
d'onde\ image est d\'{e}termin\'{e}e par la projection constante $%
r_{T}^{\prime }=r^{\prime }$cos $\delta ^{\prime }$ de \textbf{r'} sur la
direction de propagation (\textbf{figure 7}, OT')$.$ Et avec (2-18) on a 
\begin{equation}
\phi ^{\prime }=\nu ^{\prime }(t^{\prime }-r^{\prime }cos\delta ^{\prime
})=\nu ^{\prime }(t^{\prime }-r_{T}^{\prime })
\end{equation}
la phase constante $\phi ^{\prime }$ implique un temps image\textit{\ }$%
t^{\prime }$\textit{\ constant }$\forall $\textit{\ }$\delta ^{\prime }$. La 
\textbf{double transversalit\'{e} (simultan\'{e}it\'{e}) einsteinienne}
(2.14) est donc la seule interpr\'{e}tation possible de l'\'{e}criture
traditionnelle de la phase d'une onde plane au moyen du produit scalaire $%
\mathbf{k.r}$ (2.15).

Le produit scalaire vectoriel einsteinien $\mathbf{k.r}$ peut aussi
s'interpr\'{e}ter d'une mani\`{e}re quadrivectorielle. Il est bien connu que
l'on peut d\'{e}finir la phase \`{a} l'aide du produit scalaire entre le
quadrivecteur d'onde einsteinien (2.6) et le quadrivecteur spatio-temporel.

\begin{equation}
(\omega ,\text{ }\mathbf{k}).(t,\text{ }\mathbf{r})=\omega t-\mathbf{%
k.r=\phi }
\end{equation}

Il est \'{e}galement bien connu que l'invariance de ce produit scalaire
donne imm\'{e}diatement la formule Doppler d'Einstein (2.5)\cite[7]{4}. Le
''paquet einsteinien'' forme un ensemble coh\'{e}rent indissociable.

Il conviendra donc d'\'{e}crire la phase d'une autre mani\`{e}re chez
Poincar\'{e} car il y a un autre quadrivecteur d'onde (2.1) et une autre
formule Doppler (2.4).

Montrons que l'invariant einsteinien (2.15) n'est pas compatible avec
l'invariant relativiste $t^{2}-x^{2}-y^{2}=t^{\prime 2}-x^{\prime
2}-y^{\prime 2}$ par excellence (1.3 \& 1.6) . Pour passer du front objet au
front image einsteinien, on doit utiliser la transformation suivante, $(x,$ $%
y,$ $t)\rightarrow (x^{\prime },$ $y^{\prime },$ $t^{\prime }),$ \textit{%
pour tous les \'{e}v\'{e}nements}:

\begin{equation}
x^{\prime }=\gamma (x+\beta t)\qquad y^{\prime }=y\qquad t^{\prime }=\gamma
(t+\beta x_{T})
\end{equation}

avec laquelle on obtient 
\begin{equation*}
x^{2}+y^{2}-t^{2}=\gamma ^{2}(t+\beta x_{T})^{2}-\gamma ^{2}(x+\beta
t)^{2}+y^{\prime 2}
\end{equation*}

On voit ainsi que l'invariance relativiste n'est pas respect\'{e}e:
l'intervalle entre O et P (\textbf{voir figure 7}) n'est pas le m\^{e}me que
l'intervalle entre O' et P'.\ Autrement dit le vecteur $\mathbf{O}^{\prime }%
\mathbf{P}^{\prime }=\mathbf{r}^{\prime }$ n'est pas l'image par la TL de $%
\mathbf{OP=r}$ (l'invariant einsteinien (2.15) est en v\'{e}rit\'{e} un
invariant galil\'{e}en). Le traitement einsteinien \textit{asym\'{e}trique}
du temps et de l'espace en est manifestement l'origine\footnote{%
On obtient par diff\'{e}rentiation de la phase einsteinienne, $d\phi =d\phi
^{\prime }=0,$ la vitesse des diff\'{e}rents points du front d'onde: $\frac{%
dr}{dt}=\frac{\omega }{k\cos \delta }$ \ \ et \`{a} \ \ \ $\frac{dr^{\prime }%
}{dt^{\prime }}=\frac{\omega ^{\prime }}{k^{\prime }\cos \delta ^{\prime }}.$%
La vitesse de phase \'{e}tant respectivement $\frac{\omega }{k}$ $=$ $\frac{%
\omega ^{\prime }}{k^{\prime }}=c=1,$ la vitesse des points du front d'onde
einstein est toujours plus grande que la vitesse de la lumi\`{e}re sauf
\'{e}videmment pour $\delta =\delta ^{\prime }=0$). Ce qui laisse supposer
qu'Einstein aurait utilis\'{e} une jauge o\`{u} le potentiel ne se propage
pas \`{a} la vitesse de la lumi\`{e}re (voir par contraste \S 2.4).}.

\section{Phase d'une onde plane et formule Doppler de Poincar\'{e}}

\bigskip Comment \'{e}liminer l'asym\'{e}trie entre la fixation du temps
image $(t^{\prime })$ et la non-fixation de l'espace image $(r^{\prime })$
sous-jacente \`{a} l'\'{e}criture einsteinienne de l'onde plane?
L'\'{e}quation (2.21) montre qu'on semble avoir le choix entre fixer les
deux en rempla\c{c}ant $x$ par $x_{T}\ \ $dans la premi\`{e}re TL \textit{ou}
ne fixer aucun des deux, en rempla\c{c}ant $x_{T}$ par $x$ dans la seconde
TL.

En fait il s'agit d'un \textit{''ou inclusif''} car dans le premier cas on
trouve le point de tangence T'($r_{0}^{\prime },$ $t_{0}^{\prime })$ et dans
le second cas on retrouve un point $P(r^{\prime },$ $t^{\prime })$
quelconque de la tangente \`{a} l'ellipse (\textbf{figure 8})$.$ Un
\'{e}v\'{e}nement $P(r,t)$ quelconque appartenant au front se transforme en $%
P(r^{\prime },t^{\prime })$ selon la TL (1.2):

\begin{equation}
t^{2}-r^{2}=t^{\prime 2}-r^{\prime 2}  \tag{intervalle non-nul}
\end{equation}

Cet intervalle du genre spatial sera \'{e}tudi\'{e} au paragraphe suivant.
Concentrons nous \`{a} pr\'{e}sent sur le \textbf{point de tangence} T
d\'{e}fini par le ''quadrivecteur d'onde'' de Poincar\'{e} avec

\begin{equation}
t^{2}-r_{T}^{2}=t^{\prime 2}-r_{T^{\prime }}^{\prime 2}=0 
\tag{intervalle
nul}
\end{equation}

Nous devons \'{e}crire\textit{\ la phase (invariante) d'une onde plane} sans
utiliser le produit scalaire ''$\mathbf{k.r"}$ qui est \`{a} l'origine de la
violation de la relativit\'{e} de la simultan\'{e}it\'{e}. En d'autres
termes il faut \textit{d\'{e}coupler} ''invariance de la phase $\phi
^{\prime }"$ et la ''fixation du temps $t^{\prime }$''. \ Cette
op\'{e}ration d\'{e}licate semble relever de la chirurgie de la plus haute
pr\'{e}cision. Inspirons-nous de (2.18) $\nu (t-r\cos \delta )$ $\mathbf{=}$ 
$\phi =\nu ^{\prime }(t^{\prime }-r^{\prime }\cos \delta ^{\prime })=\phi
^{\prime }$ d\'{e}termin\'{e}e par le seul point de tangence T (T').

\begin{equation}
\nu (t-r_{T})\text{ }\mathbf{=}\text{ }\Phi =\nu ^{\prime }(t^{\prime
}-r_{T^{\prime }}^{\prime })=\Phi ^{\prime }
\end{equation}

Le comptage (2.2) de la phase est donc le m\^{e}me $(\Phi ^{\prime }=\phi
^{\prime })$ mais l'invariant ne s'\'{e}crit plus de la m\^{e}me mani\`{e}re
puisque ce n'est plus $\phi =\nu (t-r\cos \delta )$ $=\nu (t-x\cos \theta
-y\sin \theta )$ (2.16)$\ $qui est invariant mais bien $\Phi =\nu (t-r_{T})$
(2.19) qui est invariant.

Dans le premier cas la fixation de la phase $\phi ^{\prime }$ est
coupl\'{e}e avec la fixation de t' (ou avec la tansversalit\'{e} du front
image); dans le second cas la fixation de la phase $\Phi ^{\prime }$ ne fixe
pas t' sur le front (pour tout point sur le front image, tangente \`{a}
l'ellipse on a $\nu ^{\prime }(t^{\prime }-r^{\prime }))$. L'invariant de
phase de Poincar\'{e} (2.22) ne fait intervenir que la norme du vecteur
radial (et non pas l'angle $\delta $, \textbf{figure 8}$)$ car l'invariant
fondamental de la relativit\'{e} (1.3 \& 1.6) ne fait intervenir que la
norme du vecteur radial.

Nous pouvons d\'{e}sormais d\'{e}finir le quadrivecteur d'onde de
Poincar\'{e} (sans guillemets) pour une onde plane puisque la phase $\Phi $
(ou $\Phi ^{\prime })$ de cette derni\`{e}re est univoquement
caract\'{e}ris\'{e}e par le point de tangence T (ou T') \`{a} l'ellipse. Ce
qui \'{e}tait g\'{e}om\'{e}triquement \'{e}vident (fin du \S 2.2) est
maintenant physiquement \'{e}vident ($cqfd$).

L'expression (2.22) est donc l'\'{e}criture poincar\'{e}enne de la phase
invariante d'une ONDE PLANE. Il est donc \'{e}vident que l'on doit retrouver
une transformation de la fr\'{e}quence qui constitue une formule Doppler au
sens rigoureux du terme. Toutefois nous proposons une d\'{e}duction directe
\`{a} partir de l'invariance (2.22) \'{e}tant donn\'{e} le caract\`{e}re
in\'{e}dit de cette formule qui \'{e}tablit l'existence d'une ''structure
fine''. Comme on a: 
\begin{equation}
(t-r_{T})(t+r_{T})=(t^{\prime }-r_{T^{\prime }}^{\prime })(t^{\prime
}+r_{T^{\prime }}^{\prime })  \tag{1.3 \& 1.6}
\end{equation}

il suffit de chercher comment $(t\pm r)$ se transforme par la TL en $%
(t^{\prime }\pm r^{\prime }).$ Le cas longitudinal est tr\`{e}s simple \`{a}
traiter ($x=r$ et $x^{\prime }=r^{\prime }).$ En effet avec $\theta =0$ on
trouve une forme bien connue de la TL: 
\begin{equation}
t^{\prime }-x^{\prime }=\sqrt{\frac{1+\beta }{1-\beta }}(t-x)\qquad \qquad
t^{\prime }+x^{\prime }=\sqrt{\frac{1-\beta }{1+\beta }}(t-x)
\end{equation}
et donc imm\'{e}diatement la formule Doppler longitudinale avec (11) qui est
identique dans les deux cin\'{e}matiques (Einstein et Poincar\'{e}).
Traitons maintenant le cas quelconque o\`{u} la direction de propagation $%
r_{T}$ du signal lumineux fait un angle $\theta $ avec la direction du
mouvement de l'observateur. Et donc par (1.4 \& 1.5):

\begin{equation}
t^{\prime }-r_{T^{^{\prime }}}^{\prime }=\gamma (1+\beta \cos \theta
)(t-r_{T})\qquad \qquad t^{\prime }+r_{T^{\prime ^{\prime }}}^{\prime
}=\gamma (1-\beta \cos \theta )(t+r_{T})
\end{equation}
on a alors l'invariant (1.3 \& 1.6) \`{a} condition que la formule
g\'{e}n\'{e}rale pour la transformation des fr\'{e}quences soit: 
\begin{equation*}
\frac{\nu }{\nu ^{\prime }}=\gamma (1+\beta \cos \theta )\text{ ou }\nu
^{\prime }=\nu \frac{1}{\gamma (1+\beta \cos \theta )}
\end{equation*}
C'est la formule Doppler ''redshift'' (voir ''synth\`{e}se'' chapitre 4) de
Poincar\'{e} (2.4)

\begin{equation*}
\frac{\nu }{\nu ^{\prime }}=\gamma (1-\beta \cos \theta )\text{ ou }\nu
^{\prime }=\nu \frac{1}{\gamma (1-\beta \cos \theta )}
\end{equation*}
C'est la formule Doppler ''blueshift'' de Poincar\'{e} (voir
''synth\`{e}se'' 1.4). On peut \'{e}galement \'{e}crire les formules
''redshift et blueshift'' avec l'angle de l'observateur (2.4) 
\begin{equation}
\text{ }\nu ^{\prime }=\gamma \nu (1-\beta \cos \theta ^{\prime })\qquad 
\text{ }\nu ^{\prime }=\gamma \nu (1+\beta \cos \theta ^{\prime })
\end{equation}

On d\'{e}duit ainsi pour la quatri\`{e}me fois la formule Doppler de
Poincar\'{e} (2.4, ''sans guillemets''). Nous pouvons maintenant lever la
restriction qui \'{e}tait sous-jacente \`{a} la d\'{e}duction de la formule
Doppler avec des ondes sph\'{e}riques (2.4, ''avec guillemets'') puisque la
direction donn\'{e}e par le point de tangence T \textit{est maintenant
constante par d\'{e}finitio}n (onde plane, \textbf{sources lointaines}, voir
l'\'{e}quation corrolaire 2.28).

L'invariance de la phase ''de Poincar\'{e} peut encore s'\'{e}crire: 
\begin{equation*}
\frac{t^{\prime }}{T^{\prime }}-\frac{r_{T^{\prime }}^{\prime }}{\lambda
^{\prime }}=\Phi ^{\prime }=\frac{t}{T}-\frac{r_{T}}{\lambda }=\Phi
\end{equation*}
En notation standard on a l'invariance de Poincar\'{e} dans le cas
g\'{e}n\'{e}ral non-longitudinal $k=\left| \mathbf{k}\right| =\frac{2\pi }{%
\lambda } $ et $\omega =2\pi \nu ,$ $k^{\prime }=\left| \mathbf{k}^{\prime
}\right| =\frac{2\pi }{\lambda ^{\prime }}$ et $\omega ^{\prime }=2\pi \nu
^{\prime }.$ 
\begin{equation}
\omega ^{\prime }t^{\prime }-k^{\prime }r_{T^{\prime }}^{\prime }=\Phi
^{\prime }=\omega t-kr_{T}=\Phi
\end{equation}
Selon la direction longitudinale $\theta =\theta ^{\prime }=0,$ on a: 
\begin{equation}
\omega ^{\prime }t^{\prime }-k^{\prime }x^{\prime }=\omega t-kx
\end{equation}
On retrouve ainsi des formules identiques pour les deux cin\'{e}matiques
comme dans le cas des formules Doppler. On voit ainsi que l'expression\
scalaire $kr_{T}$ \ ne peut pas \^{e}tre, comme chez Einstein, pour tout
point du front, le \textit{produit} scalaire de deux \textit{vecteurs}%
\textbf{\ }$\mathbf{k.r}=kr$ $\cos \delta .$ \bigskip\ 

La coh\'{e}rence de la cin\'{e}matique poincar\'{e}enne est parfaite puisque
avec (2.22 ou 2.26) on a la formule Doppler (2.4) et donc le quadrivecteur
d'onde\footnote{%
On voit imm\'{e}diatement qu'en prenant $d\Phi =d\Phi ^{\prime }=0$ (voir
note 7)$,$ le point T'\ \ $\frac{dr}{dt^{\prime }}=\frac{\omega ^{\prime }}{%
k^{\prime }}=c=1$ ainsi que tous les autres points P' $\frac{dr_{P^{\prime
}}^{\prime }}{dt^{\prime }}=\frac{\omega ^{\prime }}{k^{\prime }}=c=1$ du
front\ image poincar\'{e}en se d\'{e}placent avec la vitesse de la
lumi\`{e}re. Ce qui laisse supposer que Poincar\'{e} aurait utilis\'{e} une
jauge o\`{u} le mode de propagation du potentiel est d\'{e}termin\'{e} par
la vitesse de la lumi\`{e}re (voir par contraste \S 2.3)} (2.1).

\textbf{RESUME} \bigskip La phase est invariante chez Poincar\'{e} (17)
comme chez Einstein mais:

1) Si t' est fix\'{e} sur le front (spatial ou rigide) il faut \'{e}crire la
phase avec Einstein $\phi =\omega t-\mathbf{k.r}$

2) Si t' est fix\'{e} sur le front (spatio-temporel) il faut \'{e}crire la
phase avec Poincar\'{e} $\phi =\omega t-k.r_{T}$

Dans les deux cin\'{e}matiques, la phase est invariante: $\phi =\phi
^{\prime }$ (30, Einstein) \& $\Phi =\Phi ^{\prime }($37, Poincar\'{e}). La
diff\'{e}rence c'est que, dans la cin\'{e}matique sous-jacente \`{a}
l'ellipse, ''phase invariante'' et ''temps constant'' sont \textbf{%
d\'{e}coupl\'{e}s}: c'est la phase (37 ou 41) qui est le v\'{e}ritable
invariant, pas le caract\`{e}re spatial (ou rigide) du front (\`{a} un
instant donn\'{e}).

Une objection vient alors \`{a} l'esprit: \`{a} quoi sert un tel
d\'{e}couplage? Quel est le but d'une telle op\'{e}ration chirurgicale?

1) D\'{e}duire une formule Doppler rigoureuse (2.4 avec $n_{2}=n_{2}^{\prime
}\rightarrow \infty $) pour les sources lointaines (ondes planes), par
exemple des galaxies.

2) D\'{e}duire comme \textit{corrolaire }de cette formule Doppler, \'{e}tant
donn\'{e} l'invariance de la phase (2.26), la relation sp\'{e}cifiquement
poincar\'{e}enne (1.13bis \& 2.1 ter, ou directement le quadrivecteur d'onde
2.1 proportionnel \`{a} au quadrivecteur spatio-temporel 1.13).

\begin{equation}
\frac{r_{\infty }^{\prime }}{r_{\infty }}=\frac{\lambda ^{\prime }}{\lambda }
\end{equation}

\bigskip Cette relation est incompatible avec la cin\'{e}matique
einsteinienne (voir quadrivecteur d'Einstein, 2.6). Nous proposons une
interpr\'{e}tation\ de cette relation corrolaire dans le paragraphe
''synth\`{e}se'' (chapitre 4).

\section{Composante longitudinale du vecteur tangent \`{a} l'ellipse et
4-vecteur potentiel de Poincar\'{e}}

Consid\'{e}rons $T$\ comme origine du \textbf{vecteur} $\mathbf{d}$ et un
point quelconque P comme extr\'{e}mit\'{e} du vecteur $\mathbf{d}$ (\textbf{%
figure 9}) et cherchons la relation entre le vecteur d et l'\'{e}cart \`{a}
la simultan\'{e}it\'{e}

Les coordonn\'{e}es des deux points T' et P'($\mathbf{d}^{\prime }\mathbf{=T}%
^{\prime }\mathbf{P}^{\prime })$ sont alors donn\'{e}es directement par la
TL: 
\begin{eqnarray*}
x_{T^{\prime }}^{\prime } &=&\gamma (x_{T}+\beta t)\qquad \qquad
y_{T^{\prime }}^{\prime }=y_{T}\qquad \qquad t_{T^{\prime }}^{\prime
}=\gamma (t+\beta x_{T}) \\
x_{P^{\prime }}^{\prime } &=&\gamma (x_{P}+\beta t)\qquad \qquad
y_{P^{\prime }}^{\prime }=y_{P}\qquad \qquad t_{P^{\prime }}=\gamma (t+\beta
x_{P})
\end{eqnarray*}
On a alors:

\begin{equation*}
\Delta x^{\prime }=\gamma \Delta x\qquad \Delta t^{\prime }=\gamma \beta
\Delta x\qquad \ et\qquad \Delta y^{\prime }=\Delta y
\end{equation*}

L'\'{e}cart \`{a} la simultan\'{e}it\'{e} 
\begin{equation}
\Delta t^{\prime }=\gamma \beta d_{x}
\end{equation}
n'est nul ($d_{x}=0)$ que dans le cas de la propagation longitudinale,
autrement dit lorsque les fronts sont transversaux dans les deux
syst\`{e}mes (\textbf{figure 6bis}). On v\'{e}rifie que l'on a pour tous les
points de la tangente \`{a} l'ellipse, $\Delta t^{\prime 2}-\Delta r^{\prime
2}$ $=\Delta t^{2}-\Delta r^{2}=-\Delta r^{2}$ (pour P'= T' on a $\Delta
r=0),$ autrement dit l'invariant fondamental (1.3-1.6). D\'{e}terminons
''l'\'{e}cart la simultan\'{e}it\'{e}'' en fonction des normes d et d': 
\begin{equation}
d^{2}=\Delta x^{2}+\Delta y^{2}=d_{x}^{2}+d_{y}^{2}\qquad \qquad d^{^{\prime
}2}=\Delta x^{\prime 2}+\Delta y^{\prime 2}=d_{x}^{\prime 2}+d_{y}^{\prime 2}
\end{equation}

et donc avec $d^{^{\prime }2}=\gamma ^{2}\Delta x^{2}+\Delta y^{2}=(\gamma
^{2}-1)\Delta x^{2}+\Delta x^{2}+\Delta y^{2}=\beta ^{2}\gamma ^{2}\Delta
x^{2}+d^{2}$, on obtient: 
\begin{equation}
\Delta t^{\prime 2}=d^{\prime 2}-d^{2}
\end{equation}

La diff\'{e}rence de temps (''l'\'{e}cart \`{a} la simultan\'{e}it\'{e}'')
d\'{e}pend donc directement de la diff\'{e}rence entre la distance d dans K
et la distance d' dans K'. On a donc selon Einstein 
\begin{equation*}
\Delta t^{\prime }=0\Longleftrightarrow d^{\prime }=d
\end{equation*}
En projetant maintenant $\mathbf{d}$\ et $\mathbf{d}^{\prime }$ sur les
syst\`{e}me d'axes perpendiculaires form\'{e}s par la direction de
propagation et la perpendiculaire \`{a} cette direction respectivement dans
K et K' (l'origine de chaque syst\`{e}me est au point de tangence,
respectivement au cercle T et \`{a} l'ellipse T'), on aura alors dans K:

\begin{equation}
\mathbf{d=d}_{\parallel }\mathbf{+}\text{ }\mathbf{d}_{\perp }\qquad \qquad
d=d_{\perp }\qquad d_{\parallel }=0
\end{equation}

Et donc on aura selon Einstein dans K': 
\begin{equation*}
(d=d^{\prime }=d_{\perp }^{\prime })_{einstein}
\end{equation*}

La composante transversale poincar\'{e}enne $d_{\perp }^{\prime }$ (en
traits discontinus) est identique \`{a} l'image einsteinienne $d_{E}^{\prime
}$ (en pointill\'{e}s) du front d'onde (\textbf{figure 10}). Mais il
appara\^{i}t g\'{e}om\'{e}triquement dans l'image poincar\'{e}enne une 
\textbf{composante longitudinale} (en gris fonc\'{e}) $d_{\parallel
}^{\prime }$ qu'il convient de d\'{e}terminer alg\'{e}briquement:

\begin{equation}
\mathbf{d}^{\prime }\mathbf{=d}_{\parallel }^{\prime }\mathbf{+}\text{ }%
\mathbf{d}_{\perp }^{\prime }\mathbf{\qquad }\qquad d^{\prime
2}=d_{\parallel }^{\prime 2}+d_{\perp }^{\prime 2}
\end{equation}

\begin{equation*}
\gamma ^{2}d_{x}^{2}+d_{y}^{2}=d_{\parallel }^{\prime 2}+d^{2}\qquad \gamma
^{2}d_{x}^{2}+d_{y}^{2}=d_{\parallel }^{\prime 2}+d_{x}^{2}+d_{y}^{2}
\end{equation*}

et finalement pour la composante longitudinale de la tangente \`{a}
l'ellipse: 
\begin{equation}
d_{\parallel }^{\prime }=\gamma \beta d_{x}
\end{equation}

On trouve alors\ en vertu de (2.29)

\begin{equation}
(d_{\parallel }^{\prime }=\Delta t^{\prime })_{poincar\acute{e}}
\end{equation}
\textbf{Th\'{e}or\`{e}me de Poincar\'{e}}\footnote{%
Nous avons baptis\'{e} l'\'{e}quation (51) th\'{e}or\`{e}me de Poincar\'{e}
car cet \'{e}cart \`{a} la simultan\'{e}it\'{e}, autrement dit la
diff\'{e}rence entre le temps aller t$^{+}$ et le temps retour t$^{-}$ (voir
annexe 1) joue un r\^{o}le historique pr\'{e}pond\'{e}rant dans la
construction de son ellipse allong\'{e}e: \textit{''la diff\'{e}rence des
temps est rigoureusement proportionnelle \`{a} la diff\'{e}rence des
abscisses'', }$\Delta t^{\prime }=\frac{t^{^{\prime }+}-t^{\prime -}}{2}%
=\beta \gamma \Delta x$\textit{''}$.$ L'ellipse allong\'{e}e (dans le
syst\`{e}me de la source) n'est rien d'autre que la synchronisation
poincar\'{e}enne des horloges compte tenu de la contraction.\ C'est ce point
que nous examirons en profondeur dans la partie II).}: ''\textit{L'\'{e}cart
\`{a} la simultan\'{e}it\'{e} ne d\'{e}pend que de la composante
longitudinale'' du vecteur }$\mathbf{d}$\textbf{'}\textit{\ sur la tangente
\`{a} l'ellipse.}

On voit ici clairement la diff\'{e}rence irr\'{e}ductible avec Einstein%
\textbf{\ qui annule la composante longitudinale poincar\'{e}enne}: 
\begin{equation}
(d_{\parallel }^{\prime }=\Delta t^{\prime }=0)_{einstein}
\end{equation}

pour la r\'{e}duire \`{a} sa composante transversale sur la tangente au
cercle ($\mathbf{d}$\textbf{'}$\mathbf{=}$ $\mathbf{d}_{\perp }^{\prime }).$
Einstein op\`{e}re donc un choix implicite d\'{e}cisif\footnote{%
Le choix einsteinien d\'{e}cisif se retrouve dans l'interpr\'{e}tation
standard de la contraction (3.14bis) et dans la convention standard de
synchronisation (annexe 1, equations (1 \& 2) ou (4bis).} (2.37) qui est
inscrit non seulement dans sa d\'{e}finition du front d'onde image (2.15)
mais aussi dans son quadrivecteur d'onde et sa formule Doppler (2.5). A quoi
correspond \textbf{physiquement} la composante longitudinale 2.36)? Il est
dors et d\'{e}j\`{a} clair qu'il faudra trouver une grandeur physique qui
dans le premier cas (Einstein) doit \^{e}tre d\'{e}termin\'{e}e par un type
d'\'{e}quation associ\'{e} \`{a} la simultan\'{e}it\'{e} absolue (par
exemple un laplacien ou une propagation instantan\'{e}e, voir fin du
chapitre 2) et dans le second cas doit \^{e}tre d\'{e}termin\'{e} par un
type d'\'{e}quation associ\'{e} \`{a} la simultan\'{e}it\'{e} relative (par
exemple un dalembertien ou une vitesse de propagation finie, voir fin du
chapitre 2).

Nous avons ainsi montr\'{e} que la propagation du front \`{a} la vitesse de
la lumi\`{e}re \'{e}tait caract\'{e}ris\'{e}e par les grandeur\textbf{\ }$d$%
\textbf{\ }(la distance sur le front) et $\Delta t$ (l'\'{e}cart \`{a} la
simultan\'{e}it\'{e}). Montrons que l'on peut mettre ces r\'{e}sultats $(%
\mathbf{d},$ $\Delta t)$ sous forme quadrivectorielle. Autrement dit ($%
d_{x},d_{y}$, 0, $\Delta t)$ forment les composantes d'un quadrivecteur. On
a alors dans K le quadrivecteur

\begin{equation}
(d_{x},\qquad d_{y}\ ,\ \ \ \ \ 0,\qquad \text{\ }\Delta t=0)
\end{equation}

dont la norme peut s'\'{e}crire en fonction de $d$ ou $d_{\perp }$:

\begin{equation}
\left\| (d_{x},\qquad d_{y}\ ,\ \ \ \ \ 0,\qquad 0)\right\| =d_{\perp
}^{2}=d^{2}
\end{equation}

Le 4-vecteur s'\'{e}crit dans Ox'y':

\begin{equation}
(d_{x}^{\prime },\qquad d_{y}^{\prime },\ \ \ \ \ \ \ 0,\qquad \Delta
t^{\prime })
\end{equation}

avec par la TL, on retrouve (2.29)

\begin{equation*}
d_{x}^{\prime }=\gamma d_{x}\qquad d_{y}^{\prime }=d_{y}\text{ \ \ \ \ \ \ }%
\Delta t^{\prime }=\gamma \beta d_{x}
\end{equation*}

Et avec la norme invariante,

\begin{equation}
\left\| (d_{x}^{\prime },\qquad d_{y}^{\prime },\ \ \ \ \ \ \ 0,\qquad
\Delta t^{\prime })\right\| =d^{\prime 2}-\Delta t^{\prime 2}=d_{\perp
}^{\prime 2}+d_{\parallel }^{\prime 2}-\Delta t^{\prime 2}=d_{\perp
}^{\prime 2}
\end{equation}

on retrouve le th\'{e}or\`{e}me de Poincar\'{e} (2.36) pour la propagation
de l'onde plane 
\begin{equation*}
d_{\parallel }^{\prime }=\gamma \beta d_{x}=\Delta t^{\prime }
\end{equation*}

On a ainsi montr\'{e} que le vecteur spatial $\mathbf{d}$ et l'\'{e}cart
\`{a} la simultan\'{e}it\'{e} $\Delta t$ ont une structure \textbf{%
quadrivectorielle }($d_{x},$ $d_{y},$ $0,$ $\Delta t)$\textbf{\ }du genre
espace: on peut toujours annuler la 4\`{e}me composante dans un des
syst\`{e}mes. Si on annule la quatri\`{e}me composante dans les deux
syst\`{e}mes, on retrouve la convention d'Einstein (voir convention de
synchronisation d'Einstein, annexe 1 o\`{u} le terme poincar\'{e}en $\Delta
t^{\prime }=0$ est aussi annul\'{e})

\begin{equation}
\Delta t=\Delta t^{\prime }=0\text{ et }d_{\parallel }=d_{\parallel
}^{\prime }
\end{equation}

\bigskip On voit ainsi tr\`{e}s clairement que la probl\'{e}matique des
fronts d'onde spatiotemporel (Poincar\'{e}) ou spatiaux (Einstein) se
ram\`{e}ne \`{a} la discussion des composantes d'un \textbf{quadrivecteur}.

Si on \'{e}tablit un lien maintenant entre les ondes trait\'{e}es de fa\c{c}%
on abstraites et la th\'{e}orie de l'\'{e}lectromagn\'{e}tisme, c'est le
quadrivecteur \textbf{potentiel} qu'il faudra prendre en compte et non pas
le tenseur du champ. On voit ainsi tr\`{e}s clairement que la
probl\'{e}matique des fronts d'onde spatiotemporel ou spatiaux se ram\`{e}ne
\`{a} la discussion des composantes d'un \textbf{quadrivecteur}. Il est
clair que le quadrivecteur ainsi d\'{e}fini g\'{e}om\'{e}triquement 
\begin{equation}
\ (\mathbf{d},\text{ \ }\Delta t)
\end{equation}
ne peut \^{e}tre physiquement (\'{e}lectromagn\'{e}tiquement) que le \textbf{%
quadrivecteur potentiel } 
\begin{equation}
(\mathbf{A},\text{ \ }V)
\end{equation}
\textbf{\ \'{e}crit pour la premi\`{e}re fois par Poincar\'{e} en 1905 }\cite
{18}.

\subsection{Cin\'{e}matique de Poincar\'{e} en jauge de Lorenz}

Poincar\'{e} a explicitement adopt\'{e} la jauge (relativiste) de Lorenz
indissociablement coupl\'{e}e (\'{e}quation de continuit\'{e}) dans son
analyse au quadrivecteur potentiel. La jauge de Lorenz s'\'{e}crit
longitudinalement au moyen de la d\'{e}composition de Helmholtz
d\'{e}montr\'{e}e invariante puisque $A^{\prime 2}=A_{\parallel }^{\prime
2}+A_{\perp }^{\prime 2}$ (2.34 \& 2.35) \cite{17bis}:

\begin{equation}
div\mathbf{A}_{\parallel }+\frac{1}{c^{2}}\partial _{t}V=0\qquad
divA_{_{\parallel }}^{\prime }+\frac{1}{c^{2}}\partial _{t}V^{\prime }=0
\end{equation}

dans le premier syst\`{e}me par annulation $A_{\parallel }=V=0$ mais pas
dans le syst\`{e}me image K' o\`{u} $A_{_{\parallel }}^{\prime }=V^{\prime
}\neq 0.$

Nous arrivons ainsi au but car l'ellipse de Poincar\'{e} n'est pas en
contradiction avec la transversalit\'{e} de l'onde \'{e}lectromagn\'{e}tique
(exp\'{e}riences de polarisation) \textit{au niveau des champs}. En effet la
composante longitudinale du champ \'{e}lectrique

\begin{equation}
\mathbf{E}_{\parallel }=-\partial _{t}\mathbf{A}_{\parallel }-\mathbf{grad}%
V\qquad \mathbf{E}_{\parallel }^{\prime }=-\partial _{t}\mathbf{A}%
_{_{\parallel }}^{\prime }-\mathbf{grad}V^{\prime }
\end{equation}
s'annule dans le syst\`{e}me prim\'{e} (image) \textit{par compensation )} 
\cite{17bis}\textit{\ }\'{e}tant donn\'{e} que nous avons d'apr\`{e}s le
th\'{e}or\`{e}me de Poincar\'{e} (2.36):

\begin{equation*}
V=A_{\parallel }=0\qquad V^{\prime }=A_{\parallel }^{\prime }=\gamma \beta
A_{x}
\end{equation*}

La formule Doppler de Poincar\'{e} (2.4) se trouve ainsi l\'{e}gitim\'{e}e
puisqu'elle d\'{e}coule du choix de la jauge relativiste. L'analyse de
Rousseaux \cite{25} sur le r\^{o}le fondamental des potentiels se trouve
aussi l\'{e}gitim\'{e}e puisque le choix de jauge peut se traduire par un
effet qui est en principe mesurable.

\subsection{Cin\'{e}matique d'Einstein en jauge transverse (Coulomb
compl\'{e}t\'{e}e)}

Einstein a implicitement adopt\'{e} la jauge (non-relativiste) transverse
(double transversalit\'{e} de l'onde plane, chapitre 2): $A^{\prime }=A=$ $%
A_{\perp }$ avec $A_{\parallel }=A_{\parallel }^{\prime }=0$. Avec
l'invariance de la norme du quadrivecteur potentiel: 
\begin{equation}
V^{2}-A^{2}=V^{^{\prime }2}-A^{\prime 2}  \notag
\end{equation}
on a n\'{e}cessairement 
\begin{equation}
V=V^{\prime }=0
\end{equation}
On voit ainsi que la repr\'{e}sentation einsteinienne de l'onde plane ne
revient pas seulement \`{a} ignorer l'ellipse (et sa tangente) mais bien
plut\^{o}t \`{a} ''l'annuler''(puisqu'elle engendre une composante
longitudinale non-nulle). Il est clair que cela correspond \`{a} la \textbf{%
jauge de Coulomb dite transverse}: 
\begin{equation}
div\text{ }\mathbf{A}=div\text{ }\mathbf{A}^{\prime }=0  \label{60bis}
\end{equation}
car accompagn\'{e}e de ''l'annulation de potentiel scalaire 
\begin{equation}
V=V^{\prime }=0=A_{\parallel }=A_{\parallel }^{\prime }  \tag{2.48bis}
\end{equation}
\ \textit{pour les ondes}''.

Maxwell avait adopt\'{e} la jauge de Coulomb \cite{2bis} et Einstein l'a
suivi sur ce point en ajoutant l'annulation du potentiel scalaire.
L'adoption de celle-ci (2.42bis) par Einstein rend invisible l'adoption de
la jauge de Coulomb. En effet il est bien connu que cette derni\`{e}re
implique la propagation instantan\'{e}e du potentiel scalaire (\'{e}quation
laplacienne). Si on annule la grandeur qui se propage instantan\'{e}ment,
plus rien ne semble se propager instantan\'{e}ment. Mais qu'a-t-on
annul\'{e} avec puisque V n'est pas le potentiel \'{e}lectrostatique car
nous sommes en th\'{e}orie des ondes (la densit\'{e} de charge est nulle $%
\rho =0$ $!).$

L'annulation du ''potentiel des ondes'', $V^{\prime }=0,$ peut donc \^{e}tre
interpr\'{e}t\'{e}e comme la suppression einsteinienne de l'\'{e}ther. 
\textit{Pour la premi\`{e}re fois on peut mettre une \'{e}quation
derri\`{e}re la suppression einsteinienne de l'\'{e}ther }$V^{\prime }=0$.
On n'est pas oblig\'{e} de faire $V^{\prime }=0$. On a alors $V^{\prime
}\neq 0$ et \textbf{l'\'{e}ther relativiste} de Poincar\'{e} (2\`{e}me
partie de la pr\'{e}sente \'{e}tude) selon lequel peut toujours
consid\'{e}rer que l'\'{e}ther est au repos dans l'un des syst\`{e}mes ($%
V=0) $ et alors il ne l'est pas dans l'autre ($V^{\prime }\neq 0)$ (mais
jamais les deux \`{a} la fois, Einstein). Nous avions conjectur\'{e} \cite
{13} que le photon einsteinienne \'{e}tait ins\'{e}parable de la suppression
einsteinienne de l'\'{e}ther. C'est maintenant rigoureusement
d\'{e}montr\'{e} puisque l'annulation de la composante longitudinale
d\'{e}finit le photon einsteinien en jauge de Coulomb en m\^{e}me temps que
la cin\'{e}matique einsteinienne.

Il est \'{e}vident que ce choix einsteinien (2.42bis) ne concerne pas
seulement les ondes planes. Il ne saurait \^{e}tre question de d\'{e}couper
la cin\'{e}matique einsteinienne (en ''tranches coulombiennes'' et
''tranches non-coulombiennes''). Les fronts plans einsteiniens sont tangents
aux fronts sph\'{e}riques einsteiniens (annexe 1). Ces derniers sont
ins\'{e}parables de la convention einsteinienne de synchronisation et des
tiges rigides einsteiniennes (et de sa d\'{e}finition de la contraction).
Remarquons \`{a} cet \'{e}gard qu'on retrouve bien l'annulation du terme
poincar\'{e}en ($V^{\prime }=$ $\Delta t^{\prime }=0).$ C'est bien donc
toute la cin\'{e}matique d'Einstein qui est induite par un choix (implicite)
de jauge, celle de la jauge transverse.

Ceci est d'ailleurs un r\'{e}sultat extraordinaire car les Absolutistes en
tout genre ont toujours essay\'{e} de relier la notion d'Ether \`{a}
l'Absolu (ce qui est contraire \`{a} la lettre et \`{a} l'esprit de
Poincar\'{e} \cite{20}, qui n'a pas cess\'{e} de vouloir les dissocier). Il
s'av\`{e}re dans notre analyse que la th\'{e}orie avec l'\'{e}ther est plus
nettement relativiste (jauge de Lorenz !) que celle fond\'{e}e sur sa
suppression (jauge de Coulomb compl\'{e}t\'{e}e). Il convient \`{a}
pr\'{e}sent de d\'{e}velopper la cin\'{e}matique de Poincar\'{e} en jauge de
Lorenz.

\chapter{D\'{e}finition de la distance sous-jacente \`{a} l'ellipse
allong\'{e}e}

\bigskip\ Dans l'ellipse allong\'{e}e, rien n'est \`{a} jeter. Ainsi nous
avons largement utlis\'{e} la tangente (\S 1.2) mais nous pouvons aussi
utliliser le second foyer.

\section{Dilatation de la dur\'{e}e et de la distance}

Raisonnons tout d'abord \textit{en terme de distance parcourue r' }en
consid\'{e}rant le front d'onde ''avant'' et le front d'onde ''arri\`{e}re''
de l'onde sph\'{e}rique (le cas standard, O' est au premier foyer de
l'ellipse dans K', figure 2 et 3, avec $r=SM_{1}=SM_{2}$). Pour la distance
parcourue ''avant'' $r_{1\text{ }}^{\prime }=O^{\prime }M_{1}^{\prime }$ par
la lumi\`{e}re on a dans K' (1er quadrant $\theta \prec \frac{\pi }{2})$

\begin{equation*}
r_{1}^{\prime }=k(1+\beta \cos \theta )\ \ \ 
\end{equation*}
Pour avoir la distance parcourue ''arri\`{e}re'', $r_{2}^{\prime }=O^{\prime
}M_{2}^{\prime }$, par la lumi\`{e}re dans K, on effectue une rotation de 180%
$°$ . On a (\textbf{figure 10})

\begin{equation*}
r_{2}^{\prime }=k(1-\beta \cos \theta )
\end{equation*}

La distance parcourue ''arri\`{e}re'' est donn\'{e}e dans K par la distance
diam\'{e}tralement ($M_{1}M_{2})$ oppos\'{e}e sur le cercle dans K$.$ Il en
est de m\^{e}me pour la distance ''arri\`{e}re'' $r_{2}^{\prime }$ dans K',
qui est donn\'{e} par la distance $r_{2}^{\prime }$ du point
''diam\'{e}tralement'' ($M_{1}^{\prime }M_{2}^{\prime })$ oppos\'{e}e sur
l'ellipse dans K'.\ D'un point de vue math\'{e}matique, il suffit donc de
consid\'{e}rer la distance $r_{2}^{\prime }$ \`{a} \textit{l'autre foyer F}$%
_{2}$. Etant donn\'{e} que dans une ellipse on a: $r_{1}+r_{2}=2\gamma $. On
a donc une \textit{distance moyenne parcourue} $r_{M}^{\prime }$ par l'onde
''avant'' et l'onde ''arri\`{e}re'': 
\begin{equation}
r_{M}^{\prime }=\frac{r_{1}^{\prime }+r_{2}^{\prime }}{2}=\gamma r
\end{equation}

\textit{Remarquons que cette \'{e}longation de la distance moyenne
r\'{e}sulte imm\'{e}diatement de ce que }$M_{1}$\textit{\ et }$M_{2}$\textit{%
\ sont des \'{e}v\'{e}nements simultan\'{e}s dans K alors que leurs images
dans K' ne sont pas des \'{e}v\'{e}nements simultan\'{e}s.}

Utilisons \`{a} pr\'{e}sent la tangente et le second foyer de l'ellipse.
Installons un miroir parfaitement r\'{e}flechissant M dans une direction
donn\'{e}e dans le syst\`{e}me de la source tangentiellement au cercle. Sur
la figure nous donnons deux exemples (\textbf{figure 11}) $\theta =\frac{\pi 
}{4}$ et $\theta =\frac{\pi }{2}$.

Au temps $t$, l'onde aller est r\'{e}fl\'{e}chie en M$_{1}$ du point de vue
de K et en M'$_{1}$ du point de vue de K'. Le temps aller $t^{\prime +}$
sera donc

\begin{equation}
t^{\prime +}=\gamma t(1+\beta \cos \theta )
\end{equation}

Au temps $2t$, l'observateur O' sera \`{a} la distance $2\beta \gamma $ de
la source situ\'{e}e en F$_{2}=F^{-}$.

Le temps retour sera $t^{\prime -}$

\begin{equation}
t^{\prime -}=\gamma t(1-\beta \cos \theta )
\end{equation}

On donne ainsi un sens physique au \textbf{second foyer }$\mathbf{F}_{2}$
d\'{e}fini comme le point autour duquel l'onde retour se concentre
(propri\'{e}t\'{e} \'{e}l\'{e}mentaire de la r\'{e}flexion sur un miroir
tangent \`{a} l'ellipse).

Cette fois nous obtenons un temps de parcours ''aller- retour moyen'' $%
t_{M}^{\prime }$ pour une onde lumineuse.

\begin{equation}
t_{M}^{\prime }=\frac{t^{^{\prime }+}+t^{\prime -}}{2}=\gamma t
\end{equation}

\bigskip \textit{Remarquons que cette \'{e}longation de la dur\'{e}e moyenne
de parcours r\'{e}sulte imm\'{e}diatement de ce que }$M_{1}$\textit{\ et }$%
M_{2}$\textit{\ sont des \'{e}v\'{e}nements isotopes (au m\^{e}me endroit)
dans K (le syst\`{e}me de la source) alors que leurs images dans K'} (le
syst\`{e}me de l'observateur O')\textit{\ ne sont pas des \'{e}v\'{e}nements
isotopes.} L'ellipse de Poincar\'{e} constitue alors une sorte d'oscillateur
qui g\'{e}n\'{e}ralise ainsi \`{a} toutes les directions (la tangente en
tout point est un miroir id\'{e}al\footnote{%
Cette ''mat\'{e}rialisation'' du front d'onde nous ram\`{e}ne alors \`{a} la
question de l'apparence des objets en mouvement. On sait que cette question
a \'{e}t\'{e} r\'{e}solue par Lampa \cite{8bis}, Penrose \cite{12bis} et
Terrel \cite{27} \textit{sur la base de l'axiomatique einsteinienne}
(notamment la conception einsteinienne de la contraction): les sph\`{e}res
restent des sph\`{e}res et la contraction de Lorentz est invisible. La
solution de Poincar\'{e}, bas\'{e}e sur son hypoth\`{e}se de contraction est
fondamentalement diff\'{e}rente (\cite[annexe]{16}) puisque cette
derni\`{e}re se manifeste par une expansion de l'espace corr\'{e}l\'{e}e
\`{a} un ph\'{e}nom\`{e}ne ''Doppler'' (voir ''synth\`{e}se chapitre 4)}) ce
qu'on peut appeler ''l'aller-retour standard'' o\`{u} un miroir est
plac\'{e} dans la direction du mouvement ou dans la direction
perpendiculaire au mouvement (exemple 2 sur la \textbf{figure 11},
oscillateur d'Einstein, 1913, annexe 1). Ceci correspond aussi \`{a} \textbf{%
l'exp\'{e}rience de Michelson} (mutatis mutandis, \'{e}tant donn\'{e} que
l'image elliptique, 1.15, dans le syst\`{e}me de la source se ram\`{e}ne au
cas standard) ou l'on consid\`{e}re le temps aller-retour longitudinalement (%
$\theta =0)$ et transversalement ($\theta =\frac{\pi }{2})$.

Nous voulons mentionner ici les travaux d'optique exp\'{e}rimentale de N.
Abramson \cite{00} qui observe l'oscillateur de Poincar\'{e} \`{a} 3
dimensions (mutatis mutandis selon son expression, l'exp\'{e}rience de
Michelson) \`{a} l'aide de methodes d'holographie. Il a ainsi construit les
''ellipso\"{i}des d'observation'' au moyen d'une m\'{e}thode graphique
(alors que nous utlisons plut\^{o}t une m\'{e}thode de g\'{e}om\'{e}trie
analytique bas\'{e}e sur la TL). Il a \'{e}galement soulign\'{e},
ind\'{e}pendamment des travaux de Poincar\'{e}, que ce qui est observable
est bien plut\^{o}t une extension qu'une contraction.

Poincar\'{e} conclut \`{a} propos de l'ellipse (en italique dans le texte):

\begin{quotation}
Cette hypoth\`{e}se de Lorentz et FitzGerald para\^{i}tra au premier abord\
fort extraordinaire; tout ce que nous pouvons dire pour le moment en sa
faveur est qu'elle est la traduction imm\'{e}diate du r\'{e}sultat
exp\'{e}rimental de Michelson, \textit{si l'on d\'{e}finit les longueurs par
des temps que la lumi\`{e}re met \`{a} les parcourir}.
\end{quotation}

Quelle que soit la direction, le temps moyen aller-retour sera toujours le
m\^{e}me et \`{a} fortiori le temps moyen aller-retour dans deux directions
perpendiculaires. L'ellipse isotrope de 1906 (istrope cette fois au sens des
temps de parcours aller-retour) explique donc imm\'{e}diatement le
r\'{e}sultat nul de l'exp\'{e}rience de Michelson:

\begin{equation}
\frac{r_{M}^{\prime }}{t_{M}^{\prime }}=\frac{\gamma r}{\gamma t}=c=1
\end{equation}

On voit ainsi que l'on obtient, \`{a} l'aide d'une propri\'{e}t\'{e}
\'{e}l\'{e}mentaire de l'ellipse (utilisation du second foyer), une
dilatation moyenne de la dur\'{e}e \textbf{et de la distance }(3.4 \& 3.1)
(\`{a} 3 dimensions une ''expansion de l'espace'', l'introduction d'une
troisi\`{e}me coordonn\'{e}e z ne pose aucun probl\`{e}me). Cette \textit{%
proportionalit\'{e}} typiquement poincar\'{e}enne r\'{e}sulte
imm\'{e}diatement de la \textit{sym\'{e}trie spatio-temporelle parfaite}
entre le traitement par la TL de deux \'{e}v\'{e}nements isochrones
(simultan\'{e}s) non-isotopes (l'espace) et deux \'{e}v\'{e}nements
non-isochrones isotopes (le temps): dans les deux cas ces deux
\'{e}v\'{e}nements deviennent \textit{deux \'{e}v\'{e}nements non-isochrones
non-isotopes}. Et donc on a

\begin{equation}
r_{M}^{\prime }=\frac{r^{\prime +}+r^{\prime -}}{2}=\gamma r
\end{equation}

Remarquons que cette dilatation moyenne est \textbf{ISOTROPE} aussi bien
pour l'espace que le temps (3.4 \& 3.1): la distance $r_{M}^{\prime }$ est
dilat\'{e}e de la m\^{e}me mani\`{e}re comme $t_{M}^{\prime }$ dans toutes
les directions. C'est la raison pour laquelle Poincar\'{e} propose comme
d\'{e}finition relativiste de la distance, \textbf{''le temps moyen
(aller-retour) de la lumi\`{e}re pour la parcourir''}.

\bigskip Nous n'avons cependant pas encore d\'{e}duit la \textit{contraction
purement longitudinale} des longueurs puisque nous avons obtenu ''au
contraire'' une \textit{dilatation isotrope} des longueurs. Signalons \`{a}
cet \'{e}gard que d'un point de vue historique, Poincar\'{e} insiste non pas
tant sur la somme des temps aller et retour mais plut\^{o}t sur leur \textit{%
diff\'{e}rence (th\'{e}or\`{e}me de Poincar\'{e}, 2.36)}, autrement dit

\begin{equation}
\Delta t^{\prime }=\frac{t^{+}-t^{-}}{2}=\gamma \beta r\cos \theta =\gamma
\beta x
\end{equation}

Cette derni\`{e}re expression d\'{e}finit '' la diff\'{e}rence entre le
temps aller et le temps retour'', autrement dit 'l'\'{e}cart \`{a} la
simultan\'{e}it\'{e}'' qui ne d\'{e}pend que de la diff\'{e}rence d'abscisse 
$r\cos \theta $ entre les deux stations (voir th\'{e}or\`{e}me de
Poincar\'{e}). L'interpr\'{e}tation physique de cette grandeur $\Delta
t^{\prime }$ \`{a} ce stade du raisonnement n'est pas \'{e}vidente mais il
est dors et d\'{e}j\`{a} clair que, mutatis mutandis, elle sera d\'{e}finie
de mani\`{e}re semblable pour l'ellipse dans le syst\`{e}me de la source, en
d'autres termes pour la proc\'{e}dure de synchronisation de Poincar\'{e}.
L'\'{e}cart la simultan\'{e}it\'{e} (3.7) est aussi purement \textbf{%
longitudinal} \textbf{comme la contraction de Lorentz} (voir 2\`{e}me
partie, ce terme d\'{e}finit la diff\'{e}rence entre la synchronisation
d'Einstein et celle de Poincar\'{e}). Remarquons \`{a} pr\'{e}sent que la
dilatation poincar\'{e}enne peut aussi \^{e}tre projet\'{e}e au niveau des
abscisses (voir annexe 4, les directrices de l'ellipse).

\begin{equation}
x_{M}^{\prime }=\frac{x^{\prime +}+x^{\prime -}}{2}=\gamma x  \label{67bis}
\end{equation}

Contrairement \`{a} (3.1) o\`{u} r est fix\'{e}, dans (3.8) x est variable
sur le front objet. Nous obtenons ainsi une \textbf{proportionalit\'{e}
poincar\'{e}enne directe, }observ\'{e}e depuis le syst\`{e}me en mouvement
(ici prim\'{e})\textbf{\ }entre la dilatation du temps et de l'espace; alors
qu'il est bien connu que le traitement einsteinien implique une \textbf{%
proportionalit\'{e} inverse} depuis le syst\`{e}me en mouvement: \textit{%
dilatation} des dur\'{e}es et \textit{contraction }de longueurs.

\section{\protect\bigskip Utilisation asym\'{e}trique de la TL et
d\'{e}finition de la contraction par Einstein}

Nous devons d'abord trouver la source profonde de l'asym\'{e}trie
einsteinienne (compatible avec l'asym\'{e}trie d\'{e}j\`{a} rep\'{e}r\'{e}e
au niveau des fronts d'onde einsteiniens).

\subsection{\textbf{Dilatation de la dur\'{e}e propre selon Einstein}}

En reprenant les notations gr\'{e}co-romaines d'Einstein, consid\'{e}rons
tout d'abord la dilatation du temps.

\ Le temps propre, $T_{0}=\tau _{2}-\tau _{1},$ est la dur\'{e}e entre deux
\'{e}v\'{e}nements au m\^{e}me endroit ($\xi _{1}$ = $\xi _{2}=\xi )$ dans
K. Nous trouvons la dur\'{e}e T dans K' par la seconde TL\textit{:}

\begin{equation*}
t_{1}=\gamma (\tau _{1}-\frac{v}{c^{2}}\xi )\qquad \qquad \qquad
t_{2}=\gamma (\tau _{2}-\frac{v}{c^{2}}\xi )
\end{equation*}

On obtient par simple soustraction la dur\'{e}e, $T=t_{2}-t_{1}$,
observ\'{e}e dans le syst\`{e}me en mouvement K' est 
\begin{equation}
T=\gamma T_{0}
\end{equation}

Avec la premi\`{e}r TL nous remarquons que les deux \'{e}v\'{e}nements
consid\'{e}r\'{e}s ne sont pas au m\^{e}me endroit dans K':

\begin{equation}
x_{1}=\gamma (\xi -v\tau _{1})\qquad \qquad \qquad x_{2}=\gamma (\xi -v\tau
_{2})
\end{equation}

\bigskip Ceci est une cons\'{e}quence du fait bien connu que la dilatation
r\'{e}sulte de ce que nous devons utiliser deux horloges \`{a} des places
diff\'{e}rentes 
\begin{equation}
\Delta x=\gamma vT_{0}
\end{equation}
dans le syst\`{e}me en mouvement. Soulignons qu'il n'y \`{a} une parfaite
similitude entre la dilatation poincar\'{e}enne (3.4) et la dilatation
einsteinienne du temps (3.9).

\subsection{\textbf{Contraction de la longueur propre selon Einstein}}

Consid\'{e}rons \`{a} pr\'{e}sent la contraction einsteinienne. On trouve
dans le texte de 1905 (ainsi que dans tous les livres sur la cin\'{e}matique
relativiste le raisonnement suivant:

La longueur propre $L_{0}=$ $\xi _{2}-\xi _{1}$ est la longueur d'une tige
au repos dans K $(\xi _{2}$, $\xi _{1}$ sont les coordon\'{e}es des deux
extr\'{e}mit\'{e}s de la tige dans K). La longueur de la tige en mouvement
(observ\'{e}e dans K') est alors d\'{e}finie par la distance entre les deux
extr\'{e}mit\'{e}s de la tige simltan\'{e}ment dans K'\textit{\ } ($%
t=t_{1}=t_{2})$. Nous d\'{e}duisons alors imm\'{e}diatement la longueur
impropre $L=$ $x_{2}-x_{1}$, par \textit{inversion} de la \textit{%
premi\`{e}re} TL:

\begin{equation*}
\xi _{1}=\gamma (x_{1}+vt)\qquad \qquad \qquad \xi _{2}=\gamma (x_{2}+vt)
\end{equation*}

On obtient alors par simple soustraction la fameuse contraction
einsteinienne (d\'{e}duite de la premi\`{e}re TL):

\begin{equation}
L=\gamma ^{-1}L_{0}
\end{equation}

\bigskip Les deux d\'{e}ductions einsteiniennes (cf., texte original \cite{4}%
) sont pr\'{e}sent\'{e}es dans tous les livres sur la relativit\'{e} comme
\'{e}tant parfaitement (?) sym\'{e}triques: deux \'{e}v\'{e}nements \`{a} la
m\^{e}me place (dans K) pour la dilatation du temps et deux
\'{e}v\'{e}nements en m\^{e}me temps (\textit{dans K'}) pour la contraction
des longueurs. Cette sym\'{e}trie \'{e}tant proclam\'{e}e parfaite, rien ne
doit nous emp\^{e}cher de consid\'{e}rer la deuxi\`{e}me TL (3.13) dans le
cas de la contraction de la longueur exactement comme nous avons
consid\'{e}r\'{e} la deuxi\`{e}me TL (3.10) dans le cas de la dilatation du
temps:

\begin{equation}
\tau _{1}=\gamma (t+\frac{v}{c^{2}}x_{1})\qquad \qquad \qquad \tau
_{2}=\gamma (t+\frac{v}{c^{2}}x_{2})
\end{equation}

Autrement dit, la seconde TL d\'{e}termine les temps $\tau _{1}$ $et$ $\tau
_{2}$ des deux extr\'{e}mit\'{e}s de la tige $\xi _{1}$et $\xi _{2}$ dans le
syst\`{e}me propre K. Les deux \'{e}v\'{e}nements d\'{e}finis par Einstein
simultan\'{e}s $\Delta t=0$ dans K' $(x_{1},t)$ et $(x_{2},t)$ ne sont pas,
par la TL, des \'{e}v\'{e}nements simultan\'{e}s ($\xi _{1,}$ $\tau _{1})$
and ($\xi _{2}$,$\tau _{2})$ dans K, AUTREMENT DIT DANS LE SYSTEME PROPRE.
Ceci est manifestement en contradiction avec le paragraphe 1 du travail
d'Einstein \cite{4} o\`{u} la simultan\'{e}it\'{e} \`{a} distance est
d\'{e}finie sur une tige (rigide) AB dans K; ce qui signifie \'{e}videmment
que, r\'{e}ciproquement, la tige (rigide) est d\'{e}finie (de fa\c{c}on
relativiste) par deux \'{e}v\'{e}nements SIMULTANES dans le syst\`{e}me
propre (voir annexe 1, la convention de synchronisation d'Einstein).

La m\'{e}trique einsteinienne implique donc une double simultan\'{e}it\'{e}
: la longueur propre (la DISTANCE) est d\'{e}finie par deux
\'{e}v\'{e}nements simultan\'{e}s dans K, $\Delta t=0,$ et la longueur
impropre (d'une tige en mouvement) par deux \'{e}v\'{e}nements simutan\'{e}s
dans K'. Mais \ $(x_{1},t)$ et $(x_{2},t)$ \textit{n'est pas} l'image par la
TL de \ $(\xi _{1},\tau )$ et $(\xi _{2},\tau ).$

D'apr\`{e}s (3.13), ''l'\'{e}cart la simultan\'{e}it\'{e}'' vaut par
sym\'{e}trie avec ''l'\'{e}cart \`{a} l'isotopie'' (3.11) 
\begin{equation}
\Delta \tau =\gamma \frac{v}{c^{2}}L_{0}
\end{equation}
dans le syst\`{e}me propre. On retrouve ainsi le terme poincar\'{e}en (2.36
ou 3.7). \textit{C'est ce terme qui est ANNULE par Einstein, } 
\begin{equation}
(\Delta \tau )_{einstein}=0  \tag{3.14bis}
\end{equation}
\textit{dans son interpr\'{e}tation de la contraction}; et cela en
coh\'{e}rence parfaite avec non seulement de sa th\'{e}orie des fronts
d'onde plans (2.36 jauge transverse de Coulomb) mais aussi avec sa
convention de synchronisation (autrement dit des fronts sph\'{e}riques de
synchronisation, voir annexe 1 et \S 2). Toute coh\'{e}rente qu'elle soit,
la cin\'{e}matique einsteinienne n'en est pas moins fond\'{e}e sur un
traitement asym\'{e}trique (voir conclusion) du terme $vt$ dans la 1\`{e}re
TL (3.11 qui n'est pas annul\'{e}) et du terme $\frac{v}{c^{2}}x$ dans la
seconde TL qui est annul\'{e} (3.14bis) (conclusion: la sym\'{e}trie de
Poincar\'{e}).

La contraction einsteinienne n'est donc pas imm\'{e}diatement d\'{e}duite de
la TL: elle constitue \textit{une hypoth\`{e}se suppl\'{e}mentaire} de 
\textbf{d\'{e}finition de la distance} bas\'{e}e sur l'utilisation
asym\'{e}trique d'une seule TL. Cette d\'{e}finition einsteinienne de la
distance impropre est incompatible avec celle de Poincar\'{e}. Remarquons
toutefois que l'annulation 3.14bis constitue une approximation valable pour
les petites distances ($\frac{v}{c^{2}}x<<vt)$.

\section{Utilisation sym\'{e}trique de la TL et d\'{e}finition cosmologique
de la distance par Poincar\'{e}}

Montrons tout d'abord que la dilatation poincar\'{e}enne des longueurs
d\'{e}duite de l'ellipse est directement inscrite dans une utilisation
parfaitement sym\'{e}trique de la TL. Insistons sur le fait que le
probl\`{e}me ne se situe pas au niveau de la dur\'{e}e impropre ''en
mouvement'' mais de la distance impropre ''en mouvement''. La dur\'{e}e
impropre \'{e}tant donn\'{e}e par deux \'{e}v\'{e}nements non-isotopes, il
suffit d\`{e}s lors de consid\'{e}rer \ que la distance impropre sera
d\'{e}finie comme une distance entre deux \'{e}v\'{e}nements non-isochrones.

Pour \'{e}liminer l'asym\'{e}trie einsteinienne, il suffit d\`{e}s lors de
bien d\'{e}finir les \'{e}v\'{e}nements consid\'{e}r\'{e}s (comme pour
Einstein, nous ne consid\'{e}rons que la situation longitudinale, pour la
g\'{e}n\'{e}ralisation \`{a} toute direction, voir deuxi\`{e}me partie).

Pour le temps, on consid\`{e}re la TL des deux \'{e}v\'{e}nements isotopes
non-isochrones ($0,0)$ et ($0,T_{0})$

\bigskip Un calcul imm\'{e}diat donne ($0,0)$ et ($\gamma \beta T_{0},\gamma
T_{0})$ 
\begin{equation}
T=\gamma T_{0}
\end{equation}
\textbf{DEFINITION 1:} \textit{la dur\'{e}e dans K' est la dur\'{e}e
dilat\'{e}e entre deux \'{e}v\'{e}nements non-isotopes.}

L'\'{e}cart \`{a} l'isotopie \'{e}tant:

\begin{equation}
\Delta x^{\prime }=\gamma \beta T_{0}  \tag{3.15bis}
\end{equation}

Pour l'espace, on consid\`{e}re la TL les deux \'{e}v\'{e}nements
non-isotopes isochrones ($0,0)$ et ($L_{0},0)$

\bigskip Un calcul imm\'{e}diat donne ($0,0)$ et ($\gamma L_{0},\gamma \beta
L_{0})$

\begin{equation}
L=\gamma L_{0}
\end{equation}
\textbf{DEFINITION 2:} \textit{la distance dans K' est la distance
dilat\'{e}e entre deux \'{e}v\'{e}nements non-isochrones.}

L'\'{e}cart \`{a} la simultan\'{e}it\'{e} \'{e}tant: 
\begin{equation}
\Delta t^{\prime }=\gamma \beta L_{0}  \tag{3.16bis}
\end{equation}

La d\'{e}finition 2 est le sym\'{e}trique parfait de la d\'{e}finition 1.
Cette d\'{e}finition poincar\'{e}enne relativiste de la distance impropre
est donc profond\'{e}ment non-einsteinienne. Dans le syst\`{e}me
d'unit\'{e}s de Poincar\'{e}, les deux \'{e}carts (3.15bis \& 3.16bis) sont
par d\'{e}finition \textbf{du m\^{e}me ordre de grandeur}. Remarquons qu'il
s'agit de la distance impropre au sens bien connu des astronomes ou des
cosmologistes. Citons \`{a} cet \'{e}gard Lachi\`{e}ze-Rey (nous mettons en
italique) \cite{8}

\begin{quotation}
Une difficult\'{e} fondamentale impr\`{e}gne la cosmologie et il est capital
de la garder \`{a} l'esprit. Toute l'information sur telle ou telle partie
de l'univers nous parvient par l'interm\'{e}diaire du rayonnement
\'{e}lectromagn\'{e}tique qui se propage \`{a} une vitesse finie, celle de
la lumi\`{e}re c. \ Aucun des objets que nous observons ne nous est donc
contemporain mais se r\'{e}v\`{e}le tel qu'il \'{e}tait \`{a} une \'{e}poque
vieille de L/c (...). L'ensemble des objets qui nous sont accessibles par
l'observation n'est donc pas situ\'{e} dans l'espace pris \`{a} un instant
donn\'{e}. Au contraire, on peut dire que chacun se situe dans une tranche
d'espace-temps \textit{dont l'\'{e}loignement dans le temps est
proportionnel \`{a} l'eloignement dans l'espace}; la relation entre les
intervalles de temps T et d'espace L nous s\'{e}parant de lui est donn\'{e}e
par la loi de propagation de la lumi\`{e}re c=L/T.
\end{quotation}

Lachi\`{e}ze-Rey ajoute une note p 72 (et ensuite une deuxi\`{e}me note p73)

\begin{quotation}
Une difficult\'{e} suppl\'{e}mentaire provient d'ailleurs du fait que L
n'est pas une \textit{VRAIE } distance puisque c'est un intervalle de
longueur entre la position occup\'{e}e par nous 'l'observateur) aujourd'hui
et la galaxie observ\'{e}e \`{a} un instant diff\'{e}rent.

Rappelons-nous n\'{e}anmoins que la distance ainsi consid\'{e}r\'{e}e n'est
pas une distance au sens propre du mot puisqu'elle mesure un intervalle
spatial entre deux \'{e}v\'{e}nements- l'\'{e}mission et la r\'{e}ception
d'un signal lumineux- non contemporains.
\end{quotation}

\bigskip Il est donc \'{e}vident que la d\'{e}finition poincar\'{e}enne
relativiste de la distance est la distance utilis\'{e}e par les
cosmologistes, laquelle devient donc une VRAIE distance au sens de la TL. Au
sujet de la seconde note de Lachi\`{e}ze-Rey, il est non moins \'{e}vident
que la distance consid\'{e}r\'{e}e par les cosmologistes est une \textbf{%
distance impropre} au sens de Poincar\'{e} (entre deux \'{e}v\'{e}nements
non-simultan\'{e}s 3.16) \textbf{compl\`{e}tement incompatible avec une
distance impropre au sens d'Einstein (3.12)}. Cette harmonie parfaite entre
la distance au sens de Poincar\'{e} et la distance au sens cosmologique
signifie peut-\^{e}tre que la m\'{e}trique spatio-temporelle de
Poincar\'{e}, sous-jacente \`{a} la cin\'{e}matique en jauge de Lorenz, a
quelque chose d'important (voire m\^{e}me de tr\`{e}s important) \`{a}
apporter \`{a} la cosmologie.

\section{D\'{e}finition sym\'{e}trique des unit\'{e}s spatiotemporelles $%
"c=1"$ et m\'{e}trique de Poincar\'{e}}

Rappelons que le fondement m\^{e}me de l'id\'{e}e d'une ''structure fine''
de la relativit\'{e} \cite{13} est que, \'{e}tant donn\'{e} l'isotropie de
la vitesse de la lumi\`{e}re, on peut d\'{e}finir deux m\'{e}triques
spatio-temporelles distinctes. L'utilisation compl\`{e}tement sym\'{e}trique
de la TL est en harmonie parfaite avec l'ellipse allong\'{e}e isotrope (3.1,
3.4, 3.5). A ce stade, nous pouvons enfin comprendre dans les moindres
d\'{e}tails\textbf{\ les unit\'{e}s }$\mathbf{c=1}$ de Poincar\'{e}
(chapitre 1)) ''les unit\'{e}s d'espace et du temps sont d\'{e}finies de
telle mani\`{e}re que la vitesse que $c=1$''. 
\begin{equation}
\frac{L}{T}=\frac{\gamma L_{0}}{\gamma T_{0}}=c=1.
\end{equation}

Les unit\'{e}s dans le syst\`{e}me en mouvement sont par d\'{e}finition les
''moyennes'' dilat\'{e}es aller-retour'' du signal lumineux: $\gamma L_{0}$
et $\gamma T_{0}.$ Une telle d\'{e}finition des unit\'{e}s n'aurait aucun
sens avec la convention d'Einstein puisque la vitesse de la lumi\`{e}re
serait \'{e}gale \`{a} $\gamma ^{-2}c$ (voir 2\`{e}me partie). Un
\'{e}tudiant de premi\`{e}re candidature qui aurait fait le rapport entre la
longueur contract\'{e}e et le temps dilat\'{e} pour trouver l'invariance de
la vitesse de la lumi\`{e}r aurait eu ''z\'{e}ro'' sur la base de
l'axiomatique einsteinienne (voir 2\`{e}me partie le terme $\gamma ^{2}$
dans le raisonnement de Poincar\'{e}). Soulignons trois points essentiels de
la m\'{e}trique de Poincar\'{e}:

$1°)$ Le caract\`{e}re\textbf{\ physiquement} concret d'une telle
d\'{e}finition des unit\'{e}s spatio-temporelles uniquement en termes de
trajet aller-retour de la lumi\`{e}re.

2$°)$ La m\'{e}trique de Poincar\'{e} est fond\'{e}e sur un traitement
parfaitement sym\'{e}trique des termes de m\^{e}me ordre de grandeur
''\'{e}cart \`{a} l'isotopie'' (3.15bis) et ''\'{e}cart \`{a} la
simultan\'{e}it\'{e}'' (3.16bis).

$3°)$ Rien ne change au niveau de la dilatation du temps. Les
exp\'{e}riences qui mesurent la dilatation du temps ne permettent pas de
distinguer les deux cin\'{e}matiques.

La ''structure fine'' se situe au niveau des distances: dilatation de
Poincar\'{e} (3.5 avec 3.16) et contraction d'Einstein (3.12).

Une seule objection peut encore \^{e}tre prise en compte: la contraction de
Lorentz. Est-ce que les unit\'{e}s de Poincar\'{e} s'articulent de fa\c{c}on
coh\'{e}rente avec la contraction des longueurs et la TL? Plus
pr\'{e}cis\'{e}ment encore: comment articuler d'une part la contraction
anisotrope (purement longitudinale) des longueurs, au sens lorentzien
autrement dit dans le syst\`{e}me de la source (2\`{e}me partie), et d'autre
part l'expansion isotrope de l'espace? Ceci nous am\`{e}nera \`{a} la
deuxi\`{e}me partie du pr\'{e}sent travail: la contraction et la
synchronisation de Poincar\'{e} dans le syst\`{e}me de la source.

\chapter{\protect\bigskip Expansion isotrope de l'espace et formule Doppler
relativiste de Poincar\'{e}}

Tentons une premi\`{e}re synth\`{e}se de nos r\'{e}sultats (expansion
isotrope de l'espace et formule Doppler pour les objets lointains) \`{a}
propos de l'ellipse d'observation. La formule (2.28)

\begin{equation}
\frac{r_{\infty }^{\prime }}{r_{\infty }}=\frac{\lambda ^{\prime }}{\lambda }%
=\gamma (1+\beta \cos \theta )
\end{equation}
signifie qu'il y a un lien relativiste non pas seulement entre la longueur
d'onde (ou la fr\'{e}quence) et la vitesse, comme dans le cas d'un Doppler
pr\'{e}relativiste, mais entre la distance r et la longueur d'onde $\lambda $%
. Ou plus exactement il doit exister (pour les objets lointains) un lien
relativiste entre le rapport des distances et le rapport des longueurs
d'onde. En cas de redshift $\lambda ^{\prime }>\lambda $, par exemple, cette
relation fondamentale de la cin\'{e}matique poincar\'{e}enne, signifie non
pas seulement un \textit{\'{e}loignement} (au sens pr\'{e}relativiste) mais
une \textit{extension }de la distance $r^{\prime }>r$ entre l'observateur et
l'objet lointain (d\'{e}finition cosmologique de la distance et onde plane).
Nous avons par ailleurs montr\'{e} que la formule Doppler relativiste
d'Einstein avait un s\'{e}rieux probl\`{e}me \`{a} propos de
l'\'{e}loignement pr\'{e}relativiste\cite{17}

\begin{quotation}
We note that Einstein's relativistic Doppler formula presents a strange
aspect.\ For incident light received under a fixed non zero angle, the
Doppler shift will change from blueshift to redshift (or vice-versa) for
some critical relative velocity of the observer and the source.
\end{quotation}

\bigskip Pour avoir un effet Doppler nul, il suffit de poser: 
\begin{equation}
D=\frac{1-\frac{v}{c}\cos \theta }{\sqrt{1-\frac{v^{2}}{c^{2}}}}=1 
\tag{doppler 1}
\end{equation}

et donc on trouve pour toute vitesse un angle pour lequel il n'y a pas de
Doppler $D=1$ 
\begin{equation}
\cos \theta _{t}=\frac{1-\sqrt{1-\frac{v^{2}}{c^{2}}}}{\frac{v}{c}}\approx 
\frac{1}{2}\frac{v}{c}  \tag{doppler 2}
\end{equation}

Il est \'{e}crit partout dans la litt\'{e}rature scientifique sur la
relativit\'{e} restreinte que la\ principale diff\'{e}rence entre la formule
pr\'{e}relativiste et la formule relativiste serait que celle-ci pr\'{e}voit
un effet transversal non nul alors que celle-l\`{a} prevoit un effet
transversal nul. Il s'agit certes d'une diff\'{e}rence mais certainement pas
de la diff\'{e}rence principale. La diff\'{e}rence \textbf{principale} est
que \textbf{l'effet nul ne se produit plus pour la direction transversale} $%
\theta =\frac{\pi }{2}$ ind\'{e}pendamment de \textbf{l'intensit\'{e} }de la
vitesse mais pour un angle $\theta $ quelconque qui d\'{e}pend de \textbf{%
l'intensit\'{e}} de la vitesse. L'effet relativiste nul ne correspond plus
\`{a} un changement de sens de la vitesse (le cosinus change de signe,
passage de l'\'{e}loignement au rapprochement). Pour toute direction (sauf
longitudinale) on passe d'un blueshift $(D>1)$ \`{a} un redshift $(D<1)$ (ou
inversement) pour une vitesse non-nulle bien d\'{e}termin\'{e}e par la
formule (doppler1). Nous avions conclu:

\begin{quotation}
We will prove in another communication that this very serious problem is
connected with \textit{Poincar\'{e}'s elongated ellipse} from which we
obtain another definition of the units of space and time and also \textit{%
another relativistic Doppler formula} (reference 2).
\end{quotation}

\bigskip La correction relativiste $\gamma $ dans la formule einsteinienne
suppose donc une r\'{e}volution bien plus importante\ que la simple
pr\'{e}diction d'un effet transversal puisque le lien pr\'{e}relativiste
redshift-\'{e}loignement et blueshift-rapprochement dispara\^{i}t sauf pour
la direction longitudinale\footnote{%
Exp\'{e}rimentalement on v\'{e}rifie toujours la formule pour un angle
tr\`{e}s petit (cf. Mandelberg).}. Or la formule Doppler de Poincar\'{e} est
structurellement la m\^{e}me que celle d'Einstein. \textbf{Tout est donc
dans l'interpr\'{e}tation de l'effet sp\'{e}cifiquement relativiste du
second ordre} $\gamma .$ Chez Einstein, la correction relativiste du second
ordre est un effet de dilatation du temps.

\bigskip Qu'en est-il chez Poincar\'{e}? Consid\'{e}rons \`{a} cet \'{e}gard
la d\'{e}finition poincar\'{e}enne aller-retour des unit\'{e}s de mesure, on
a imm\'{e}diatement

\begin{equation}
\frac{r_{M}^{\prime }}{r_{M}}=\frac{\lambda _{M}^{\prime }}{\lambda _{M}}%
=\gamma
\end{equation}

Le terme du second ordre, dans la cin\'{e}matique de Poincar\'{e}, traduit
une \textbf{extension de la distance} ou une expansion \textit{isotrope}
(r\'{e}ciproque) de l'espace traduite n\'{e}cessairement par un \textbf{%
redshift} pour les objets lointains (par exemple des galaxies) \`{a}
condition d'admettre le syst\`{e}me d'unit\'{e}s de Poincar\'{e} et donc que
la longueur d'onde \'{e}mise par un atome suppose un ''aller-retour'',
autrement dit une oscillation subatomique\footnote{%
Selon l'\'{e}tat de la physique th\'{e}orique actuelle la question de savoir
pourquoi l'espace se dilate \`{a} grande \'{e}chelle $r_{M}^{\prime }$ et ne
se dilate pas \`{a} petite \'{e}chelle $\lambda _{M}^{\prime }$ (les atomes
ne se dilatent pas) n'a qu'une explication n\'{e}gative (pas de gravitation
au niveau atomique). La ''structure fine'' de la Relativit\'{e} Restreinte
donne une explication positive: il y a deux choix (jauges) possibles pour la
d\'{e}finition de la distance.}. La cin\'{e}matique de Poincar\'{e}
appara\^{i}t ainsi, \`{a} travers les relations 4.1 \& 4.2, comme une
v\'{e}ritable th\'{e}orie ondulatoire ($\lambda ,$ $\nu )$ de l'espace-temps 
$(r,$ $t)$.\ \textbf{''L'expansion ou le Redshift''} est donc un \textbf{%
effet relativiste du second ordre} qui corrige le Doppler-\'{e}loignement
(le Doppler est une relation fr\'{e}quence-vitesse) pr\'{e}relativiste. Nous
sugg\'{e}rons d'appeler cet effet relativiste d'expansion \textbf{''effet
Bondi''} en hommage \`{a} la ''m\'{e}thode du ''Doppler-Radar'' dite du
facteur k'' de Milne et de Bondi\footnote{%
La d\'{e}fintion poincar\'{e}enne de la distance avec des temps de parcours
de la lumi\`{e}re est le point de d\'{e}part de la m\'{e}thode du RADAR
(d\^{i}te du ''facteur k'') de Milne (relativit\'{e} cin\'{e}matique) et de
Bondi (steady state). Mais ce dernier ne va pas jusqu'au bout de son
raisonnement (purement longitudinal, il ne connaissait pas l'ellipse de
Poincar\'{e}) en ne proposant pas une nouvelle interpr\'{e}tation de la
contraction et donc une nouvelle d\'{e}finition rigoureuse de la distance.}).

A\ trois dimensions cet ''effet Bondi'' signifie que la relativit\'{e}
cin\'{e}matique de Poincar\'{e} pr\'{e}voit une expansion isotrope de
l'espace directement traduite par une formule Doppler qui est elle-m\^{e}me
directement li\'{e}e \`{a} la \textbf{structure de l'espace-temps}
(4-vecteur d'onde de Poincar\'{e}, 2.1). En d'autres termes, la dilatation
du temps dans la formule Doppler d'Einstein n'a rien \`{a} voir avec le
redshift. Mais la dilatation du temps traduite en terme d'expansion de
l'espace dans la cin\'{e}matique de Poincar\'{e} est directement li\'{e}e au
redshift\footnote{%
Ceci est en rapport avec le fait (voir annexe 2.1) que la longueur d'onde
einsteinienne ne se tranforme pas comme la longueur de la tige
(contract\'{e}e).}.

Ouvrons ici une parenth\`{e}se cosmologique. D'une certaine mani\`{e}re la
relation corrolaire (4.1) \'{e}voque irr\'{e}sistiblement la formule de
Lema\^{i}tre o\`{u} on aurait remplac\'{e} le facteur d'\'{e}chelle de
dilatation, $a^{\prime }(t^{\prime })$ et $a(t),$ respectivement dans le
syst\`{e}me de l'observateur et de la source (la galaxie), par $r^{\prime }$
et $r$, es distances \'{e}tant d\'{e}finies selon l'usage en astronomie et
en cosmologie (definition 2). \ On sait que notre estim\'{e} coll\`{e}gue
Lema\^{i}tre, d'une universit\'{e} g\'{e}ographiquement voisine (Louvain)
\`{a} la n\^{o}tre (Bruxelles), avait introduit, \`{a} la suite de Friedman,
ce facteur ad hoc d'expansion ($a^{\prime }(t^{\prime })$ et $a(t))$
irr\'{e}m\'{e}diablement coupl\'{e} avec l'hypoth\`{e}se d'une cr\'{e}ation
de l'univers. Gr\^{a}ce \`{a} la relativit\'{e} cin\'{e}matique de
sous-jacente \`{a} l'ellipse de Poincar\'{e} (2005), nous pouvons maintenant
lui r\'{e}pondre, comme Laplace, que \textit{''nous n'avons pas eu besoin de
cette hypoth\`{e}se''} (suppl\'{e}mentaire) pour expliquer l'expansion
isotrope de l'espace. Toutefois l'id\'{e}e de Lema\^{i}tre selon laquelle le
redshift des galaxies traduit une expansion de l'espace et non pas un
\'{e}loignement pr\'{e}relativiste (ce qui \'{e}tait l'opinion de Hubble)
demeure intacte. Fermons ici la parenth\`{e}se cosmologique.

\part{Ellipse dans le syst\`{e}me de la source (source au foyer)}

\chapter{\protect\bigskip Contraction de Lorentz et synchronisation des
horloges ''en onde ellipso\"{i}dale''}

Ce qu'il reste \`{a} \'{e}claircir c'est le\ statut physique de l'ellipse
allong\'{e}e dans le syst\`{e}me de la source (rappelons que
math\'{e}matiquement 1.15 elle r\'{e}sulte imm\'{e}diatement de la TL
inverse) ainsi que le statut de la contraction des longueurs chez
Poincar\'{e}. Certains auteurs (\cite{12}) n'ont pas h\'{e}sit\'{e} \`{a}
\'{e}crire que Poincar\'{e} n'avait pas compris que la contraction \'{e}tait
une cons\'{e}quence directe de la TL (comme chez Einstein). Une telle
affirmation est pour le moins h\^{a}tive puisque nous avons montr\'{e} que
la contraction einsteinienne n'\'{e}tait compatible qu'avec une seule TL.

Il faut cependant reconna\^{i}tre que le lien entre contraction et TL n'a
jamais \'{e}t\'{e} clairement explicit\'{e} par Poincar\'{e}: il s'est
toujours content\'{e} de r\'{e}p\'{e}ter (en 1906, 1908, 1909, 1912) que
''\'{e}tant donn\'{e} la contraction des unit\'{e}s, les sph\`{e}res
para\^{i}tront des ellipso\"{i}des allong\'{e}s''.

L'articulation entre la TL et la contraction est donc plus ''subtile'' que
ce que Pais imaginait. A ce stade du raisonnement la physique th\'{e}orique
ne dispose donc plus d'une th\'{e}orie compl\`{e}tement relativiste de la
contraction . Nous devons donc encore montrer que c'est l'inverse de
l'affirmation de Pais qui est vraie: c'est la contraction selon Poincar\'{e}
qui est profond\'{e}ment inscrite dans la TL (compatible avec ''toute la
TL'' et ''rien que la TL'').

\section{S\'{e}quence historique (1900-1912)}

D\`{e}s 1900, Poincar\'{e} traite de fa\c{c}on approch\'{e}e (au premier
ordre) le probl\`{e}me de synchronisation des horloges par \'{e}change de
signaux lumineux (''La th\'{e}orie de Lorentz et le principe de
r\'{e}action''). Il annonce que pour obtenir une synchronisation exacte, il
faudra tenir compte de la contraction de Lorentz (au deuxi\`{e}me ordre).
Pr\'{e}cisons d'embl\'{e}e que c'est l'exp\'{e}rience de Michelson (pourtant
sans horloge) qui induit chez Poincar\'{e} cette r\'{e}flexion sur la
n\'{e}cessit\'{e} de r\'{e}gler les horloges de telle fa\c{c}on que la
vitesse de la lumi\`{e}re soit la m\^{e}me dans les deux syst\`{e}mes. En
effet l'observateur Michelson, la source de lumi\`{e}re S, le miroir de
l'interf\'{e}rom\`{e}tre M et la Terre sont suppos\'{e}s entra\^{i}n\'{e}s
dans le m\^{e}me systeme (K) \guillemotleft en mouvement \guillemotright\ de
translation uniforme par rapport \`{a} \guillemotleft\ l'\'{e}ther par
d\'{e}finition au repos \guillemotright\ (dans le syst\`{e}me K').

Il est clair que le probl\`{e}me de synchronisation des horloges se pose
selon Poincar\'{e} dans le syst\`{e}me entra\^{i}n\'{e} K car dans le
syst\`{e}me K' la solution du probl\`{e}me est \'{e}vidente puisqu'une
source plac\'{e}e en O' (origine de K') est alors au repos par rapport au
milieu de propagation des ondes (le temps aller $t^{\prime +}=$ le temps
retour $t^{\prime -}$ ). Le couple de syst\`{e}mes inertiels de Poincar\'{e}
est donc bien d\'{e}fini: \'{e}tant donn\'{e} que les ondes sont
sph\'{e}riques dans K', comment va-t-on synchroniser dans K?

Le probl\`{e}me de synchronisation, tel qu'il est pos\'{e} par Poincar\'{e}
entre 1900 et 1912, poss\`{e}de une solution einsteinienne imm\'{e}diate.\
En effet, comme le concept de vitesse relative par rapport \`{a} l'\'{e}ther
n'a aucun sens selon Einstein, on a dans le syst\`{e}me propre source-
Michelson (voir annexe 1): $t^{+}=$ $t^{-}$($\Delta t=0).$ Nous tentons ici
de comprendre pourquoi Poincar\'{e} a travaill\'{e} pendant 12 ans sur un
probl\`{e}me qui poss\`{e}de une solution einsteinienne imm\'{e}diate.

En 1904, Poincar\'{e} reformule la synchronisation des horloges au premier
ordre et il introduit (dans une autre partie de la conf\'{e}rence faite
\`{a} St Louis) pour la premi\`{e}re fois l'ellipse allong\'{e}e mais comme
une solution \textit{alternative} \`{a} la contraction de Lorentz. Il faut
alors abandonner l'isotropie de la vitesse de la lumi\`{e}re et les ondes de
lumi\`{e}re se propageraient plus vite dans la direction longitudinale que
dans la direction transvsersale au mouvement de K par rapport \`{a} K'. Dans
l'esprit de Poincar\'{e} ''l'invariance de la vitesse de la lumi\`{e}re''
est donc indissociablement coupl\'{e}e avec ''la contraction de Lorentz''.

Dans le travail de \textit{dynamique} relativiste de 1905 de Poincar\'{e} 
\cite{18bis}, il n'est nulle part question de synchronisation des horloges
(et a fortiori, du temps vrai et du temps local).

En 1906 \cite{19}, Poincar\'{e} d\'{e}crit pour la premi\`{e}re fois
l'ellipse allong\'{e}e comme \'{e}tant la solution exacte au probl\`{e}me de
la synchronisation des horloges : il ne consid\`{e}re plus l'allongement des
ondes de lumi\`{e}re comme une alternative \`{a} la contraction mais comme
une \textit{cons\'{e}quence} de la contraction de Lorentz. C'est dans son
cours \`{a} la Sorbonne qu'il \'{e}nonce pour la premi\`{e}re fois ce que
nous avons appel\'{e} ''le th\'{e}or\`{e}me de Poincar\'{e}'' (2.35) en
utilisant les deux foyers de l'ellipse ($t^{+}$ et $t^{-}).$

En 1908 \cite{21}, sous la rubrique du principe de relativit\'{e}, il traite
essentiellement de l'ellipse allong\'{e}e mais en ne consid\'{e}rant plus
qu'un seul foyer ($t^{+})$. Et, comme T. Damour l'a mis en \'{e}vidence \cite
{2ter}, Poincar\'{e} se trompe\footnote{%
L'\'{e}quation de l'ellipse dans les notations de Poincar\'{e} est $%
AB+\varepsilon AB^{\prime }=ct.\sqrt{1-\varepsilon ^{2}}$ (avec le
param\`{e}tre de l'ellipse $p=\gamma ^{-1}=k^{-1}).$ Traduite avec nos
pr\'{e}sentes notations cela donne: $t=t^{\prime }\sqrt{1-\beta ^{2}}-\beta
x^{\prime }$ (notre \'{e}quation 7 invers\'{e}e puisque la source est au
foyer), ce qui revient \`{a} \'{e}crire la TL (les deux TL)$.$ Mais juste
apr\`{e}s dans le texte, Poincar\'{e} \'{e}crit \textit{''Supposons que la
diff\'{e}rence entre le temps vrai et le temps local en un point quelconque
soit \'{e}gale \`{a} l'abscisse de ce point multipli\'{e} par la
constante...''} Cela donne alors selon Poincar\'{e} $t=t^{\prime }-\beta 
\frac{x^{\prime }}{\sqrt{1-\beta ^{2}}}.$ Comme le souligne tr\`{e}s
justement T. Damour \textit{cette \'{e}quation est fausse}. Il eut fallu
\'{e}crire \`{a} partir de l'ellipse: $\frac{t}{\sqrt{1-\beta ^{2}}}%
=t^{\prime }-\varepsilon \frac{x^{\prime }}{\sqrt{1-\beta ^{2}}}$. En fait
Poincar\'{e} semble penser qu'il peut exprimer son th\'{e}or\`{e}me en
soustrayant ($t-t^{\prime })$ de la m\^{e}me fa\c{c}on que ($t^{+}$ - $%
t^{-}),$ ce qui n'est pas juste pr\'{e}cis\'{e}ment en fonction de
l'\'{e}quation de l'ellipse. En 1912 Poincar\'{e}, quelques semaines avant
sa mort, \'{e}liminera cette erreur malencontreuse en se servant de la
directrice de l'ellipse \cite{24}.} en formulant son th\'{e}or\`{e}me non
pas en fonction de $t^{+}$ $-t^{-}$ (1906)\ mais en fonction de $t-t^{\prime
},$ ce qui est incompatible avec l'\'{e}quation de l'ellipse.

En 1912, il corrige son erreur \`{a} l'aide de la directrice de l'ellipse
devant un auditoire d'\'{e}l\`{e}ves-ing\'{e}nieurs t\'{e}l\'{e}graphistes%
\footnote{%
Ce qui, soit dit en passant souligne l'extr\^{e}me importance de la
synchronisation des horloges (pour Poincar\'{e} comme pour Einstein) \cite
{5bis} avec un signal \textit{physique }(la lumi\`{e}re) si l'on veut donner
un sens \textit{physique} au symbole ''t''. Si ''c'' n'est plus la vitesse
de la lumi\`{e}re, ''t'' n'est plus le temps et nous obtenons alors une
m\'{e}tath\'{e}orie (une m\'{e}taphysique) de la relativit\'{e} (Ignatovski
1911).}.

La synth\`{e}se relativiste de Poincar\'{e} est magistrale : \'{e}tant
donn\'{e} que la vitesse de la lumi\`{e}re est invariante (hypoth\`{e}se 1)
et que les m\`{e}tres sont contract\'{e}s (hypoth\`{e}se 2) dans K,
l'observateur O entra\^{i}n\'{e} avec la source dans K ''voit'' des ondes
ellipsoidales (alors que O' ''voit'' des ondes sph\'{e}riques non
concentriques \'{e}mises par S). Les ondes ellipsoidales constituent donc
selon Poincar\'{e} ''l'essence m\^{e}me de la relativit\'{e}'' car d'un
point de vue pr\'{e}relativiste les ondes sont toujours sph\'{e}riques dans
les deux syst\`{e}mes.

\section{Contraction interne et propagation externe selon Poincar\'{e}}

Il est maintenant possible de d\'{e}finir de fa\c{c}on pr\'{e}cise la
diff\'{e}rence entre \textit{la contraction selon Einstein} et \textit{la
contraction selon Poincar\'{e}}. On trouve partout dans la litt\'{e}rature
scientifique l'opposition entre une contraction qui serait r\'{e}elle (et
donc non-r\'{e}ciproque) ''\`{a} la Lorentz-Poincar\'{e}'' et qui serait
apparente (et donc r\'{e}ciproque) ''\`{a} la Einstein''. Cette opposition
entre ''r\'{e}alit\'{e}'' et ''apparence'' est plus m\'{e}taphysique que
physique. La physique se situe au niveau de la mise en mouvement (donc de
l'acc\'{e}l\'{e}ration \textit{adiabatique} des tiges). Lorsque les deux
syst\`{e}mes K et K' sont au repos, les tiges sont identiques aussi bien
pour Poincar\'{e} que pour Einstein. Si on acc\'{e}l\`{e}re le syst\`{e}me K
jusqu'\`{a} une vitesse uniforme, la tige est identique dans K selon
Einstein (elle est observ\'{e}e plus courte par K') tandis qu'elle est
CONTRACTEE DANS K selon Poincar\'{e}.

Ce qui est physique \`{a} travers les effets nuls ou non nuls de
l'acc\'{e}l\'{e}ration adiabatique\footnote{%
Nous avons montr\'{e} que les transformations adiabatiques de Poincar\'{e}
\'{e}taient isobariques tandis que celles d'Einstein-Planck \'{e}taient
isentropiques et isochoriques (''La RR avec entropie invariante
d'Einstein-Planck et la RR avec action invariante de Poincar\'{e}'', Annales
de la fondation de Broglie).}, c'est bien plut\^{o}t l'opposition, \textit{%
contraction interne} \`{a} K chez Poincar\'{e}, et \textit{contraction
externe }\`{a} K chez Einstein (la tige de K vue de K'). C'est donc
pr\'{e}cis\'{e}ment parce que l'observateur dans K utilise ses propres tiges
contract\'{e}es qu'il doit observer une ellipse allong\'{e}e dans K. Dans
les deux cas (Einstein et Poincar\'{e}) il y a \textbf{r\'{e}ciprocit\'{e}}
mais les tiges rigides (identiques) einsteiniennes ont incontestablement une
plus grande proximit\'{e} avec la conception g\'{e}oMETRIQUE habituelle
o\`{u} les m\`{e}tres sont d\'{e}finis ind\'{e}pendamment de leur \'{e}tat
de mouvement DANS chaque syst\`{e}me du couple relatiste K-K'. On comprend
d\`{e}s lors que Poincar\'{e} se trouve dans une situation plus difficile
qu'Einstein puisqu'\textbf{il doit raisonner dans un des deux syst\`{e}me du
couple relativiste} (ici K), entra\^{i}n\'{e} dans un mouvement de
translation (ici par rapport \`{a} K'), o\`{u} la tige est contract\'{e}e
afin de d\'{e}finir une m\'{e}trique.

Comment la contraction de Lorentz \textbf{purement longitudinale} selon
Poincar\'{e} se transforme t-elle en une \textbf{dilatation isotrope}? Le
raisonnement de Poincar\'{e} se fait en deux \'{e}tapes:

\begin{quotation}
(1\`{e}re \'{e}tape: inobservabilit\'{e} de la contraction sans$"c=1"$).

Un corps sph\'{e}rique lorsqu'il est en repos, prendra ainsi la forme d'un
ellipso\"{i}de de r\'{e}volution aplati lorsqu'il sera en mouvement; mais
l'observateur le croira toujours sph\'{e}rique, parce qu'il a subi
lui-m\^{e}me une d\'{e}formation analogue, ainsi que tous les objets qui lui
servent de points de rep\`{e}re

(2\`{e}me \'{e}tape: observabilit\'{e} de la contraction avec $"c=1"$ ).

Au contraire, les surfaces d'onde de la lumi\`{e}re, qui sont rest\'{e}es
rigoureusement sph\'{e}riques lui para\^{i}tront des ellipso\"{i}des
allong\'{e}s.
\end{quotation}

Poincar\'{e} constate d'abord que cette contraction de tous les corps selon
la direction longitudinale doit \^{e}tre inobservable\ puisque les
unit\'{e}s longitudinales sont elles aussi contract\'{e}es ($1_{x}=\gamma
^{-1}1_{y})$. Il n'y a aucun moyen purement INTERNE de mettre en
\'{e}vidence une contraction qui est d\'{e}finie comme \'{e}tant interne au
syst\`{e}me K. A premi\`{e}re vue on est donc ramen\'{e} \`{a} la position
d'Einstein car, entre une contraction inobservable ou une tige identique, il
ne peut y a avoir aucune diff\'{e}rence physique.

Mais que se passe-t-il alors (2\`{e}me \'{e}tape) si des observateurs dans K
utilisaient la vitesse de la lumi\`{e}re? Une onde sph\'{e}rique \textit{%
d\'{e}finie dans K'} mesur\'{e}e par un \textit{m\`{e}tre longitudinalement
contract\'{e} dans K} doit donner une onde ellipso\"{i}dale allong\'{e}e.
Ici nous sommes au coeur de la cin\'{e}matique poincar\'{e}enne car il
s'agit d'une \textit{synth\`{e}se entre le principe d'invariance de la
vitesse de la lumi\`{e}re et le principe de contraction}. Cette synth\`{e}se
est profond\'{e}ment non-einsteinienne puisque la contraction y joue un
r\^{o}le crucial. Le PRODUIT DE CETTE SYNTH\`{e}SE doit \^{e}tre la TL.
Essayons de formuler ceci math\'{e}matiquement.

La proc\'{e}dure de synchronisation est toujours d\'{e}crite de la m\^{e}me
mani\`{e}re par Poincar\'{e} entre 1900 et 1912. Il suppose deux stations
fix\'{e}es dans K, A et B, qui \'{e}changent des signaux lumineux et qui
sont entra\^{i}n\'{e}es dans un mouvement de translation par rapport \`{a}
K'(o\`{u} l'\'{e}ther est d\'{e}fini au repos). La source en A (avec
l'observateur O), les horloges en A et B, le miroir en B sont dans le
m\^{e}me syst\`{e}me K (''translation commune'' ici K, non prim\'{e}). En
bref, tout le syst\`{e}me optique se trouve dans K (voir aussi
exp\'{e}rience de Sagnac, annexe 3). Nous allons traiter d'abord le cas
longitudinal et ensuite le cas non-longitudinal.

Soulignons que selon Poincar\'{e} le processus (purement) ondulatoire de
propagation de la lumi\`{e}re N'est PAS un processus purement INTERNE. A la
diff\'{e}rence d'Einstein qui d\'{e}finit \`{a} priori des ondes
sph\'{e}riques dans les deux syst\`{e}mes K et K' du couple (les
syst\`{e}mes d'Einstein sont des \textbf{bo\^{i}tes \`{a} photons}),
Poincar\'{e} compose la vitesse de la lumi\`{e}re avec celle du
r\'{e}f\'{e}rentiel puisque dans le cas de figure consid\'{e}r\'{e}
l'observateur (avec la source) est suppos\'{e} en mouvement par rapport
\`{a} K' et donc du point de vue d'une th\'{e}orie purement ondulatoire
(pr\'{e}relativiste) de la lumi\`{e}re il faut ''composer'' avant
d'introduire l'hypoth\`{e}se relativiste de contraction.

\section{Synchronisation longitudinale et d\'{e}duction de la TL}

Supposons une propagation purement longitudinale (AB est dans la direction
du mouvement). La longueur de r\'{e}f\'{e}rence A'B' est d\'{e}finie dans le
syst\`{e}me au repos K'. La longueur contract\'{e}e inobservable dans K est
par hypoth\`{e}se 
\begin{equation}
AB_{inobs}=\gamma ^{-1}A^{\prime }B^{\prime }  \tag{hypoth\UNICODE{0xe8}se}
\end{equation}
il faut d\'{e}montrer que la longueur dilat\'{e}e observable est donn\'{e}e 
\begin{equation}
x_{obs}=\gamma A^{\prime }B^{\prime }  \tag{th\UNICODE{0xe8}se}
\end{equation}
$.$ La th\`{e}se de Poincar\'{e} se traduit donc par $x_{obs}=\gamma
^{2}AB_{inobs}.$ puisqu'on on passe d'une contraction $\gamma ^{-1}$ \`{a}
une dilatation $\gamma $. D\'{e}montrons cette derni\`{e}re formule sur la
base de la synchronisation de Poincar\'{e}.

Selon Poincar\'{e} les montres doivent \^{e}tre r\'{e}gl\'{e}es de telle fa%
\c{c}on que la vitesse de la lumi\`{e}re c est invariante ($c\neq 1)$ dans
les deux syst\`{e}mes. Poincar\'{e} calcule de fa\c{c}on tout \`{a} fait
classique en composant la vitesse de la lumi\`{e}re avec celle du
syst\`{e}me entra\^{i}n\'{e} K les deux expressions suivantes (qui sont
explicit\'{e}es dans son cours \`{a} la Sorbonne, voir 3.4 \& 3.7).

\begin{equation}
t_{M}=\frac{t^{+}+t^{-}}{2}=\frac{1}{2}(\frac{AB}{c+v}+\frac{AB}{c-v})=\frac{%
1}{2}AB\text{ }\frac{2c}{c^{2}-v^{2}}=\gamma ^{2}\frac{AB}{c}\approx \frac{%
A^{\prime }B^{\prime }}{c}
\end{equation}

\begin{equation}
\Delta t=\frac{t^{+}-t^{-}}{2}=\frac{1}{2}(\frac{AB}{c+v}-\frac{AB}{c-v})=%
\frac{1}{2}AB\text{ }\frac{2v}{c^{2}-v^{2}}=\gamma ^{2}\frac{v}{c}\frac{AB}{c%
}\approx \frac{vA^{\prime }B^{\prime }}{c^{2}}
\end{equation}
On retrouve ainsi le terme poincar\'{e}en $\Delta t$ qui avait \'{e}t\'{e}
mise en \'{e}vidence par la composante longitudinale sur la tangente \`{a}
l'ellipse (mutatis mutandis puisqu'ici nous sommes dans le cas
non-standard). Pour obtenir le temps aller et le temps retour, il suffira de
calculer:

\begin{equation}
t^{\pm }=t_{M}\pm \Delta t
\end{equation}

Poincar\'{e} obtient ainsi d\`{e}s 1900 le temps local t au premier ordre%
\footnote{%
en d\'{e}tails Poincar\'{e} constate \textit{que} la formule de
synchronisation physique\textbf{\ }des horloges correspond \`{a} la formule
math\'{e}matique du temps local (ici t) de Lorentz lequel s'\'{e}crit $%
t=t^{\prime }+\frac{v}{c^{2}}x^{\prime }$ \`{a} condition d'identifier au
premier ordre $t_{M}^{{}}\approx t^{\prime }$ et surtout \`{a} poser que $%
\overline{AB}$ = $\overline{A^{\prime }B^{\prime }}$. On a pour le temps
aller-retour $t_{A}^{\ast }=2t^{\prime },$ et par d\'{e}finition le temps
local $(t_{A}=0)$ en B $t_{B}=t^{\prime }+\frac{v}{c^{2}}x_{B}^{\prime }$ $\ 
$\ et donc $t^{+}=t_{B}$ $=$ $t^{\prime }+\frac{v}{c^{2}}x^{\prime }$ et $%
t^{-}=2t_{M}-t_{B}=2t^{\prime }-t^{\prime }-\frac{v}{c^{2}}x^{\prime }).$ $\ 
$ Pour les temps ''locaux'' de parcours aller et retour de la lumi\`{e}re,
on a respectivement\ $t^{+}=t^{\prime }+\frac{v}{c^{2}}x^{\prime }$et $%
t^{-}=t^{\prime }-\frac{v}{c^{2}}x^{\prime }$.} sans contraction ($%
AB=A^{\prime }B^{\prime })$ avec $x^{\prime }=A^{\prime }B^{\prime }$ et $\ 
\frac{A^{\prime }B^{\prime }}{c}=t^{\prime }$:

\begin{equation*}
t^{\pm }=t_{M}\pm \Delta t\approx t_{M}\pm \frac{vA^{\prime }B^{\prime }}{%
c^{2}}\approx t^{\prime }\pm \frac{vx^{\prime }}{c^{2}}\qquad \qquad
\end{equation*}

Faisons maintenant le calcul exact au second ordre \textit{et donc avec AB
contract\'{e} dans K}. Etant donn\'{e} que (5.3)

\begin{equation*}
t^{\pm }=t_{M}\pm \Delta t=\gamma ^{2}\frac{AB_{inobs}}{c}(1\pm \frac{v}{c})
\end{equation*}

avec $AB_{inobs}=\gamma ^{-1}A^{\prime }B^{\prime }$ on a: 
\begin{equation}
t^{\pm }=\gamma (t^{\prime }\pm \frac{v}{c^{2}}x^{\prime })
\end{equation}

Autrement dit le temps local au 2\`{e}me ordre, et donc la TL temporelle.

Les longueurs parcourues $x^{^{\pm }}$ (respectivement aller et retour) sont
telles que la vitesse de la lumi\`{e}re est invariante. On multiplie par c
l'\'{e}quation (5.4) et on obtient:

\begin{equation}
ct^{\pm }=\gamma (ct^{\prime }\pm c\frac{v}{c}\frac{x^{\prime }}{c})=x^{\pm
}=\gamma (x^{\prime }\pm vt^{\prime })
\end{equation}
\ 

\bigskip Autrement dit, la 1\`{e}re TL pour la coordonn\'{e}e x spatiale
(5.5). \textbf{Les deux TL sont ainsi d\'{e}duites au moyen d'une
synth\`{e}se directe entre la contraction interne de Poincar\'{e} et
l'invariance de la vitesse de la lumi\`{e}re}. Il faut maintenant
d\'{e}finir la distance observable AB dans K. Eh bien cette distance\ $%
x_{obs}=AB$ est \textit{d\'{e}finie }par le temps aller-retour de parcours:

\begin{equation*}
x_{M}=\frac{x^{+}+x^{-}}{2}=\gamma A^{\prime }B^{\prime }
\end{equation*}
et donc on retrouve la th\`{e}se de Poincar\'{e} 
\begin{equation}
x_{M}=x_{obs}=\gamma ^{2}AB_{inobs}  \tag{cqfd}
\end{equation}

On a donc d\'{e}duit (en conformit\'{e} avec 3.4 \& 3.7) \textit{%
sym\'{e}triquement} \textbf{les deux TL} \`{a} partir de la contraction.
R\'{e}ciproquement, \textbf{l'interpr\'{e}tation poincar\'{e}enne de la
contraction }est la seule qui soit\textbf{\ compl\`{e}tement compatible }%
avec \textbf{''rien que la TL\ et toute la TL} (l'inverse de l'affirmation
de Pais \cite{12}): \textbf{la distance observ\'{e}e par \'{e}change de
signaux lumineux est dilat\'{e}e parce que les unit\'{e}s de mesure sont
contract\'{e}es}. Comme (5.5) est d\'{e}duit de (5.4), cela revient \`{a}
d\'{e}finir les unites d'espace et de temps de telle mani\`{e}re que $c=1$
(1.1).

Ecrivons les deux expressions de Poincar\'{e} (5.1 \& 5.2) dans le cas
relativiste deviennent:

\begin{equation}
t_{M}=\gamma t^{\prime }\qquad \qquad \Delta t=\gamma \frac{v}{c^{2}}%
x^{\prime }
\end{equation}

\section{Synchronisation non-longitudinale et d\'{e}duction de l'ellipse
''source''}

Maintenant supposons que B ne soit pas situ\'{e} sur la direction du
mouvement. Autrement dit on introduit une ordonn\'{e}e $y_{B}\neq 0$ avec le
vecteur radial $\mathbf{r=AB}$ \ et $\mathbf{r}^{\prime }\mathbf{=A}^{\prime
}\mathbf{B}^{\prime }$ dont la composante longitudinale est respectivement $%
x=r\cos \theta $ et $x^{\prime }=r^{\prime }\cos \theta ^{\prime }$.
L'\'{e}quation (5.6) s'\'{e}crit alors:

\begin{equation}
(t_{M})_{isotrope}=\gamma t^{\prime }\qquad \qquad (\Delta
t)_{longitudinal}=\gamma \frac{v}{c^{2}}r^{\prime }\cos \theta ^{\prime
}=\gamma \frac{v}{c^{2}}x^{\prime }
\end{equation}
Cette double \'{e}quation (5.7) DEFINIT \textbf{l'ellipse de synchronisation}
(\textbf{figure 4}, 15) avec, mutatis mutandis, (3.4, 3.7) (voir annexe 4,
la ''somme'' d\'{e}finit les foyers et la ''diff\'{e}rence'' d\'{e}finit les
directrices de l'ellipse).

La premi\`{e}re relation ''aller-retour'' ne d\'{e}pend pas de la direction
tandis que la seconde relation ne d\'{e}pend que de l'abscisse de B.
Appelons B$_{\perp }$ la projection de B sur la direction du mouvement:
l'\'{e}change de signaux lumineux entre A\ et B se ram\`{e}ne \`{a}
l'\'{e}change de signaux lumineux entre A\ et B$_{\perp }.\;$On est donc
ramen\'{e} \`{a} la description purement longitudinale, si ce n'est que la
longueur contract\'{e}e n'est plus AB mais AB$_{\perp }$.

La contraction (inobservable) purement anisotrope implique une expansion
isotrope (observable) des distances puisque ces derni\`{e}res sont
d\'{e}finies sur un aller-retour $(cqfd).$

Le r\'{e}sultat (5.7) \`{a} deux dimensions spatiales (et donc \`{a} trois
dimensions spatiales) est absolument d\'{e}cisif car il donne aussi un sens
\`{a} la th\'{e}orie de la dilatation au niveau des \'{e}lectrons classiques
sph\'{e}riques au repos de Poincar\'{e}: les sph\`{e}res observ\'{e}es en
mouvement seront dilat\'{e}es d'un facteur $\gamma $ comme cons\'{e}quence
d'un applatissement inobservable d'un facteur $\gamma ^{-1}$ (voir
conclusion finale).

Quelles sont maintenant les diff\'{e}rences \textbf{irr\'{e}ductibles }entre
la synchronisation einsteinienne ''en ondes sph\'{e}riques'' et la
synchronisation de Poincar\'{e} ''en onde ellipso\"{i}dale'' (annexe 1)? Non
seulement on peut mettre en \'{e}vidence le $\Delta t$ non null de
Poincar\'{e} dans\textbf{\ l'exp\'{e}rience de Sagnac} (voir annexe 3) mais
aussi et surtout, la synchronisation de Poincar\'{e} sur la base de la
contraction sym\'{e}triquement articul\'{e}e avec les deux TL l\'{e}gitimise
la d\'{e}finition de la distance impropre dilat\'{e}e.

\section{Formule Doppler relativiste, \'{e}ther relativiste de Poincar\'{e}
et fond de rayonnement cosmologique}

Rappelons que l'ellipse dans le syst\`{e}me de la source (15) est
math\'{e}matiquement introduite par la TL inverse. On peut maintenant
l'interpr\'{e}ter physiquement\footnote{%
M\^{e}me si l'ellipse de Poincar\'{e} est impos\'{e}e par la logique
math\'{e}matique et la physique-math\'{e}matique sous-jacente \`{a} la TL,
elle n'en demeure pas moins \'{e}trange d'un point de vue physique.
Pourquoi? Le physicien associe en effet imm\'{e}diatement ''cercle'' \`{a}
isotropie de la vitesse de la lumi\`{e}re et ''ellipse'' \`{a}
''non-isotropie de la vitesse de la lumi\`{e}re''. Une \'{e}ventuelle
anisotropie de la vitesse de la lumi\`{e}re renverrait ainsi au fant\^{o}me
non-relativitste par excellence l'\'{e}ther au repos dans l'espace absolu
(pourtant totalement rejet\'{e} par Poincar\'{e}). Mais ce n'est pas de
l'ellipse non-relativiste (et non-isotrope) de 1904 (note 2) qu'il s'agit
mais de l'ellipse de 1906 avec l'isotropie de la vitesse de la lumi\`{e}re
(10 ter). L'ellipse de Poincar\'{e} est isotrope au sens o\`{u} la vitesse
de la lumi\`{e}re est isotrope (\'{e}v\'{e}nements d'intervalles nuls, voir
\S 1.4).}. Si on peut toujours choisir, selon Poincar\'{e}, le syst\`{e}me
dans lequel on met l'\'{e}ther au repos, ce choix n'est possible que parce
que les deux situations sont math\'{e}matiquement compl\`{e}tement
\'{e}quivalentes (m\^{e}me formule Doppler si la source est en mouvement ou
si l'observateur est en mouvement par rapport \`{a} l'\'{e}ther, annexe 2).
La \textbf{sym\'{e}trie ''source-observateur''} poincar\'{e}enne est
particuli\`{e}rement importante dans le cas de la formule Doppler car d'un
point de vue pr\'{e}relativiste ces \textit{deux cas }sont distincts. En
effet si la source est en mouvement par rapport \`{a} l'\'{e}ther, le
principe de Fresnel (la vitesse de la lumi\`{e}re dans l'\'{e}ther ne
d\'{e}pend pas de la vitesse de la source) s'applique mais ce n'est pas le
cas si l'observateur est en mouvement (il faut composer...). On voit ainsi
qu'avec l'ellipse de Poincar\'{e}, le principe de Fresnel devient
relativiste (sym\'{e}trisation du r\^{o}le de l'\'{e}ther): peu importe si
l'on place l'\'{e}ther au repos dans l'un (ondes sph\'{e}riques dans K, \S
1) ou dans l'autre (ondes sph\'{e}rique dans K' \S 2) syst\`{e}me du couple,
les \'{e}quations qui relient les fr\'{e}quences respectives deux
syst\`{e}mes seront les m\^{e}mes (voir Doppler \S 1.2 et annexe 2).

La terminologie \textbf{temps vrai-temps local} de Poincar\'{e} est ainsi
compl\`{e}tement \'{e}claircie: si on choisit le temps vrai ''circulaire''
(sph\`{e}res ou \'{e}ther au repos) dans un syst\`{e}me, on aura un temps
local ''elliptique'' (ellipso\"{i}des ou \'{e}ther en mouvement) dans
l'autre:

\begin{equation}
x^{\prime }=k(x+\varepsilon t)\qquad \qquad \qquad y^{\prime }=y\qquad
\qquad t^{\prime }=k(t+\varepsilon x)  \tag{1direct}
\end{equation}

o\`{u} t est le temps vrai et t' est le temps local. Et inversement :

\begin{equation}
x=k(x^{\prime }-\varepsilon t^{\prime })\qquad \qquad \qquad y^{\prime
}=y\qquad \qquad t=k(t^{\prime }-\varepsilon x^{\prime })  \tag{1inverse}
\end{equation}

o\`{u} t' est le temps vrai et t est le temps local.

Dans la logique einsteinienne la mod\'{e}lisation non-standard (l'ellipse
dans le syst\`{e}me de la source) est impossible, tandis qu'elle est
\'{e}quivalente\ (\`{a} la mod\'{e}lisation standard) dans la logique
poincar\'{e}enne. Comme en physique, il faut adapter le mod\`{e}le au type
d'exp\'{e}rience envisag\'{e}, il se pourrait que cette mod\'{e}lisation
non-standard soit math\'{e}matiquement possible mais physiquement inutile
(hormi l'exp\'{e}rience de Michelson). Nous montrons en annexe\ 3 qu'elle
est notamment tr\`{e}s utile pour l'exp\'{e}rience de Sagnac.

Dans la perspective de Poincar\'{e} le crit\`{e}re de sph\'{e}ricit\'{e} de
l'onde est li\'{e} \`{a} l'\'{e}ther au repos: \textit{il suffit de prendre
la TL inverse pour le mettre au repos dans l'autre syst\`{e}me}. Il s'agit
donc d'un \'{e}ther au repos relatif. La \textit{sym\'{e}trie cin\'{e}matique%
} \textbf{''mati\`{e}re-\'{e}ther''} de Poincar\'{e} signifie que l'on ne
peut attribuer \`{a} tout corps (source, observateur...) \textit{comme \`{a}
l'\'{e}ther} qu'une vitesse... relative:

\begin{quotation}
\guillemotleft\ Quoi qu'il en soit, il est impossible d'\'{e}chapper \`{a}
cette impression que le principe de relativit\'{e} est une loi
g\'{e}n\'{e}rale de la Nature, qu'on ne pourra jamais par aucun moyen
imaginable, mettre en \'{e}vidence que des VITESSES RELATIVES, et j'entends
par l\`{a} non pas seulement des vitesses des corps PAR RAPPORT A L'ETHER
mais les vitesses des corps les uns par rapport aux autres (Poincar\'{e},
1908, le principe de relativit\'{e}$°$\ \guillemotright
\end{quotation}

{\footnotesize \bigskip }Avec la convention d'Einstein-Minkowski, cette
sym\'{e}trie de Poincar\'{e} est supprim\'{e}e avec l'\'{e}ther (''on ne
peut lui attribuer AUCUN vecteur vitesse'', \cite[inrroduction]{4}); le
milieu de la propagation de l'onde \'{e}tant supprim\'{e}, le crit\`{e}re de
repos (ondes sph\'{e}riques) ne peut \^{e}tre que li\'{e} \`{a} la source de
l'onde.

\begin{quotation}
\guillemotleft\ On verra que l'introduction d'un \'{e}ther lumineux devient
superflue par le fait que notre conception ne fait aucun usage d'un ''espace
absolu au repos'' dou\'{e} de propri\'{e}t\'{e}s particuli\`{e}res, et ne
fait correspondre \`{a} un point de l'espace vide, o\`{u} ont lieu des
procesus \'{e}lectromagn\'{e}tiques, AUCUN VECTEUR VITESSE (Einstein, 1905,
introduction)\guillemotright
\end{quotation}

Comment d\`{e}s lors peut-on distinguer alors la sph\`{e}re einsteinienne et
l'ellipsoide poincar\'{e}en dans le syst\`{e}me de la source? Remarquons que
lorsqu'on mesure par effet Doppler la vitesse (relative) de la Terre
(environ 300km/s) par rapport au ''fond de rayonnement cosmologique'', ce
''fond'' est bien un \'{e}ther (relativiste) au sens de Poincar\'{e}.
D'apr\`{e}s la cin\'{e}matique en jauge de Lorenz, ce ''fond'' (qui n'a 
\textit{pas de source identifiable}, $\rho _{elecstat}=\rho
_{elecstat}^{\prime }=0$ mais $V^{\prime }\neq 0)$ doit donc \^{e}tre en
rapport avec l'expansion de l'espace (distance impropre au sens de
Poincar\'{e}, 78). La cin\'{e}matique poincar\'{e}enne de l'expansion de
l'espace \'{e}tablit donc un lien relativiste in\'{e}dit (avec 1-direct et
1-inverse, voir ci-dessus) entre les mesures de Hubble et les mesures de
Penzias-Wilson.

\part{Conclusion: r\'{e}flexion sur la sym\'{e}trie spatio-temporelle de
Poincar\'{e}}

La d\'{e}marche que nous avons suivie dans le pr\'{e}sent travail est
exactement l'inverse du cheminement historique suivi par Poincar\'{e}. Nous
avons commenc\'{e} par d\'{e}duire l'ellipse\ allong\'{e}e de la TL: elle
n'est rien d'autre qu'une repr\'{e}sentation concr\`{e}te de la
relativit\'{e} de la simultan\'{e}it\'{e} au moyen d'un front d'onde \textit{%
spatiotemporel}.\ Ensuite une fois que la l\'{e}gitimit\'{e}
physico-math\'{e}matique de l'ellipse allong\'{e}e a \'{e}t\'{e} \'{e}tablie
nous n'avons fait qu'interpr\'{e}ter physiquement toutes les
caract\'{e}ristiques, les propri\'{e}t\'{e}s, les sym\'{e}tries
math\'{e}matiques de l'ellipse. Avec la tangente \`{a} l'ellipse nous avons
montr\'{e} comment d\'{e}finir l'image spatio-temporelle d'un front d'onde
plane et en utilisant l'invariance de la phase nous avons d\'{e}duit
rigoureusement pour les objets lontains une formule Doppler relativiste
(2.4) qui n'est pas la m\^{e}me que celle d'Einstein (2.5).

En utilisant le second foyer et la tangente nous avons retrouv\'{e} les
intuitions de d\'{e}part de Poincar\'{e} notamment la d\'{e}finition de la
distance, en accord parfait avec la TL, avec le temps dilat\'{e} de parcours
aller-retour de la lumi\`{e}re. En utilisant enfin l'autre ellipse
d\'{e}finie par sym\'{e}trie dans le syst\`{e}me de la source, nous avons
montr\'{e} que la contraction selon Poincar\'{e} (les unit\'{e}s
contract\'{e}es allongent les sph\`{e}res en ellipso\"{i}des...) est la
seule interpr\'{e}tation compl\`{e}tement compatible avec la TL. Dans le
m\^{e}me mouvement nous avons prouv\'{e} dans les moindres d\'{e}tails (en
utilisant l'ellipse dans le syst\`{e}me de la source, c-\`{a}-d, \textit{%
l'ellipse de synchronisation}) que la convention de synchronisation ''en
ondes sph\'{e}riques'' d'Einstein $(\Delta t=\Delta t^{\prime }=0)$
n'\'{e}tait pas la m\^{e}me que la convention de synchronisation ''en onde
ellipso\"{i}dale'' de Poincar\'{e} ($\Delta t^{\prime }=0$ et $\Delta t\neq
0)$. La seule exp\'{e}rience de physique sensible \`{a} cette diff\'{e}rence
de temps de parcours ''one way'' de la lumi\`{e}re $\Delta t\neq 0$ non
nulle est l'exp\'{e}rience de Sagnac.

La transformation de LorenTz est inscrite dans l'ellipse et il n'est pas
\'{e}tonnant qu'on y retrouve finalement la jauge de LorenZ. Toute la
cin\'{e}matique de Poincar\'{e} est donc ''Lorenz-Lorentzienne''. Il est
d\'{e}sormais incontestable que le jeune Einstein en 1905 a fait un travail
original mais qu'il a simplifi\'{e} le probl\`{e}me en prenant des
libert\'{e}s aussi bien avec la transformation de Lorentz qu'avec la jauge
de Lorenz. Peu avant sa mort, Poincar\'{e} \cite{20} \'{e}crivait que
l'Ecole allemande de relativit\'{e}
(Einstein-Planck-Minkowski-Laue-Sommerfeld-Born) avait choisi une autre 
\textit{convention} que le sienne. Un si\`{e}cle apr\`{e}s l'ann\'{e}e
miraculeuse (des deux relativit\'{e}s) 1905 nous montrons que \textit{ce
choix de convention n'est autre qu'un }\textbf{choix de jauge}. Si ce choix
est neutre (libert\'{e} de jauge) au niveau de la th\'{e}orie
\'{e}lectromagn\'{e}tique des champs transversaux qui d\'{e}crivent les
ondes de lumi\`{e}re, il ne l'est pas \`{a} un niveau purement \textit{%
relativiste} puisque les quadrivecteurs d'onde (2.1 \& 2.6), les formules
Doppler (2.4 \&2.5), les d\'{e}finitions de la phase d'une onde plane (2.15
\& 2.26) sont diff\'{e}rentes. En principe dans une cin\'{e}matique
relativiste, il faut travailler avec une jauge relativiste comme la jauge de
Lorenz. \ Il n'est cependant pas si simple de se passer de la jauge de
Coulomb dite transverse car il faut alors passer de la conception
einsteinienne purement spatiale du front d'onde \`{a} la conception \textit{%
spatiotemporelle} du front d'onde sous-jacente \`{a} l'ellipse allong\'{e}e.

C'est Max Born \cite{1} qui nous a mis sur la piste de l'existence d'une
''structure fine'' de la relativit\'{e} \cite{13} car il est le seul
physicien \`{a} avoir soulign\'{e} qu'Einstein avait fait une hypoth\`{e}se
tacite qui va au-del\`{a} de la TL (1921):

\begin{quotation}
A fixed rod that is at rest in the system K and is of length 1 cm, will, of
course, also have the length 1 cm, when it is at rest in the system k. We
may call this \textbf{tacit assumption} of Einstein's theory the \textit{%
principle of the physical identity of the units of measure}.
\end{quotation}

Une v\'{e}ritable cha\^{i}ne logique unit les diff\'{e}rents maillons de la
th\'{e}orie einsteinienne: tiges rigides identiques, fronts sph\'{e}riques
de synchronisation, fronts rigides (spatiaux) d'onde plane, quadrivecteur
d'onde, formule doppler avec phase vectorielle, double transversalit\'{e}
(simultan\'{e}it\'{e} absolue), jauge de Coulomb dite transverse. Il est
impossible de s\'{e}parer ces maillons car l'encha\^{i}nement conduit au
photon einsteinien (ou le quadrivecteur d'onde einsteinien, 21, ce qui
revient au m\^{e}me).

On ne peut r\'{e}parer une injustice (celle qui a \'{e}t\'{e} commise
historiquement envers Poincar\'{e}) en commettant une autre injustice. En
montrant que toute la cin\'{e}matique einsteinienne repose sur un choix de
jauge, loin de l'avoir d\'{e}valoris\'{e}e, nous l'avons en quelque sorte
''r\'{e}actualis\'{e}e''. La cin\'{e}matique d'Einstein est au fond une
th\'{e}orie quantique de la lumi\`{e}re (et des horloges identiques \cite
{24ter}) dont on est loin d'avoir mesur\'{e} toute la port\'{e}e (notamment
au niveau de l'infiniment petit). La jauge de Coulomb sur laquelle elle
repose est la jauge de quantification par excellence et la question se pose
d\'{e}sormais de savoir quel est le lien entre la violation de la
relativit\'{e} de la simultan\'{e}it\'{e} et la fameuse non-localit\'{e}
quantique? D'ailleurs si on quantifie en jauge de Lorenz on trouve le photon
''longitudinal et temporel'' dont on se demande ce qu'ils viennent faire
dans cette gal\`{e}re et dont la raison d'\^{e}tre v\'{e}ritable se trouve
bien entendu dans la cin\'{e}matique de Poincar\'{e} (voir th\'{e}or\`{e}me
de Poincar\'{e} li\'{e} \`{a} la composante longitudinale du vecteur
potentiel, 51).

La fronti\`{e}re classique-quantique passe donc entre les deux
relativit\'{e}s, comme nous l'avions supput\'{e} il y a 7 ans \cite{13}. La
cin\'{e}matique einsteinienne r\`{e}gne, et r\`{e}gnera s\^{u}rement encore
longtemps, dans le domaine atomistique et plus g\'{e}n\'{e}ralement sur ce
qu'il est convenu d'appeler ''l'infiniment petit''.

Il n'est donc pas \'{e}tonnant que Max Born soit aussi le seul membre de
l'Ecole allemande de relativit\'{e} \`{a} avoir directement affront\'{e} le
front ellipso\"{i}dal. Apr\`{e}s une vive discussion avec Keswani (voir note
2), o\`{u} Poincar\'{e} n'est jamais cit\'{e}, Born r\'{e}fute l'existence
d'un front ellipso\"{i}dal de la mani\`{e}re suivante (voir \S 1-1) \cite{2}:

\begin{quotation}
It is obvious that the radii (and the corresponding transmission times) are
not the same for the longitudinal point and the two transverse points. But
this does not, of course, mean that the wavefront in K' is ellipsoidal ;
only its time of expansion is different for the three points when observed
in K'.
\end{quotation}

Ainsi Max Born, orthodoxe einsteinien a parfaitement aper\c{c}u ce qui
\'{e}tait compatible avec la m\'{e}trique d'Einstein-Minkoswski et ce qui ne
l'\'{e}tait pas. Il est clair que l'on peut admettre une expansion du temps
de parcours (c'est la dilatation du temps) mais pas une expansion de \textit{%
l'espace} (et donc des fronts \textit{spatio}temporels ellipso\"{i}daux), ce
qui serait en opposition totale avec la th\'{e}orie einsteinienne de la
contraction des tiges rigides (3.12) ($\Delta \tau =$ $\Delta t=0$). Or,
cette \textbf{expansion isotrope de l'espace,} directement li\'{e}e au
ph\'{e}nom\`{e}ne Doppler, c'est ce qui fait pr\'{e}cis\'{e}ment le dernier
mot (la signification la plus profonde) de l'ellipse allong\'{e}e de
Poincar\'{e} \cite{22} Cette derni\`{e}re suppose une autre d\'{e}finition
de la contraction, une autre d\'{e}finition de la distance et \textbf{donc
une autre m\'{e}trique spatiotemporelle }(que celle d'Einstein).

En consid\'{e}rant la \textbf{figure 6} (avec une source \`{a} l'infini),
supposons une s\'{e}rie de sources \`{a} l'infini dans K dans des directions
diff\'{e}rentes par rapport \`{a} O qui \'{e}mettent alors des ondes planes.
La th\'{e}orie d'Einstein pr\'{e}voit que la fr\'{e}quence de ces sources
sera modifi\'{e}e pour l'observateur O' et que cela n'a strictement rien
\`{a} voir avec la m\'{e}trique minkowskienne de l'espace-temps. La
th\'{e}orie de Poincar\'{e} pr\'{e}voit aussi que la fr\'{e}quence sera
modifi\'{e}e mais que cette modification est directement li\'{e}e \`{a} la 
\textbf{structure }de l'espace-temps (le quadrivecteur d'onde de
Poincar\'{e} est li\'{e} au quadrivecteur isotrope spatiotemporel par la
formule 4.1 ou 4.2). L'effet Doppler pour ces objets lointains est donc
directement li\'{e}e \`{a} l'expansion de l'espace dans la m\'{e}trique de
Poincar\'{e}. Les mod\`{e}les anglo-saxons qui utilisent la m\'{e}trique $%
ds^{2}$ de Robertson-Walker sont tous bas\'{e}s sur un facteur d'\'{e}chelle
introduit de fa\c{c}on adhoc. Ils deviendraient alors obsol\`{e}tes s'il se
confirme que l'expansion de l'univers est un effet relativiste indirect de
la contraction de Lorentz \`{a} condition d'adopter la m\'{e}trique de
Poincar\'{e} (qui est la seule vraiment compl\`{e}tement relativiste
puisqu'en jauge de Lorenz).

La diff\'{e}rence entre les deux cin\'{e}matiques ne se situe pas au niveau
de l'invariance la vitesse de la lumi\`{e}re (c=1 dans les deux syst\`{e}mes
de Poincar\'{e}) mais bien au niveau de la \textbf{sym\'{e}trisation} des
r\^{o}les respectifs de l'espace et du temps. Et c'est peut \^{e}tre l\`{a}
que le d\'{e}veloppement de la cin\'{e}matique poincar\'{e}enne apporte la
plus grande surprise: c'est que cette sym\'{e}trisation spatiotemporelle est
pouss\'{e}e beaucoup plus loin dans cette derni\`{e}re que dans la
cin\'{e}matique d'Einstein. D'o\`{u} la remise en question relativiste de la
rigidit\'{e}... des fronts d'onde einsteiniens (bas\'{e}e sur la double
simultan\'{e}it\'{e}). D'o\`{u} la remise en question de la contraction
einsteinienne de la tige rigide (bas\'{e}e aussi sur une double
simultan\'{e}it\'{e})

D'o\`{u} provient en derni\`{e}re analyse la \textbf{sym\'{e}trisation
spatiotemporelle} sous-jacente \`{a} la th\'{e}orie de Poincar\'{e}?

Nous avons d\`{e}s le d\'{e}part soulign\'{e} le caract\`{e}re
compl\`{e}tement sym\'{e}trique en l'espace et le temps de l'\'{e}criture
poincarienne de la TL (''unit\'{e}s $c=1$''). Nous avons commenc\'{e} avec
la TL \'{e}crite par Poincar\'{e} (1) et nous terminerons par cette
derni\`{e}re (87):

\begin{equation}
x^{\prime }=k(x+\varepsilon t)\qquad \qquad \qquad y^{\prime }=y\qquad
\qquad t^{\prime }=k(t+\varepsilon x)  \tag{1.1}
\end{equation}

Cette \'{e}criture poincar\'{e}enne pose imm\'{e}diatement la question de la
limite galil\'{e}l\'{e}enne, $x^{\prime }=x+\varepsilon t,$ $t^{\prime }=t$,
de la TL. Lorsqu'Einstein \'{e}tablit la limite ''standard'' galil\'{e}enne
de la TL, le terme $k\varepsilon x$ dispara\^{i}t car dans son syst\`{e}me
d'unit\'{e}s $\frac{v}{c^{2}}$ s'\'{e}vanouit d\`{e}s que la vitesse v est
faible avec k$\rightarrow 1.$ Il y a donc une asym\'{e}trie entre le
traitement \textit{\`{a} la limite }des termes $k\varepsilon t$ et $%
k\varepsilon x.$ Nous avons montr\'{e} que cette asym\'{e}trie einsteinienne
n'est pas seulement ''\`{a} la limite galil\'{e}enne'' mais qu'elle se
retrouve \textit{au coeur} m\^{e}me de sa cin\'{e}matique. Donnons en
conclusion un recensement (non-exhaustif) des lieux o\`{u} Einstein \textbf{%
annule}\footnote{%
Sauf bien entendu pour d\'{e}montrer l'invariance des \'{e}quations de
Maxwell. C'est la raison pour laquelle au niveau des champs, il n'y a pas de
''structure fine''. Cette derni\`{e}re se situe au niveau des jauges mais
l'utilisation d'une jauge non-relativiste ou relativiste au niveau d'une
th\'{e}orie de la relativit\'{e} est \'{e}videmment d'un int\'{e}r\^{e}t
immense.} le terme poincar\'{e}en $\Delta t^{\prime }=$ $k\varepsilon x=0$:

\begin{tabular}{l}
$1°)$ repr\'{e}sentation sph\'{e}rique des fronts ou ce qui revient au
m\^{e}me, convention de synchronisation (annexe 1, \'{e}quation 6) \\ 
$2°)$ \ d\'{e}finition de la contraction (3.14bis) \\ 
$3°)$ repr\'{e}sentation de l'onde plane la composante longitudinale est
annul\'{e}e (2. 14, 2.36) et donc au niveau de la jauge adopt\'{e}e
(2.48bis).
\end{tabular}

\bigskip Certains physiciens, comme F. Selleri, sugg\`{e}rent m\^{e}me de se
passer du terme $k\varepsilon x.$ Mais alors la TL ne forme plus un groupe.
On retrouve ainsi un lien tr\`{e}s \'{e}troit entre les sym\'{e}tries
spatiotemporelles de l'ellipse et les \textbf{propri\'{e}t\'{e}s groupales}
de la TL \cite{24bis}. La d\'{e}finition poincar\'{e}enne de la \textbf{%
m\'{e}trique }$"c=1"$ implique une sym\'{e}trie spatio-temporelle \textbf{%
parfaite} dans le traitement de $k\varepsilon x$ et $k\varepsilon t.$ \ Nous
avons montr\'{e} que cette sym\'{e}trie poincar\'{e}enne spatiotemporelle,
qui suppose $\Delta t^{\prime }=$ $k\varepsilon x\neq 0,$ se retrouvait
\`{a} tous les niveaux: description spatiotemporelle de l'image des fronts
d'onde, aussi bien sph\'{e}riques que plans, sym\'{e}trie source-\'{e}ther
pour le ph\'{e}nom\`{e}ne Doppler, sym\'{e}trie pour la dilatation du temps
et de l'espace, sym\'{e}trie dans \textit{la relativit\'{e} de la
simultan\'{e}it\'{e} et de la relativit\'{e} de la synchronisation}: il y a
toujours un des deux syst\`{e}mes DANS\ lequel les horloges ne sont pas
synchrones au sens d'Einstein: elles sont r\'{e}gl\'{e}es ($c=1$) de telle
mani\`{e}re que le temps aller N'est PAS \'{e}gal au temps retour de la
lumi\`{e}re $\Delta t^{\prime }=$ $k\varepsilon x\neq 0$. En derni\`{e}re
analyse, cette sym\'{e}trie spatiotemporelle parfaite suppose que le concept
de vitesse relative s'applique aussi bien aux corps qu'\`{a} l'\'{e}ther.

Il est vrai que pour que le terme $k\varepsilon x$ soit du m\^{e}me ordre%
\footnote{%
Remarquons que ce terme non-galil\'{e}en\ $\varepsilon x$ peut \^{e}tre
interpr\'{e}t\'{e} comme un effet relativiste au premier ordre pour des
vitesses faibles.} de grandeur que $k\varepsilon t$, dans un cadre de
comparaison entre des temps aller et retour, il faut consid\'{e}rer des
distances $x.$... astronomiques. Mais justement (voir chapitre 3 \S 3), nous
avons montr\'{e} que la d\'{e}finition poincar\'{e}enne de la distance, par
contraste avec l'atomisme du jeune Einstein \cite{13}, \'{e}tait
particuli\`{e}rement adapt\'{e}e \`{a} ce qu'il est convenu d'appeler
'l'infiniment grand'' et donc \`{a} la cosmologie.

La question vient alors \`{a} l'esprit: y aurait-il un lien entre la \textbf{%
sym\'{e}trie spatiotemporelle parfaite} de Poincar\'{e}, d'une part, et le
principe cosmologique parfait de Hoyle et Bondi d'autre part \cite{0}. On
sait que ce principe \'{e}tablit une sym\'{e}trie parfaite entre l'espace et
le temps en ce qui concerne l'uniformit\'{e} de l'univers (la densit\'{e}
uniforme de mati\`{e}re $\varrho $ de l'univers) et par cons\'{e}quent, il
est imm\'{e}diatement incompatible la th\'{e}orie du big bang. Cependant le
probl\`{e}me principal de Hoyle et Bondi \'{e}tait que l'univers \'{e}tant
en expansion, il \'{e}tait imm\'{e}diatement logiquement n\'{e}cessaire
d'introduire une ''cr\'{e}ation ex nihilo de mati\`{e}re m'' (?) afin de
maintenir une densit\'{e} observ\'{e}e constante. Or, le seul
\'{e}l\'{e}ment que nous n'avons pas encore utilis\'{e} dans l'ellipse est
sa surface. Et donc dans l'ellipso\"{i}de son volume (2\`{e}me partie).

Si nous raisonnons, comme Poincar\'{e}, en terme de volume de la
mol\'{e}cule (sic) d'\'{e}lectron, il est clair que d'apr\`{e}s la partie
II, l'ellipso\"{i}de applati de la mol\'{e}cule (resic) d'\'{e}lectron'' $%
\gamma ^{-1}V$ est inobservable. Il n'en est pas de m\^{e}me pour le volume
dilat\'{e} de l'ellipso\"{i}de allong\'{e} $\gamma V$ qui devient une
grandeur \'{e}minemment observable (\S 2.1). Nous conclurons alors en
constatant simplement que la relativit\'{e} cin\'{e}matique de Poincar\'{e}
donne la densit\'{e} de mati\`{e}re suivante (laquelle correspond \`{a} une
densit\'{e} d'\'{e}nergie, ou dimensionnellement une ''pression'' pour le
fluide cosmique): 
\begin{equation*}
\varrho ^{\prime }=\frac{\gamma m}{\gamma V}=\frac{m}{V}=\varrho
\end{equation*}

autrement dit un \'{e}tat stationnaire de l'univers \textit{en expansion}
\`{a} densit\'{e} constante.

D\'{e}cid\'{e}ment, et c'est l\`{a} \ que r\'{e}side peut-\^{e}tre
essentiellement le destin tragique qui a \'{e}t\'{e} historiquement
r\'{e}serv\'{e} au Grand Oeuvre de Poincar\'{e} sur la relativit\'{e}, ''Le
lit en \ Expansion'' qu'il d\'{e}crit dans ''La relativit\'{e} de l'espace''
(1907) \cite{20} est arriv\'{e} (au moins) ... 20 ans trop t\^{o}t.

\section{Remerciements}

Je remercie Jean Reignier sans lequel la pr\'{e}sente \'{e}tude n'aurait
jamais pu voir le jour. Je remercie \'{e}galement Thomas Durt, Pierre Marage
et Germain Rousseaux (Universit\'{e} de Nice). Ce dernier m'a communiqu\'{e}
un document exceptionnel \cite{17} o\`{u} Poincar\'{e} utilise la directrice
de l'ellipse allong\'{e}e. Germain Rousseaux est en possession d'un
exemplaire original de ce document o\`{u} Poincar\'{e}, quelques jours avant
sa mort, persiste et signe en soulignant (sur la m\^{e}me page 46-47!) le
r\^{o}le pr\'{e}pond\'{e}rant de l'ellipse allong\'{e}e en cin\'{e}matique
relativiste, et, de la th\'{e}orie des potentiels de Riemann-Lorenz en
\'{e}lectromagn\'{e}tisme. A\ une ''structure fine'' de la relativit\'{e}
(Yves Pierseaux) correspond donc une ''structure fine'' de
l'electromagn\'{e}tisme (Germain Rousseaux). Un travail commun (YP et GR)
\`{a} ce sujet est propos\'{e} \`{a} la publication.

\chapter{Annexes}

\section{Convention de synchronisation d'Einstein ''en ondes
sph\'{e}riques'' et param\`{e}tre de Reichenbach}

\bigskip La convention de synchronisation d'Einstein est d\'{e}finie au
paragraphe de son article fondamental de 1905 \cite{4}. Einstein
consid\`{e}re d'abord le syst\`{e}me K $(x,y,z,t)$

\begin{quotation}
Supposons qu'un rayon lumineux parte \`{a} l'instant $t_{A}$ de A\ vers B
qu'il soit r\'{e}fl\'{e}chi \`{a} l'instant $t_{B}$ et qu'il soit de retour
en A \`{a} l'instant $t_{A}^{\ast }$. Les deux horloges sont par
d\'{e}finition synchrones si 
\begin{equation}
t_{B}-t_{A}=t_{A}^{\ast }-t_{B}  \tag{ 1}
\end{equation}
\end{quotation}

Avec les notations de Poincar\'{e}, on a $t^{+}=t^{-},$ autrement dit le
temps aller = le temps retour. Einstein d\'{e}finit donc la
simultan\'{e}it\'{e} de deux \'{e}v\'{e}nements A\ et B situ\'{e}s \`{a} la
distance AB. Remarquons que, r\'{e}ciproquement, lorsqu'on consid\`{e}re
deux \'{e}v\'{e}nements A\ et B qui ne sont pas au m\^{e}me endroit dans K, 
\textit{leur distance (propre) est par d\'{e}finition (relativiste)
d\'{e}finie par la simultan\'{e}it\'{e} de ces \'{e}v\'{e}nements dans K}.
Einstein conclut son premier paragraphe:

\begin{quotation}
Ce qui est essentiel c'est d\'{e}finir le temps au moyen d'horloges au repos
dans un syst\`{e}me au repos; \`{a} cause de cette relation \'{e}troite avec
le syst\`{e}me au repos, nous appelerons le temps que nous venons de
d\'{e}finir ''temps du syst\`{e}me au repos''.
\end{quotation}

La d\'{e}finition einsteinienne ''temps du syst\`{e}me au repos'' est donc
valable pour les deux syst\`{e}mes K et K'. Autrement dit, les horloges au
repos dans l'autre syst\`{e}me K' ($\xi ,$ $\eta ,\varsigma ,\tau )$ en
translation uniforme par rapport \`{a} K) seront synchronis\'{e}es
exactement de la m\^{e}me mani\`{e}re et qu'elles definiront le temps au
repos du syst\`{e}me K'. Einstein pr\'{e}cise d'ailleurs cela explicitement
\`{a} deux reprises au d\'{e}but de son paragraphe 3:

\begin{quotation}
\bigskip Nous supposons de m\^{e}me que le temps $\tau $ du syst\`{e}me en
mouvement K' soit d\'{e}termin\'{e} au moyen de la m\'{e}thode des signaux
lumineux d\'{e}crite au paragraphe 1, pour tous les points de ce syst\`{e}me
K' o\`{u} se trouve des horloges au repos relativement \`{a} ce dernier.

A cette fin nous devons exprimer par des \'{e}quations que $\tau $ n'est
rien d'autre que l'ensemble des indications des horloges au repos dans le
syst\`{e}me K' qui sont synchronis\'{e}es par la r\`{e}gle du paragraphe 1.
\end{quotation}

\bigskip \textbf{La vitesse relative (v ou }$\beta )$\textbf{\ entre les
deux syst\`{e}mes n'intervient pas} dans la proc\'{e}dure einsteinienne de
synchronisation du second syst\`{e}me (au contraire de Poincar\'{e}, partie
II, voir aussi le m\^{e}me contraste pour les ondes planes 2.12 \& 2.13). On
consid\`{e}re une horloge (avec une source\footnote{%
Remarquons qu'Einstein ne parle jamais de source dans ce paragraphe, comme
si les horloges envoyaient elles-m\^{e}mes les rayons lumineux, autrement
dit comme si elles \'{e}taient des atomes (ce qu'il \'{e}crit d'ailleurs
explicitement en 1907).}) au repos dans K' en A' et une horloge (avec un
miroir) en B' au repos dans K'. Et on a:

\begin{equation}
\tau _{B^{\prime }}-\tau _{A^{\prime }}=\tau _{A^{\prime }}^{\ast }-\tau
_{B^{\prime }}  \tag{2}
\end{equation}

Avec les notations de Poincar\'{e}, on a $t^{\prime +}=t^{\prime -},$ $%
\Delta t^{\prime }=0.$ \ De 1900 \`{a} 1912 (\S 2), Poincar\'{e}
d\'{e}veloppe sa convention de synchronisation dans le syst\`{e}me des deux
stations A et B suppos\'{e} entra\^{i}n\'{e} avec une vitesse relative par
rapport \`{a} l'\'{e}ther. La solution einsteinienne est imm\'{e}diate car 
\textbf{dans chaque syst\`{e}me de tout couple de syst\`{e}mes en
translation uniforme l'un par rapport \`{a} l'autre}, on doit avoir avec la
convention einsteinienne: le temps aller= le temps retour.

Si l'on veut d\'{e}finir de mani\`{e}re identique la simultan\'{e}it\'{e}
\`{a} distance de deux \'{e}v\'{e}nements respectivement dans K et dans K',
il faut \'{e}videmment disposer d'une longueur identique dans chaque
syst\`{e}me. C'est bien ce qu'Einstein pr\'{e}cise au paragraphe 2 du
m\^{e}me article:

\begin{quotation}
D'apr\`{e}s le principe de relativit\'{e}, la longueur \`{a} trouver par
l'op\'{e}ration a que nous voulons appeler la longueur de la tige dans le
syst\`{e}me en mouvement doit \^{e}tre identique \`{a} la longueur de la
tige au repos.
\end{quotation}

\bigskip Autrement dit, on a selon Einstein (voir citation compl\`{e}te
ci-dessous): 
\begin{equation}
AB=A^{\prime }B^{\prime }  \tag{3}
\end{equation}

Et donc des unit\'{e}s identiques d'espace et de temps dans chaque
syst\`{e}me (double normalisation \S 1.2). Il est clair que cela est
exactement \'{e}quivalent, l'\'{e}ther \'{e}tant supprim\'{e}, aux deux
sources identiques qui \'{e}mettent deux fronts sph\'{e}riques identiques
dans les deux syst\`{e}mes. Un front sph\'{e}rique \'{e}tant
caract\'{e}ris\'{e} par le temps aller = le temps retour, la
simultan\'{e}it\'{e} des \'{e}v\'{e}nements diam\'{e}tralement oppos\'{e}s
sur le front d\'{e}finit dans chaque syst\`{e}me une longueur identique.

\bigskip Comment Einstein articule maintenant la contraction des longueurs
avec sa convention de synchronisation (\S 1.3)? Lorsque nous \'{e}crivons
que la synchronisation des horloges avec des tiges identiques est PREALABLE
chez Einstein \`{a} toute utilisation de la TL, nous n'avons rien
invent\'{e} car cela est explicit\'{e} dans les moindres d\'{e}tails par
Einstein dans son paragraphe 2 (avant la d\'{e}duction de la TL, fin du
paragraphe 3). Il distingue deux op\'{e}rations a et b

\begin{quotation}
''Soit donn\'{e}e une tige rigide au repos, et supposons que sa longueur
soit $l_{0}$ d'apr\`{e}s la mesure effectu\'{e}e avec une r\`{e}gle
\'{e}galement au repos.

\textbf{Op\'{e}ration a}

L'observateur pourvu de la r\`{e}gle se trouve en mouvement ainsi que la
tige \`{a} mesurer; il mesure la longueur de celle-ci en y appliquant
directement la r\`{e}gle, exactement comme si la tige \`{a} mesurer,
l'observateur et la r\`{e}gle \'{e}taient au repos.

\textbf{Op\'{e}ration b }

L'observateur d\'{e}termine au moyen d'horloges synchrones (conform\'{e}ment
au paragraphe 1) qui sont au repos dans le syst\`{e}me au repos, avec quels
points du syst\`{e}me au repos co\"{i}ncident \`{a} une moment donn\'{e} les
deux points extr\^{e}mes de la tige \`{a} mesurer. La distance entre ces
deux points, mesur\'{e}e \`{a} l'aide de la tige d\'{e}j\`{a} utilis\'{e}e
et qui dans ce cas est au repos est \'{e}galement une longueur que l'on peut
appeler longueur de la tige.

D'apr\`{e}s le principe de relativit\'{e}, la longueur L \`{a} trouver par 
\textbf{l'op\'{e}ration a} que nous voulons appeler ''la longueur de la tige
(au repos) dans le syst\`{e}me en mouvement'' doit \^{e}tre \'{e}gale \`{a}
la longueur L de la tige au repos.

\bigskip Quant \`{a} la longueur \`{a} trouver par \textbf{l'op\'{e}ration b}%
, que nous voulons appeler ''longueur de la tige en mouvement dans le
syst\`{e}me au repos'', nous la d\'{e}terminerons en nous appuyant sur nos
deux principes et montrerons qu'elle est diff\'{e}rente de L.''
\end{quotation}

Une telle rigueur ne se retrouve dans aucun livre sur la th\'{e}orie de la
relativit\'{e}. Il est impossible de prendre en d\'{e}faut la logique du
jeune Einstein. Tout ce que nous pouvoins faire, c'est de montrer qu'il y a
une autre logique et d\'{e}couvrir \`{a} quoi elle aboutit.

L'\textbf{op\'{e}ration a }d\'{e}fini l'identit\'{e} des tiges (rigides)
dans chaque syst\`{e}me et l'\textbf{op\'{e}ration b} d\'{e}finit ''longueur
de la tige en mouvement dans le syst\`{e}me au repos'' en regardant ''avec
quels points du syst\`{e}me au repos co\"{i}ncident \`{a} une moment
donn\'{e} les deux points extr\^{e}mes de la tige \`{a} mesurer''. La
contraction einsteinienne est donc purement externe: c'est toujours quant
elle est observ\'{e}e dans le syst\`{e}me en mouvement que la tige
para\^{i}t contract\'{e}e; de surcroit elle est bas\'{e}e sur une double
simultan\'{e}it\'{e}: la longueur propre est d\'{e}finie par deux
\'{e}v\'{e}nements simultan\'{e}s dans le syst\`{e}me propre tandis que la
longueur impropre est d\'{e}finie par deux \'{e}v\'{e}nements simultan\'{e}s
dans le syst\`{e}me impropre (76bis). Encore une fois la logique
einsteinienne est implacable car dans l'interpr\'{e}tation de la contraction
(76bis) comme dans la convention de synchronisation, le m\^{e}me terme
elliptique poincar\'{e}en est annul\'{e}.

\bigskip On peut mettre en \'{e}vidence le contraste entre les deux
conventions de synchronisation en utilisant le c\'{e}l\`{e}bre param\`{e}tre 
$\epsilon $ de Reichenbach, g\'{e}n\'{e}ralement d\'{e}fini \`{a} partir des
notations einsteiniennes. On a d'apr\`{e}s (1)

\begin{equation}
t_{B}=\frac{1}{2}(t_{A}^{\ast }+t_{A})\qquad t_{B}=\epsilon (t_{A}^{\ast
}+t_{A})  \tag{4}
\end{equation}

avec 
\begin{equation}
0\leq \epsilon <1  \tag{4bis}
\end{equation}

Dans les deux syst\`{e}mes on a, dans la cin\'{e}matique einsteinienne: 
\begin{equation}
\epsilon =\frac{1}{2}  \tag{4ter}
\end{equation}

En posant $t_{A}=0,$ on a 
\begin{equation*}
t_{B}=\epsilon t_{A}^{\ast }=\frac{1}{2}t_{A}^{\ast }\qquad \tau
_{B}=\epsilon \tau _{A}^{\ast }=\frac{1}{2}\tau _{A}^{\ast }
\end{equation*}

Dans les notations poincariennes (\S 2.1) on a dans le syst\`{e}me K

\begin{equation*}
t^{+}=t_{B}\qquad t^{-}=t_{A}^{\ast }-t_{B}\qquad t_{M}=\frac{t_{A}^{\ast }}{%
2}
\end{equation*}

Selon Poincar\'{e}, on peut toujours choisir $\epsilon _{K^{\prime }}=\frac{1%
}{2}$ dans un syst\`{e}me (disons K' afin de faciliter la comparaison avec
la synchronisation en onde ellipsoidale, \S 2). Calculons alors $\epsilon $
dans l'autre syst\`{e}me K.

On a donc \'{e}tant donn\'{e} que $t_{M}=$ $\gamma t^{\prime }$ (\S 2)

\begin{eqnarray*}
t^{+} &=&2\epsilon t_{M}=2\gamma \epsilon t^{\prime }=\gamma t^{\prime
}(1+\beta \cos \theta ^{\prime }) \\
t^{-} &=&2(1-\epsilon )t_{M}=2\gamma (1-\epsilon )t^{\prime }=\gamma
t^{\prime }(1-\beta \cos \theta ^{\prime })
\end{eqnarray*}

Et donc on trouve une relation remarquable entre le param\`{e}tre de
Reichenbach $\epsilon $ et la vitesse relative $\beta $

\begin{equation}
\epsilon =\frac{1+\beta \cos \theta ^{\prime }}{2}  \tag{5}
\end{equation}

On retrouve ainsi la caract\'{e}ristique essentielle de la convention
einsteinienne $\epsilon =\frac{1}{2}\rightarrow $ $\beta =0.$ Autrement dit
les deux syst\`{e}mes einsteiniens sont synchronis\'{e}s \textit{comme si
ils \'{e}taient au repos }$\beta =0$ l'un par par rapport \`{a} l'autre
(l'\'{e}ther est supprim\'{e}). Pour $\theta ^{\prime }=\frac{\pi }{2}$, on
retrouve le seul cas o\`{u} $t^{+}=t^{-}$ (voir figure 13, voir \textbf{%
oscillateur d'Einstein}, 1913, voir \cite{5bis}). Calculons le terme
typiquement poincar\'{e}en $\Delta t$ en fonction du param\`{e}tre de
Reichenbach.

\begin{equation}
\Delta t=t_{M}(2\epsilon -1)=\beta t_{M}\cos \theta ^{\prime }  \tag{6}
\end{equation}

Ce terme $\Delta t$ est toujours nul avec la convention einsteinienne $%
t^{+}=t^{-}$ . \ On red\'{e}couvre que dans la logique einsteinienne de
synchronisation tout se passe comme si $\beta =0.$ La faiblesse de toutes
les approches que l'on trouve dans la litt\'{e}rature scientifique \`{a}
propos d'une convention alternative \`{a} celle d'Einstein $\epsilon =\frac{1%
}{2}$ appara\^{i}t alors clairement. Ces approches consistent \`{a}
introduire ce param\`{e}tre (ou un autre param\`{e}tre apparent\'{e}) DANS\
LA\ TL. Ce qui non seulement enlaidit consid\'{e}rablement cette
derni\`{e}re mais est en outre profond\'{e}ment nuisible \`{a} sa structure
groupale. En fait la TL n'impose nullement le choix einsteinien $\epsilon =%
\frac{1}{2}; $ elle est parfaitement compatible avec $0\leq \epsilon <1.$

\section{D\'{e}ductions de la formule Doppler relativiste de Poincar\'{e}}

\subsection{Sym\'{e}trisation par correction elliptique des deux formules
pr\'{e}relativistes}

Montrons que la formule de Poincar\'{e} (2.4) est la limite relativiste des 
\textit{deux} formules non-relativistes des ondes. On suppose connue la
d\'{e}duction classique des \textit{deux} formules Doppler avec des ondes
sph\'{e}riques (il faut une approximation sur la constance de l'angle $%
\theta ^{\prime }$ entre la r\'{e}ception par l'observateur (O' dans K') de 
\textit{deux} fronts d'onde successifs, voir onde plane, 5.2). On obtient
pour le cas o\`{u} la ''source est au repos'' par rapport \`{a} l'\'{e}ther
la formule non-relativiste $\nu ^{\prime }=\nu (1-\beta \cos \theta ^{\prime
})$ (avec l'angle $\theta ^{\prime }$ fait par ''l'observateur en
mouvement'' par rapport \`{a} la direction source-observateur) $\ $et pour
le cas o\`{u} la ''source est en mouvement'' par rapport \`{a} l'\'{e}ther
la formule non-relativiste $\nu ^{\prime }=\frac{\nu }{1+\beta \cos \theta }$
(avec l'angle $\theta $ fait par la source en mouvement par rapport \`{a} la
direction source-observateur)$.$

Dans le cas classique (pr\'{e}relativiste) les ondes sont toujours \textit{%
sph\'{e}riques} dans les deux r\'{e}f\'{e}rentiels (concentriques dans le
premier cas et non-concentriques dans le second).\ Il suffit d\`{e}s lors
d'introduire la c\textit{orrection relativiste} induite par le caract\`{e}re 
\textit{allong\'{e}} des fronts d'onde re\c{c}us par par l'observateur O'
(facteur $\gamma $ de dilatation pour $\lambda $ $et$ $T$) et donc aussi la
transformation relativiste de l'angle $\theta $ en $\theta ^{\prime }$
(1.11). On consid\`{e}re d'abord le premier cas ''standard'' (\textbf{figure
3}). En passant par (2.3) on retrouve alors directement $\nu ^{\prime
}=\gamma \nu (1-\beta \cos \theta ^{\prime })$. On constate ainsi que la
formule (2.4) est bien la limite relativiste de la formule non-relativiste 
\textit{avec l'angle }$\theta ^{\prime }$\textit{\ de l'observateur}. Le cas
non-standard se ram\`{e}ne imm\'{e}diatement par sym\'{e}trie
cin\'{e}matique (\textbf{figure 4}) au cas standard $v^{\prime }=\frac{\nu }{%
\gamma (1+\beta \cos \theta )}=\gamma \nu (1-\beta \cos \theta ^{\prime }).$

En r\'{e}sum\'{e}, dans un traitement non-relativiste de l'effet Doppler, il
faut distinguer (\textit{deux} formules diff\'{e}rentes!) le cas ''source au
repos par rapport \`{a} l'\'{e}ther'' et le cas ''source en mouvement par
rapport \`{a} l'\'{e}ther''. Ce dernier cas est \textit{sym\'{e}tris\'{e}}
par Poincar\'{e} avec (2.4) laquelle formule est d\`{e}s lors la limite
relativiste des \textit{deux }formules classiques tandis qu'il est \textit{%
supprim\'{e}} par\ Einstein-Minkowski avec (2.5) laquelle formule ne peut
d\`{e}s lors \^{e}tre que la limite relativiste d'\textit{une seule }formule
classique, le premier cas standard.

Remarquons \`{a} cet \'{e}gard que la correction elliptique fonctionne aussi
bien chez Poincar\'{e} pour la formule Doppler des fr\'{e}quences que celle
des longueurs (d'onde). En effet pour un redshift il faut \'{e}videmment une
longueur d'onde dilat\'{e}e. Tel n'est pas le cas des corrections
einsteiniennes qui fonctionnent avec la dilatation du temps \textit{mais pas}
avec la \textbf{d\'{e}finition einsteinienne de la contraction} des
longueurs. En effet \ si la longueur d'onde \'{e}tait une longueur comme les
autres, il faudrait alors introduire un facteur de contraction dans la
formule pr\'{e}relativiste, et on n'obtiendrait pas la formule Doppler
ad\'{e}quate. On rep\`{e}re ainsi une nouvelle asym\'{e}trie espace-temps
dans le traitement einsteinien du ph\'{e}nom\`{e}ne Doppler (voir
synth\`{e}se \S 1.4, relation \textbf{4.1 \& 4.2}).

\subsection{Couplage avec la formule d'aberration stellaire (2\`{e}me
d\'{e}duction)}

Il est bien connu qu'Einstein ''couple'' la d\'{e}duction de la formule
doppler et de la formule d'aberration \`{a} partir de l'invariance de la
phased'une onde plane. Montrons que la formule Doppler de Poincar\'{e} est
directement d\'{e}ductible des formules de transformation relativiste de la
vitesse.

Traitons le probl\`{e}me \`{a} deux dimensions spatiales. On a la
transformation de Lorentz

\begin{equation}
x^{\prime }=\gamma (x+\beta t)\qquad \qquad y^{\prime }=y\qquad \qquad
t^{\prime }=\gamma (t+\beta x)  \tag{1, 1bis, 1ter}
\end{equation}
et imm\'{e}diatement la transformation diff\'{e}rentielle de Lorentz: 
\begin{equation}
dx^{\prime }=\gamma (dx+\beta dt)\qquad \qquad dy^{\prime }=dy\qquad \qquad
dt^{\prime }=\gamma (dt+\beta dx)  \tag{1, 1bis, 1ter}
\end{equation}

Si on divise les deux premi\`{e}res relations par la troisi\`{e}me, on
obtient la loi de composition relativiste des vitesses

\begin{equation}
\text{v}_{x}^{\prime }=\frac{\text{v}_{x}+\beta }{1+\beta \text{v}_{x}}\text{
\ \ \ \ v}_{y}^{\prime }=\frac{1}{\gamma (1+\beta \text{v}_{x})}\text{v}_{y}
\tag{2, 2bis}
\end{equation}

Si la vitesse v' fait un angle $\theta ^{\prime }$ avec l'axe Ox', et donc v
fait un angle $\theta $ avec Ox on a:

\begin{equation}
\text{v}^{\prime }\cos \theta ^{\prime }=\frac{\text{v}\cos \theta +\beta }{%
1+\beta \text{v}^{\prime }\cos \theta ^{\prime }}\text{ \ \ \ \ v}^{\prime
}\sin \theta ^{\prime }=\frac{1}{\gamma (1+\beta \text{v}\cos \theta )}\text{%
v}\sin \theta  \tag{3, 3bis}
\end{equation}

Maintenant consid\'{e}rons v$^{\prime }=$ v $=c=1$, on a

\begin{equation}
\cos \theta ^{\prime }=\frac{\cos \theta +\beta }{1+\beta \cos \theta }\text{
\ \ \ \ }\sin \theta ^{\prime }=\frac{1}{\gamma (1+\beta \cos \theta )}\sin
\theta  \tag{4, 4bis}
\end{equation}

On retrouve ainsi les formules relativistes de changement de l'angle
(aberration ou effet ''headlight'') directement d\'{e}duites de l'ellipse
allong\'{e}e de Poincar\'{e}. Mais on a oubli\'{e} la troisi\`{e}me
\'{e}quation dans la TL (\textbf{1ter}).

Ecrivons la en fonction de l'angle (\textbf{5ter}): 
\begin{equation}
dr\cos \theta =\gamma (dr^{\prime }\cos \theta ^{\prime }+\beta dt^{\prime
})\qquad dr\sin \theta =dr^{\prime }\sin \theta ^{\prime }\qquad dt=\gamma
(dt^{\prime }+\beta dr^{\prime }\cos \theta ^{\prime })  \tag{5, 5bis, 5ter}
\end{equation}

On a donc pour un point mat\'{e}riel le rapport des temps de parcours d'un
point mat\'{e}riel (5ter)

\begin{equation}
\frac{dt^{\prime }}{dt}=\gamma (1+\beta \text{v}\cos \theta )  \tag{7ter}
\end{equation}

Pour un point du front d'onde lumineux avec $v=c=1$.

\begin{equation*}
\frac{dt^{\prime }}{dt}=\gamma (1+\beta \cos \theta )
\end{equation*}

Ce sont les \'{e}quations (13bis) de l'ellipse isotrope. On obtient ainsi un
rapport de temps de parcours $dt^{\prime }$ observ\'{e} dans K' en fonction
du temps de parcours $dt$ d'un point d'un front d'onde d\'{e}fini dans K. 
\textbf{C'est cette diff\'{e}rence entre les temps de parcours dans K et K'
qui d\'{e}finit le ph\'{e}nom\`{e}ne Doppler}: On a alors la variation des
fr\'{e}quences d\'{e}finies classiquement comme l'inverse du temps de
parcours ($\nu =\frac{1}{t},$ $\nu ^{\prime }=\frac{1}{t^{\prime }}):$

\begin{equation}
\frac{\nu ^{\prime }}{\nu }=\frac{1}{\gamma (1+\beta \cos \theta )} 
\tag{8ter}
\end{equation}

Autrement dit en utilisant (4) la formule Doppler de Poincar\'{e} (19,
texte) induite directement du quadrivecteur d'onde de Poincar\'{e}

\begin{equation}
\nu ^{\prime }=\gamma \nu (1-\beta \cos \theta ^{\prime })  \tag{5}
\end{equation}

La formule de Poincar\'{e} est donc structurellement inscrite dans la TL et
ne n\'{e}cessite aucune hypoth\`{e}se suppl\'{e}mentaire (une d\'{e}finition
de la fr\'{e}quence ind\'{e}pendante des temps de parcours (7ter).

\section{Exp\'{e}rience de Sagnac et ''Ecart \`{a} la convention de
synchronisation standard''}

Nous avons montr\'{e} que l'ellipse dans le syst\`{e}me de source donnait
une explication imm\'{e}diate du r\'{e}sultat nul de l'exp\'{e}rience de
Michelson. Nous voulons maintenant montrer que la m\^{e}me ellipse donne une
explication non moins imm\'{e}diate de l'exp\'{e}rience de Sagnac et cela
sans faire intervenir la gravitation (ou la Relativit\'{e}
G\'{e}n\'{e}rale). Contrairement \`{a} l'exp\'{e}rience de Michelson qui est
bas\'{e}e sur la comparaison de deux trajets aller-retour dans deux
directions diff\'{e}rentes, l'exp\'{e}rience de Sagnac (1913) permet de
comparer le temps de parcours de la lumi\`{e}re dans deux sens de rotation
oppos\'{e}s d'un plateau tournant.\ Autrement dit cette exp\'{e}rience
permet de comparer un temps aller simple t$^{+}$ avec un temps retour simple
t$^{-}$ et pourrait ainsi permettre de tester le terme poincar\'{e}en non
nul $\Delta t.$ Citons Langevin qui demeure la meilleure r\'{e}f\'{e}rence
en la mati\`{e}re:

\begin{quotation}
\bigskip On sait que M Sagnac fait interf\'{e}rer deux rayons lumineux issus
d'une m\^{e}me source apr\`{e}s leur avoir fait parcourir, gr\^{a}ce \`{a}
des miroirs convenablement plac\'{e}s, un m\^{e}me circuit ferm\'{e} dans
des \textbf{sens oppos\'{e}s}. Il constate que la mise en rotation avec la
vitesse angulaire $\omega $ de la plateforme \textbf{qui porte l'ensemble du
syst\`{e}me optique} produit un d\'{e}placement des franges qui correspond
\`{a} une diff\'{e}rence $\frac{4\omega A}{c^{2}}$ entre les dur\'{e}es de
parcours du m\^{e}me circuit dans les deux sens ($t^{+}-t^{-})$, A
repr\'{e}sentant l'aire int\'{e}rieure au circuit projet\'{e} sur un plan
normal \`{a} l'axe de rotation et c la vitesse de la lumi\`{e}re.
\end{quotation}

\bigskip L'ensemble du syst\`{e}me optique est sur le plateau tournant, cela
signifie que la source lumineuse et la plaque photographique sont
entra\^{i}n\'{e}s (par rapport au syst\`{e}me K') \`{a} une vitesse
constante en intensit\'{e} (mais pas en direction) dans le m\^{e}me
syst\`{e}me K. Sagnac a en fait op\'{e}r\'{e} un calcul purement classique
exactement comme l'\'{e}quation au premier ordre \'{e}crite par Poincar\'{e}
(partie II, 5.2)

\begin{equation}
\Delta t=\frac{t^{+}-t^{-}}{2}=\frac{1}{2}(\frac{2\pi R}{c+v}-\frac{2\pi R}{%
c-v})=\frac{1}{2}2\pi R\text{ }\frac{2v}{c^{2}-v^{2}}\approx \frac{vR}{c^{2}}
\tag{5.2}
\end{equation}

Il est clair que avec $\omega =2\pi \nu $ et $v=R\omega =2\pi R\nu ,$ on
retrouve imm\'{e}diatement le r\'{e}sultat de Sagnac pr\'{e}sent\'{e} par
Langevin (avec $A=$ $\pi R^{2})$

\begin{equation}
t^{+}-t^{-}\approx \frac{4\pi \omega R^{2}}{c^{2}}\approx \frac{4\omega A}{%
c^{2}}  \tag{5.2bis}
\end{equation}

\bigskip Langevin argumente ensuite de la mani\`{e}re suivante:

\begin{quotation}
Remarquons tout d'abord qu'il s'agit d'une exp\'{e}rience du premier
ordre... On ne saurait donc, \`{a} aucun point de vue, comparer cette
exp\'{e}rience \`{a} celle de M. Michelson. Celle-ci est du second ordre en
fonction de la vitesse de translation et son importance tient \`{a} ce
qu'elle est venue mettre en \'{e}vidence de mani\`{e}re aig\"{u}e la
n\'{e}cessit\'{e} d'introduire un cin\'{e}matique nouvelle...
\end{quotation}

Or nous avons montr\'{e} qu'on obtenait, en introduisant une contraction
longitudinale (ici dans le plateau), un \textit{r\'{e}sultat exact au second
ordre} avec 
\begin{equation}
t^{+}-t^{-}=\gamma \frac{4\omega A}{c^{2}}  \tag{5.2ter}
\end{equation}

L'exp\'{e}rience de Sagnac acquiert ainsi le statut d'exp\'{e}rience
fondamentale (du 2\`{e}me ordre) dans la perspective de Poincar\'{e}. Dans
le cas pr\'{e}sent avec les vitesses petites du plateau $\gamma \rightarrow
1.$ Le raisonnement relativiste de Poincar\'{e} (voir \S 9) \'{e}pouse
enti\`{e}rement le calcul classique et en d\'{e}composant le mouvement de
rotation en un s\'{e}rie de petits mouvements de translation $dx$ \`{a}
vitesse constante $v,$ il suffit alors d'int\'{e}grer 
\begin{equation}
\Delta t=\int dt=\int_{0}^{2\pi R}\gamma \frac{v}{c^{2}}dx.\rightarrow
\Delta t=\gamma \frac{2\pi vR}{c^{2}}  \tag{5.2quater}
\end{equation}
sur l'ensemble du contour (comme pour le calcul classique) pour obtenir le
r\'{e}sultat observ\'{e} exp\'{e}rimentalement..

\bigskip Sans vouloir entrer dans les d\'{e}tails de la d\'{e}monstration de
Langevin, contentons-nous de faire remarquer qu'il est contraint de faire
intervenir la gravitation (la relativit\'{e} g\'{e}n\'{e}rale). On ne peut
obtenir en effet le r\'{e}sultat correct avec la cin\'{e}matique d'Einstein
car l'on a en vertu m\^{e}me de la d\'{e}finition du temps propre, le temps
aller= le temps retour instantan\'{e}ment dans chaque syst\`{e}me inertiel.
On a en fait des sph\`{e}res infinit\'{e}simales \`{a} chaque instant si on
adopte le point de vue d'Einstein-Minkowski. La cin\'{e}matique
einsteinienne donnnant un r\'{e}sultat nul $\Delta t=0$, il faut donc passer
\`{a} la ''G\'{e}n\'{e}rale'' pour obtenir un r\'{e}sultat compatible avec
les exp\'{e}rience de Sagnac. La seule diff\'{e}rence entre la sph\`{e}re
dans le syst\`{e}me de la source et l'ellipso\"{i}de dans le syst\`{e}me de
la source se situe au niveau du terme $\Delta t$ (puisque l'ellipse est
isotrope c=1 et au niveau des temps aller-retour). L'exp\'{e}rience de
Sagnac montre donc clairement que l'ellipse de Poincar\'{e} DANS LE SYSTEME
DE LA SOURCE peut avoir un sens physique.

Quant \`{a} l'argument selon lequel on ne peut pas traiter ce probl\`{e}me
(o\`{u} un des syst\`{e}mes est acc\'{e}l\'{e}r\'{e}) dans le cadre de la
relativit\'{e} restreinte, il n'a pas plus de pertinence que celui selon
lequel on ne peut pas traiter les jumeaux de Langevin (o\`{u} un des jumeaux
est acc\'{e}l\'{e}r\'{e}) dans le cadre de la relativit\'{e} restreinte. Si
tel \'{e}tait le cas on ne pourrait plus tracer dans les diagrammes de
Minkowki que des droites et on ne pourrait plus fonder une dynamique
relativiste sur la base d'une cin\'{e}matique relativiste. N'oublions pas
que Minkowski d\'{e}finit le temps propre comme un invariant quel que soit
le mouvement corps consid\'{e}r\'{e} (ici un plateau tournant) par rapport
\`{a} un syst\`{e}me inertiel donn\'{e} (ici le laboratoire).

\section{Directrices et expressions rationnelles des rayons focaux}

\bigskip Quelques semaines avant sa mort, Poincar\'{e} dessine pour la
premi\`{e}re fois une des directrices de l'ellipse \cite{24}. Montrons que
la directrice est en rapport avec le th\'{e}or\`{e}me de Poincar\'{e} (51).
Les deux termes de Poincar\'{e} (somme et diff\'{e}rence) 
\begin{equation*}
\frac{r^{+}+r^{-}}{2}=\gamma \qquad \frac{r^{+}-r^{-}}{2}=\beta x_{C}
\end{equation*}

sont respectivement li\'{e}s aux \textbf{foyers} et aux \textbf{directrices }%
de l'ellipse. En effet l'\textit{expression rationnelle des rayons focaux de
l'ellips}e s'\'{e}crit

\begin{equation*}
r^{+}=\gamma +\beta x_{C}\qquad r^{-}=\gamma -\beta x_{C}
\end{equation*}

\bigskip o\`{u} l'abscisse $x_{C}$ est d\'{e}finie par rapport au centre de
l'ellipse (autrement dit par rapport \`{a} l'observateur O', consid\'{e}rons
l'ellipse de Poincar\'{e} proprement dite, le cas non-standard) d'o\`{u}
l'on extrait imm\'{e}diatement les deux directrices donn\'{e}es par 
\begin{equation*}
x_{DC}=\pm \frac{\gamma }{\beta }
\end{equation*}

On a donc en transposant au niveau des temps

\begin{equation*}
\Delta r=\Delta t=\frac{t^{+}-t^{-}}{2}=\beta x_{C}=\gamma \beta x^{\prime }
\end{equation*}

On peut donc conclure que 
\begin{equation*}
x_{C}=\gamma x^{\prime }
\end{equation*}

En outre on doit faire un changement de coordonn\'{e}es puisque les abscisses%
$\ x^{\pm }$ spnt d\'{e}finies par rapport au p\^{o}le, le foyer de
l'ellipse qui est \`{a} la distance $\gamma \beta $ du centre .\ On a donc 
\begin{equation*}
x^{\pm }\pm \gamma \beta =x_{C}
\end{equation*}
On a alors 
\begin{equation*}
t^{+}=\gamma +\beta (x-\gamma \beta )\qquad t^{-}=\gamma -\beta (x+\gamma
\beta )
\end{equation*}

\begin{equation*}
t^{+}=\gamma (1-\beta ^{2})+\beta x^{+}\qquad t^{-}=\gamma (1-\beta
^{2})-\beta x^{-}
\end{equation*}

et finalement

\begin{equation*}
t^{+}=\gamma ^{-1}+\beta x^{+}\qquad t^{-}=\gamma ^{-1}-\beta x^{-}
\end{equation*}

Ce sont les \'{e}quations \'{e}crites par Poincar\'{e} (avec t'=1)

\begin{equation*}
\Delta r=\Delta t=\frac{t^{+}-t^{-}}{2}=\beta x_{M}
\end{equation*}

avec

\begin{equation*}
x_{M}=\frac{x^{+}+x^{-}}{2}=\gamma x^{\prime }=x_{C}
\end{equation*}

On retrouve ainsi l'oscillateur einsteinien purement transversal $x_{C}$
(1913) discut\'{e} dans l'ouvrage remarquable de Galison \cite{5bis}.

\end{document}